\documentclass{bmcart}
\usepackage[pdftex]{graphicx}
\usepackage{float}
\usepackage{tikz}
\usepackage{listings}
\usepackage{amsmath}
\usepackage{bm}
\usepackage{booktabs}
\usepackage{subfigure}
\usepackage{amssymb}
\RequirePackage{hyperref}
\usepackage[utf8]{inputenc} %unicode support
\startlocaldefs

\endlocaldefs

\begin{document}

%%% Start of article front matter
\begin{frontmatter}

\begin{fmbox}
\dochead{Research}


\title{Modeling Human Mobility considering Spatial, Temporal and Social Dimensions}

%%%%%%%%%%%%%%%%%%%%%%%%%%%%%%%%%%%%%%%%%%%%%%
%%                                          %%
%% Enter the authors here                   %%
%%                                          %%
%% Specify information, if available,       %%
%% in the form:                             %%
%%   <key>={<id1>,<id2>}                    %%
%%   <key>=                                 %%
%% Comment or delete the keys which are     %%
%% not used. Repeat \author command as much %%
%% as required.                             %%
%%                                          %%
%%%%%%%%%%%%%%%%%%%%%%%%%%%%%%%%%%%%%%%%%%%%%%

\author[
   addressref={aff1},                   % id's of addresses, e.g. {aff1,aff2}
    %corref={},                      % id of corresponding address, if any
   noteref={},                        % id's of article notes, if any
   email={giuliano.cornacchia@gmail.com}   % email address
]{\inits{GC}\fnm{Giuliano} \snm{Cornacchia}}
\author[
   addressref={aff2},  
   email={}
]{\inits{GR}\fnm{Giulio} \snm{Rossetti}}
\author[
   addressref={aff2},
   corref={aff2},
   email={luca.pappalardo@isti.cnr.it}
]{\inits{LP}\fnm{Luca} \snm{Pappalardo}}

%%%%%%%%%%%%%%%%%%%%%%%%%%%%%%%%%%%%%%%%%%%%%%
%%                                          %%
%% Enter the authors' addresses here        %%
%%                                          %%
%% Repeat \address commands as much as      %%
%% required.                                %%
%%                                          %%
%%%%%%%%%%%%%%%%%%%%%%%%%%%%%%%%%%%%%%%%%%%%%%

\address[id=aff1]{%                           % unique id
  \orgname{University of Pisa}, % university, etc
%  \street{Waterloo Road},                     %
  %\postcode{}                                % post or zip code
  \city{Pisa},                              % city
  \cny{Italy}                                    % country
}
\address[id=aff2]{%
  \orgname{ISTI-CNR},
%  \street{D\"{u}sternbrooker Weg 20},
%  \postcode{24105}
  \city{Pisa},
  \cny{Italy}
}

%%%%%%%%%%%%%%%%%%%%%%%%%%%%%%%%%%%%%%%%%%%%%%
%%                                          %%
%% Enter short notes here                   %%
%%                                          %%
%% Short notes will be after addresses      %%
%% on first page.                           %%
%%                                          %%
%%%%%%%%%%%%%%%%%%%%%%%%%%%%%%%%%%%%%%%%%%%%%%

\begin{artnotes}
%\note{Sample of title note}     % note to the article
%\note[id=aff1]{Equal contributor} % note, connected to author
\end{artnotes}

\end{fmbox}% comment this for two column layout

%%%%%%%%%%%%%%%%%%%%%%%%%%%%%%%%%%%%%%%%%%%%%%
%%                                          %%
%% The Abstract begins here                 %%
%%                                          %%
%% Please refer to the Instructions for     %%
%% authors on http://www.biomedcentral.com  %%
%% and include the section headings         %%
%% accordingly for your article type.       %%
%%                                          %%
%%%%%%%%%%%%%%%%%%%%%%%%%%%%%%%%%%%%%%%%%%%%%%

\begin{abstractbox}

\begin{abstract} % abstract
Modelling human mobility is crucial in several areas, from urban planning to epidemic modelling, traffic forecasting, and what-if analysis.
On the one hand, existing models focus mainly on reproducing the spatial and temporal dimensions of human mobility, while the social aspect, though it influences human movements significantly, is often neglected.
On other hand, those models that capture some social aspects of human mobility have trivial and unrealistic spatial and temporal mechanisms.
In this paper, we propose STS-EPR, a modelling framework that embeds mechanisms to capture the spatial, temporal and social aspects together. 
Our experiments show that STS-EPR outperforms existing spatial-temporal or social models on a set of standard mobility metrics, and that it can be used with limited amount of information without any significant loss of realism. 
STS-EPR, which is open-source and tested on open data, is a step towards the design of a mechanistic models that can capture all the aspects of human mobility in a comprehensive way.

%\parttitle{First part title} %if any
%Text for this section.

%\parttitle{Second part title} %if any
%Text for this section.
\end{abstract}

%%%%%%%%%%%%%%%%%%%%%%%%%%%%%%%%%%%%%%%%%%%%%%
%%                                          %%
%% The keywords begin here                  %%
%%                                          %%
%% Put each keyword in separate \kwd{}.     %%
%%                                          %%
%%%%%%%%%%%%%%%%%%%%%%%%%%%%%%%%%%%%%%%%%%%%%%


% MSC classifications codes, if any
%\begin{keyword}[class=AMS]
%\kwd[Primary ]{}
%\kwd{}
%\kwd[; secondary ]{}
%\end{keyword}

\end{abstractbox}
%
%\end{fmbox}% uncomment this for twcolumn layout

\end{frontmatter}

%%%%%%%%%%%%%%%%%%%%%%%%%%%%%%%%%%%%%%%%%%%%%%
%%                                          %%
%% The Main Body begins here                %%
%%                                          %%
%% Please refer to the instructions for     %%
%% authors on:                              %%
%% http://www.biomedcentral.com/info/authors%%
%% and include the section headings         %%
%% accordingly for your article type.       %%
%%                                          %%
%% See the Results and Discussion section   %%
%% for details on how to create sub-sections%%
%%                                          %%
%% use \cite{...} to cite references        %%
%%  \cite{koon} and                         %%
%%  \cite{oreg,khar,zvai,xjon,schn,pond}    %%
%%  \nocite{smith,marg,hunn,advi,koha,mouse}%%
%%                                          %%
%%%%%%%%%%%%%%%%%%%%%%%%%%%%%%%%%%%%%%%%%%%%%%

%%%%%%%%%%%%%%%%%%%%%%%%% start of article main body
% <put your article body there>

%%%%%%%%%%%%%%%%
%% Background %%
%%

%%%%%%%%%%%%%%%%%%%%%%%%%%%%%%%%%%%%%%%%%%%%%%
%%              INTRODUCTION                %%
%%%%%%%%%%%%%%%%%%%%%%%%%%%%%%%%%%%%%%%%%%%%%%
\section{Introduction}
Modeling the mechanisms that govern human mobility is of fundamental importance in different disciplines, such as computational epidemiology, traffic forecasting, urban planning, and what-if analysis \cite{paper_1,ditras,im_model_song,gonzalez2008understanding, pappalardo2019human, cintia2020relationship}.
First of all, mobility data sets contain sensitive information about the individuals whose movements are described \cite{montjoye1,montjoye2, pellungrini2017data}, making them expensive and difficult to obtain from providers, and rarely available publicly for scientific research.
Individual mobility models, often referred to as generative algorithms of human mobility \cite{barbosa2018human, pappalardo2019human, karamshuk2011human}, can reproduce synthetic trajectories that are realistic in reproducing fundamental patterns of individual human mobility. 
%These synthetic trajectories can be used in what-if analysis or for training a machine learning model without the need of real data.
A significant advantage of using generative models concerns the cost and the time spent in the data collection, which is negligible with respect to the acquisition of a real dataset.
Moreover, using generative algorithms allows the simulation of the mobility for a set of agents in an unseen scenario, hence allowing complex what-if analysis.

%In last decade, the availability of large mobility datasets such as Call Detail Records (CDRs) \cite{gonzalez2008understanding,im_model_song,ditras}, traces from GPS devices embedded in smartphones and cars \cite{ditras}, and geo-tagged posts on Location Based Social Networks (LBSN) \cite{datasetFS}, allows to observe and study human movements in great detail and quantify the patterns that characterize the individual movements.
%These studies show the heterogeneous nature of travel patterns, the existence of a power-law distribution in jump lengths, namely the distance between the source and destination location of a displacement \cite{brockmann,gonzalez2008understanding}, and in the characteristic spatial spread of an individual, referred to as radius of gyration \cite{gonzalez2008understanding}.
%Humans exhibit a strong tendency to return to locations they visited before \cite{gonzalez2008understanding} and a propensity to be stationary during the night hours while they preferentially move at specific times of the day, following a circadian rhythm \cite{ditras,opp_net}.
%The time spent by an individual in a location is not uniformly distributed, but follows a power-law distribution \cite{im_model_song}.
%Moreover, the sociality shapes the displacements of the individuals, about the $10$-$30\%$ of the human movements can be explained by social purposes \cite{cho}.

Most individual models focus on capturing the spatial and temporal patterns of human mobility,
such as the existence of a power-law distribution in jump lengths \cite{brockmann,gonzalez2008understanding, pappalardo2013understanding} and in the characteristic spatial spread of an individual \cite{gonzalez2008understanding}, a strong tendency to return to locations they visited before \cite{gonzalez2008understanding} and a propensity to move following a circadian rhythm \cite{ditras,opp_net}.
However, the social dimension of human mobility is often neglected, despite the fact that  about $10$-$30\%$ of human movements can be explained by social purposes \cite{cho}. 
The only model that takes into consideration the social dimension is GeoSim \cite{geosim}, which contemplates a mechanism related to the individual preference and social influence. 
Unfortunately, the temporal and spatial mechanisms of GeoSim are not realistic, making the model incomplete and hardly usable in practice. 

In this paper, we propose STS-EPR, which combines the most realistic spatial, temporal and social mechanisms into one modelling framework.
Namely, STS-EPR includes a mechanism that takes into account the spatial distance between locations as well as the relevance of a location \cite{im_model_song, ret_exp_pappalardo}.
Second, STS-EPR includes a temporal mechanism based on a diary generator, a data-driven algorithm able to capture the tendency of individuals to follow a circadian rhythm \cite{ditras}.
Finally, our model includes social mechanisms inspired by GeoSim in order to model the social dimensions of mobility, too \cite{geosim}. Finally, we include novel mechanisms related to social popularity and constraints in the location an agent can reach in a given time.
We provide the code to reproduce our model and conduct experiments on an open data sets, hence making our work fully replicable and reproducible.

Our experiments on data describing the checkins of 1000 users in New York City show that the generated trajectories are realistic with respect to the social, temporal and mobility aspects, outperforming existing models on a set of standard mobility metrics.
We further validate the modeling ability of STS-EPR, simulating  the mobility of individuals moving in London, including in the model different levels of knowledge concerning their mobility behaviors, demonstrating that our model can be applied on different regions and with limited amount of information about that region without any significant loss of realism.

Our model is a further step towards the design of a three-dimensional generative models for human mobility, which combines the state-of-the-art spatial, temporal and social mechanisms with action-correction mechanisms specifying with borderline cases during the simulation.
\section{Related Work}\label{section:related}

\subsection{Human mobility patterns}
The study of human mobility focuses on discovering and modeling the mechanisms that rule the movements of individuals and groups of individuals. In the last decade, researchers from several disciplines showed that that human mobility, far from being random, follows well-defined statistical laws \cite{barbosa2018human, pappalardo2019human}.

The seminal work by Brockmann et al. \cite{brockmann} analyzes nation-wide trajectories of dollar bills and finds that the distribution of the distance between two consecutive positions of a banknote follows a power-law. Subsequent studies confirm this finding on nation-wide trajectories of mobile-phone users \cite{gonzalez2008understanding} and region-wide trajectories of private vehicles \cite{pappalardo2013understanding, ret_exp_pappalardo}. 
González et al. \cite{gonzalez2008understanding} find that, on nation-wide mobile phone data, the empirical distribution of the radius of gyration, the characteristic distance traveled by an individual, can be approximated with a truncated power-law; a statistical law confirmed on region-wide trajectories of private vehicles \cite{pappalardo2013understanding, ret_exp_pappalardo}.

Song et al. \cite{im_model_song} find that the distribution of time between two displacements of an individual (waiting time), can be described by a truncated power-law and that the potential predictability of an individual's mobility is high \cite{song_limits}. Other laws regard the existence of the returners and explorers dichotomy \cite{ret_exp_pappalardo} and the conservation of the number of locations visited by an individual in a period of time \cite{memory_epr}.

Another strand of research demonstrates a strong relationship between human mobility and social ties \cite{geosim,link_prediction,cho,chao,datasetFS}. 
First of all, friends have a higher probability of living or working together or having the same hobbies, increasing their mobility similarity compared with strangers \cite{chao}.
Moreover, individuals are more likely to visit a location right after a friend has visited the same location, and the probability drops off following a power-law function \cite{cho}. 
Cho et al. \cite{cho} show that the probability for an individual to visit a friend remains constant as a function of the distance; Wang et al. \cite{link_prediction} find that individuals with a similar visitation pattern are more likely to establish a social link.

\subsection{Generative models}
Building upon the above findings, many models have been proposed which try to reproduce the statistical laws of human mobility.
For this paper's sake, we are interested in \emph{mechanistic} generative models of human mobility \cite{barbosa2018human}.

The seminal paper by Song et al. \cite{song_limits}  proposes the exploration and preferential return (EPR) model. 
EPR relies on two mechanisms: exploration and preferential return.
The exploration mechanism is a random walk process with truncated power-law jump size distribution. 
The preferential return mechanism reproduces the propensity of humans to return to locations they visited before. 
If an agent returns to a previously visited location, it selects the location to visit with probability proportional to the number of times the agent visited that location.
An agent in the model selects to explore a new location with probability $P_{exp}$, which decreases as the agent visits more and more locations. With complementary probability $P_{ret}=1-P_{exp}$, the agent returns to a previously visited location. 

Several studies subsequently improved the EPR model by adding increasingly sophisticated mechanisms to reproduce statistical laws more realistically.
In the d-EPR model \cite{d_epr}, an agent visits a new location depending on both its distance from the current position and collective relevance.
In the recency-EPR model \cite{barbosa2015effect}, the preferential return phase includes information about the recency of location visits.
In the memory-EPR model \cite{memory_epr}, during the exploration mechanism, the agent selects a location with probability proportional to the number of times it visited that location in the previous $M$ days.
EPR and its extension focus on the spatial aspect of human mobility, neglecting to reproduce realistic temporal patterns. For example, the displacements of individuals are not uniformly distributed during the day but follow the circadian rhythm, a property that is not captured by EPR-like models.
Two refined models, namely TimeGeo \cite{timegeo} and DITRAS \cite{ditras}, overcome this problem by including a more sophisticated temporal mechanism.

TimeGeo \cite{timegeo} is a mechanistic modeling framework to generate individual mobility trajectories with realistic spatio-temporal properties. TimeGeo models the temporal dimension using a time-inhomogeneous Markov chain that captures the circadian propensity to travel and the likelihood of arranging short and consecutive activities \cite{timegeo}.
It integrates the temporal mechanism with a rank-based version of the EPR model ($r$-EPR), which assigns a rank to each unvisited location during the selection of a new location to visit, depending on its distance from the trip origin \cite{timegeo}.

DITRAS (DIary-based TRAjectory Simulator) \cite{ditras} generates the trajectories using two probabilistic models: a diary generator and a trajectory generator.
The diary generator is a Markov model trained on mobility trajectory data of real individuals, which captures the probability for individuals to follow or break their routine at specific times \cite{ditras}; the diary generator builds a mobility diary with abstract locations for each agent in the simulation.
The trajectory generator is an algorithm that, given a weighted spatial tessellation, translates the abstract locations in physical locations using the d-EPR model \cite{im_model_song}.

Despite the link between human mobility and social ties, the only mechanistic model that tries to reproduce the socio-mobility patterns is GeoSim \cite{geosim}. GeoSim takes into account both the mobility and the social dimension, although incorporating a trivial temporal mechanism. 
GeoSim introduces two mechanisms in addition to the explore and preferential return ones: individual preference and social influence.
The agent has to decide if its next displacements are influenced or not by its social contact, respectively, with probability $\alpha$ and $1-\alpha$.

\paragraph{Position of our work.} An overview of the literature cannot avoid noticing the lack of generative models able to reproduce the spatial, temporal, and social dimensions at the same time.
On the one hand, GeoSim can capture important patterns describing the link between mobility and sociality, but cannot reproduce realistic spatio-temporal patterns.  
On the other hand, TimeGeo and DITRAS well reproduce spatial and temporal patterns but neglect the social dimension. 
In this paper, we build the STS-EPR model combining the mechanisms of existing mechanistic models to reproduce the three dimensions of human mobility.

%%%%%%%%%%%%%%%%%%%%%%%%%%%%%%%%%%%%%%%%%%%%%%
%%       Modeling Individual Mobility       %%
%%%%%%%%%%%%%%%%%%%%%%%%%%%%%%%%%%%%%%%%%%%%%%
\section{Modeling Spatial, Temporal and Social patterns}
\label{section:model}

We define a mobility trajectory as a sequence $T =\langle(r_1,t_1),\dots,(r_n,t_n)\rangle$ where $t_i$ is a timestamp such that $\forall i \in [1,n) \; t_i < t_{i+1}$ and $r_i$ is defined as $(x_i, y_i)$ where the components are coordinates on a bi-dimensional space. 
In particular, we assume that individuals move on a weighted spatial tessellation $L$, representing the tiling of a bi-dimensional space, resulting in a non-overlapped set of locations.
Every location has a weight corresponding to its relevance at a global level \cite{ditras}, and it has a representative point; generally, the centroid of the tile expressed as a pair of coordinates.
$L =\langle(r_1,w_1),\dots,(r_n,w_n)\rangle$
where $w_j$ is the weight of the tile $j$ and $r_j$ is the representative point of the tile $j$.
We represent the visitation pattern of an individual $a$ as a vector $lv_a$ of $|L|$ elements, called location vector, where $|L|$ is the total number of locations.
The \textit{j-th} element of the location vector, $lv_a[j]$, contains the number of times $a$ visited the location $r_j$.
We also assume that the individual's network of contacts $G$ influences their movements. $G = (V, E)$ is a graph in which $V$ indicates the set of individuals and $E$ the set of social ties between individuals.
To capture the spatial, temporal, and social patterns simultaneously, we create STS-EPR. 

\subsection{STS-EPR}\label{sect:sts_epr}

The spatial, temporal and social EPR model (STS-EPR) extends the EPR model by including the social dimension, taking in to account the fact that social purposes can explain about $10$-$30\%$ of the human movements \cite{cho}; and improving the temporal dimension, to accurately reproduce the distribution of the number of movements during the day.
STS-EPR can simulate the mobility $N$ agents, based on an undirected graph $G$ modeling their sociality, a weighted spatial tessellation $L$ modeling the geographic space, and a mobility diary generator $\mbox{MDG}$ modeling daily mobility schedules.

STS-EPR consists of three phases: initialization, action selection, and location selection (Figure \ref{fig:action_schema}).
After the initialization phase, the agents execute the action selection and location selection phases until a stopping criterion is satisfied (e.g., the number of hours to simulate is reached).

\begin{figure}
    \centering
    \begin{tikzpicture}[level distance=2cm,
      level 1/.style={sibling distance=5cm},
      level 2/.style={sibling distance=5cm},
      level 3/.style={sibling distance=5cm},
      level 4/.style={sibling distance=2.5cm}]
      \node{Mobility Diary Generator}
      child { node {$\langle (ab_0,t_1), (ab_1,t_2), \dots (ab_j,t_{j+1}), (ab_0,t_{j+2}), \dots)\rangle$}
      child {node {if agent moves}
        child {node {explore}
          child {node {individual} edge from parent node [left] {$1-\alpha \;$}}
          child {node {social} edge from parent node [right] {$\; \alpha$}}
           edge from parent node [left] {$\rho S^{-\gamma} \;$}
        }
        child {node {return}
        child {node {individual} edge from parent node [left] {$1-\alpha \;$}}
          child {node {social} edge from parent node [right] {$\; \alpha$}}
          edge from parent node [right] {$\; 1-\rho S^{-\gamma}$}
        }}};
    \end{tikzpicture}
    \caption[Action selection phase of GeoSim and its extensions]{A schematic description of the initialization and action selection phase of the considered models. 
    The individual first decides whether to explore a new location or return to a previously visited one.
    Then the agent determines if its social contacts will affect or not its choice for the location to visit next.}
    \label{fig:action_schema}
\end{figure}

\paragraph{Initialization phase.}
In the initialization phase, the $N$ agents are connected in an undirected graph $G$, describing the social links between agents.
The weight assigned to each edge represents the mobility similarity between the linked agents.
For each agent in the simulation, the model assigns a mobility diary produced by the mobility diary generator $\mbox{MDG}$, a Markov model trained on mobility trajectory data of real individuals, which captures the probability for individuals to follow or break their routine at specific times \cite{ditras}. The diary generator builds a mobility diary with abstract locations for each agent in the simulation.
A mobility diary $\mbox{MD}$ for an agent $a$ is defined as:
\begin{equation}
    \mbox{MD}_a = \langle (ab_0,t_1), (ab_1,t_2), \dots (ab_j,t_{j+1}), (ab_0,t_{j+2}), \dots)\rangle
\end{equation}{}
Where $ab$ is an \textit{abstract location}, $ab_0$ denotes the home location of the agent $a$, $t_i$ is a timestamp and the visits between two home locations are called $run$.
The physical locations visited by an agent during a $run$ must be distinct from each other, but the physical location resulting from the mapping of $ab_i$ can be different in different runs.
The abstract location $ab_0$ is assigned randomly to a physical location $r_j \in L$.
Each agent will move according to the entries in its mobility diary at the time specified; if the current abstract location is $ab_0$, the agent visits the home location, otherwise converts the abstract location into a physical one through the action and location selection steps.

\paragraph{Action selection phase.}
When an agent moves, it first decides whether to explore a new location or return to a previously visited one by selecting one of two competing mechanisms: exploration and preferential return.
The exploration mechanism models the scaling law presented by Song et al. \cite{im_model_song}: the tendency to explore new locations decreases over time.
Preferential return reproduces individuals' significant propensity to return to locations they explored before \cite{im_model_song,ditras,d_epr}.
An agent explores a new location with probability $P_{exp}=\rho S^{-\gamma}$, or returns to a previously visited location with a complementary probability $P_{ret}=1-\rho S^{-\gamma}$, where $S$ is the number of unique locations visited by the individual and $\rho=0.6$, $\gamma=0.21$ are constants \cite{im_model_song}.
When the agent returns, it selects a location with a probability proportional to its visitation frequency.

At that point, independently of the spatial mechanism selected, the agent determines if the choice of the location to visit is affected or not by the other agents involved in the simulation, selecting between the individual and the social influence mechanisms. 
With a probability $\alpha=0.2$ \cite{geosim}, the agent's social contacts influence their movement. With a complementary probability of $1-\alpha$, the agent selects a location without the influence of the visitation pattern of the other agents.

\paragraph{Location selection phase}
After the agent selected one of the four possible combinations of the spatial and social mechanisms, it decides which location will be the destination of its next displacement.
For an agent $a$, we define the sets containing the indices of the locations $a$ can explore or return respectively, as follows:
\begin{equation}\label{equation:set_exp}
    exp_a = \{i\;|\;lv_a[i]=0\}
\end{equation}
\begin{equation}
    ret_a = \{i\;|\;lv_a[i] > 0\}
\end{equation}\label{equation:set_ret}
The frequency of visits of an individual $a$ relative to a location $r_i$ is referred as $f_a(r_i) = \frac{lv_a[i]}{\sum_{j=1}^{|L|} lv_a[j]}$.\\

\begin{itemize}
    \item \textbf{Spatial Exploration:}  During the spatial exploration, an agent $a$ chooses a new location to explore from the set $exp_a$.
   The power-law behavior of the probability density function of the jump length suggests that individuals are more likely to move at small rather than long distances. 
Individuals take into account also the relevance of a location at a collective level together with the distance from their current location \cite{ditras}.
The method used for coupling both the distance and the relevance is the same used in the \textit{d-EPR} model \cite{d_epr}: the use of a gravity law.
Its accuracy justifies the gravity model's usage in estimating origin-destination matrices even at the country level \cite{d_epr}.
An agent $a$ currently at location $r_j$, during the Exploration-Individual action  selects an unvisited location $r_i$, with $i \in exp_a$, with probability $p(r_i)\propto \frac{w_i w_j}{d_{ij}^2}$ where $d_{ij}$ is the geographic distance between location $r_i$ and $r_j$ and $w_i$, $w_j$ represent their relevance.
 
    \item \textbf{Social Exploration:} In the social exploration action, an agent $a$ selects an agent $c$ among its social contacts.
    The probability $p(c)$ for a social contact $c$ to be selected is directly proportional to the mobility-similarity between them: $p(c) \propto mob_{sim}(a,c)$.
    After the contact $c$ is chosen, the candidate location to explore is an unvisited location for the agent $a$ that was visited by the agent $c$, more formally the location is selected from the set $A = exp_a \cap ret_c$; the probability $p(r_i)$ for a location $r_i$, with $i\in A$, to be selected is proportional to the visitation pattern of the agent $c$, namely $p(r_i) \propto f_c(r_i)$.
    \item \textbf{Individual Return:} In the individual return action, an agent $a$ picks the return location from the set $ret_a$ with a probability directly proportional to its visitation pattern.
    The probability for a location $r_i$ with $i\in ret_a$ to be chosen is: $p(r_i) \propto f_a(r_i)$.
    \item \textbf{Social Return:} The contact $c$ is selected as in the Exploration-Social action, while the set where the location is selected from is defined as $A = ret_a \cap ret_c$; the probability $p(r_i)$ for a location $r_i$ to be selected is proportional to the visitation pattern of the agent $c$, namely $p(r_i) \propto f_c(r_i)$.
\end{itemize}{}

\subsection{Additional features}
To make STS-EPR more realistic, we include some additional features that model crucial aspects of human mobility in different and more complex scenarios.

\paragraph{Relevance-based Starting Locations.}\label{sect:rsl}
Given a weighted spatial tessellation $L$, during the initialization phase, the agents are assigned to a starting location $r_i$ with a probability $p(r_i) \propto \frac{1}{|L|}$.
With the introduction of the concept of relevance at a collective level for a location, we assign the agents at the starting location following the RSL principle (Relevance-based Starting Locations): the probability $p(r_i)$ for an agent of being assigned to a starting location $r_i$ is $\propto w_i$, where $w_i$ is the relevance of the location at a collective level.

\paragraph{Reachable locations.}
When an agent is allowed to move, it is associated with waiting time, specified in the mobility diary of the agent.
The agent associated with a waiting time $\Delta_t$ cannot physically visit every location. Realistically, the agent should consider only the locations it can reach moving at a certain speed for the picked amount of time.
We define $speed_{agent}$ as the typical speed of an individual and $I$ as the set of all the locations the agent can visit, the set $R$ of the reachable locations for an agent starting from the location $r_j$ is computed as follows:
\begin{equation}
    R = \{i \in I\;|\;dist(r_j,r_i) \leq dist_{max}\}
\end{equation}\label{equation:reachable}
where $dist_{max}= \Delta t \cdot speed_{agent}$

\paragraph{Social Choice by Degree.}
When an agent $a$ performs a social action, selects a contact $c$ with a probability $p(c) \propto mob_{sim}(a,c)$.
The choice of social contact is personal for the agent; in fact, it is determined using only individual information, and no collective information is considered.
During a social return, the selection process is the same as in the STS-EPR model.
Instead, in social exploration, the contact is determined using its popularity at the collective level.
The popularity $pop$ of an individual $u$ within a social graph $G$ is defined as:
\begin{equation}
    pop(u,G) = deg(u,G)
\end{equation}{}
where the degree of a node $n$ in the graph $G$ is denoted as $deg(n,G)$.
In the social exploration, an agent $a$ selects a social contact $c$ with probability $p(c) \propto pop(c,G)$.
When an individual decides to explore a new location with a friend's influence, a popular friend will be more likely to influence its decision than an unpopular one, even if their mobility patterns are very different.
For example, an event promoter (generally a popular node within a social graph) has a high probability of influencing one of its contacts, during the selection of the next location to explore, even though they can have different mobility behaviors.
In contrast, when an individual decides to return at an already visited location with the influence of its social contacts, it is reasonable to think that the contact's choice is conducted using individual information.
During the return action, individuals follow their routines. Consequently, they are more likely to select a contact with a similar mobility pattern.

\paragraph{Action-correction phase.}
The set of possible locations an agent can reach from the current one is limited and, in some extreme cases, can be empty. 
As an example, the agent cannot reach far away locations that would reached at unrealistic speeds.
It may also happen that all locations on the spatial tessellation have been visited at least once, and so there are no new locations to explore.
To comply with these constraints, we include an action-correction phase, which is executed after the location selection phase, if the latter is too restrictive and does not allow movements in any location.

\begin{itemize}
    \item \textbf{No new location to explore}: When an agent $a$ performs the selection action phase (Figure \ref{fig:action_schema}) and decides to explore individually an unvisited location, it selects the location from the set $exp_a$ (Equation \ref{equation:set_exp}).
    In case the agent visited all the locations on the spatial tessellation at least once, no choice can be made since $exp_a=\varnothing$.
    We deal with this case correcting the action (Figure \ref{fig:routine_1}) of the agent from Exploration-Individual to Return-Individual, preserving in this way the choice of performing the location selection without any influence of its social contacts.
    \item \textbf{No location in social choices}: If an agent $a$ decides to move with the influence of a social contact $c$, and the set $A$ computed for the relative action $A = ret_a \cap ret_c$ or $A = exp_a \cap ret_c$ is empty, we correct the action from the current to Return-Individual (Figure \ref{fig:routine_1}).
    \item \textbf{No reachable locations}: When an individual is allowed to move, it is associated with a waiting time $\Delta t$, and it can reach the locations in the set $R$ (Eq. \ref{equation:reachable}).
    The set $R$ may be empty even if $I\neq\varnothing$, meaning there is no location the agent can visit within the radius $dist_{max}$ but the set of possible choices is not empty.
    If the agent was performing an exploration, a new $\Delta t_1 > \Delta t$ is selected to expand the area the agent can cover during its displacement. After picking the new waiting time the set $R$ is computed, if $R=\varnothing$ then a new $\Delta t_2 > \Delta t_1$ is picked, (this procedure is repeated for a maximum of $n_{max}$ time) and an Exploration-Individual action is performed with the new waiting time.
    If the agent was performing a Return-Social action or in the case the incrementing of the waiting time was performed $n_{max}$ times, then the action is corrected with a Return-Individual; note that Return-Individual can not fail, since the agent can always return in its current location $r_j$ from the moment that $dist(r_j,r_j)=0$ (Figure \ref{fig:routine_2}).
    \item \textbf{Run in the Mobility Diary}: In the mobility diary the locations visited by an agent during a \textit{run} (defined as the visits between two home return) must be distinct from each other.
    Given a run $d=\langle ab_1, ab_2, \dots , ab_n \rangle$ of length $n$, all the abstract locations $ab_i \in d$ must be assigned to distinct physical locations, the mapping between the abstract locations in $d$ and the real locations in $L$ must be injective.
    The injectivity of the mapping can not always be guaranteed: the location selection can fail due to the three cases presented above. 
    With the use of the mobility diary, also the Individual-Return action can fail, since the agent can not even visit its current location.\\
    In the action correction phase (Figure \ref{fig:routine_3}), if a social choice cannot be completed, the next action executed is the action with the same mechanism performed without the influence of social contacts. In the case an individual action can not be performed, the complementary individual action is performed.
    If even the complementary individual action fails, the agent returns to the home location.
    When an agent returns to the home location due to action failure during the assignment of the abstract location $ab_j \in d$, the run $d$ is splitted in $d1=\langle ab_1,\dots,ab_j \rangle$, $d2=\langle ab_{j+1},\dots,ab_n \rangle$ and the agent start the mapping of the new run $d2$.
\end{itemize}{}

\begin{figure}
    \centering
    \tikzset{every picture/.style={line width=0.75pt}} %set default line width to 0.75pt        
\begin{tikzpicture}[x=0.75pt,y=0.75pt,yscale=-1,xscale=1]
%Shape: Rectangle [id:dp5113533713537675] 
\draw  [fill={rgb, 255:red, 126; green, 211; blue, 33 }  ,fill opacity=0.39 ][line width=0.75]  (28,28.17) -- (173,28.17) -- (173,53.17) -- (28,53.17) -- cycle ;
%Shape: Rectangle [id:dp2453357910618561] 
\draw  [fill={rgb, 255:red, 126; green, 211; blue, 33 }  ,fill opacity=0.39 ][line width=0.75]  (28,104.25) -- (173,104.25) -- (173,129.25) -- (28,129.25) -- cycle ;
%Shape: Rectangle [id:dp7554144177886372] 
\draw  [fill={rgb, 255:red, 126; green, 211; blue, 33 }  ,fill opacity=0.39 ][line width=0.75]  (28,180.33) -- (173,180.33) -- (173,205.33) -- (28,205.33) -- cycle ;
%Shape: Rectangle [id:dp22357315421503277] 
\draw  [fill={rgb, 255:red, 74; green, 144; blue, 226 }  ,fill opacity=0.24 ][line width=0.75]  (276.75,104.2) -- (421.75,104.2) -- (421.75,129.2) -- (276.75,129.2) -- cycle ;
%Straight Lines [id:da605229715687423] 
\draw    (172.6,117) -- (273.8,116.81) ;
\draw [shift={(276.8,116.8)}, rotate = 539.89] [fill={rgb, 255:red, 0; green, 0; blue, 0 }  ][line width=0.08]  [draw opacity=0] (8.93,-4.29) -- (0,0) -- (8.93,4.29) -- cycle    ;
%Curve Lines [id:da8764141750437828] 
\draw    (172.9,193.1) .. controls (263.57,192.17) and (307.81,184.51) .. (329.19,173.85) .. controls (349.18,163.87) and (348.45,146.62) .. (348.27,132.52) ;
\draw [shift={(348.25,129.63)}, rotate = 450] [fill={rgb, 255:red, 0; green, 0; blue, 0 }  ][line width=0.08]  [draw opacity=0] (8.93,-4.29) -- (0,0) -- (8.93,4.29) -- cycle    ;
%Curve Lines [id:da4489926881144929] 
\draw    (172.9,39.86) .. controls (263.57,40.79) and (307.81,48.45) .. (329.19,59.11) .. controls (349.18,69.08) and (348.45,86.33) .. (348.27,100.44) ;
\draw [shift={(348.25,103.33)}, rotate = 270] [fill={rgb, 255:red, 0; green, 0; blue, 0 }  ][line width=0.08]  [draw opacity=0] (8.93,-4.29) -- (0,0) -- (8.93,4.29) -- cycle    ;
%Straight Lines [id:da7321976587724227] 
\draw    (419,46) -- (472,46) ;
\draw [shift={(475,46)}, rotate = 180] [fill={rgb, 255:red, 0; green, 0; blue, 0 }  ][line width=0.08]  [draw opacity=0] (8.93,-4.29) -- (0,0) -- (8.93,4.29) -- cycle    ;
% Text Node
\draw (424,31) node [anchor=north west][inner sep=0.75pt]  [font=\small] [align=left] {if fails};
% Text Node
\draw  [draw opacity=0]  (298.75,105.25) -- (399.75,105.25) -- (399.75,128.25) -- (298.75,128.25) -- cycle  ;
\draw (349.25,109.25) node [anchor=north] [inner sep=0.75pt]  [font=\small] [align=left] {{\small Return-Individual}};
% Text Node
\draw  [draw opacity=0]  (59,181.38) -- (142,181.38) -- (142,204.38) -- (59,204.38) -- cycle  ;
\draw (100.5,185.38) node [anchor=north] [inner sep=0.75pt]  [font=\small] [align=left] {{\small Return-Social}};
% Text Node
\draw  [draw opacity=0]  (38,29.22) -- (163,29.22) -- (163,52.22) -- (38,52.22) -- cycle  ;
\draw (100.5,33.22) node [anchor=north] [inner sep=0.75pt]  [font=\small] [align=left] {{\small Exploration-Individual}};
% Text Node
\draw  [draw opacity=0]  (47,105.3) -- (154,105.3) -- (154,128.3) -- (47,128.3) -- cycle  ;
\draw (100.5,109.3) node [anchor=north] [inner sep=0.75pt]  [font=\small] [align=left] {{\small Exploration-Social}};
\end{tikzpicture}
    \caption[Action correction phase of GeoSim for handling cases where the agent can neither explore nor perform a social choice]{A description of the action correction routine in the case the agent can neither explore nor perform a social choice.
    The green rectangle denotes a starting state while the light blue a final state.
    When one of the already mentioned action fails, a Return-Individual is executed; the individual return at an already visited location can always be performed since the set $ret_a$ contains always at least one element, the starting location of the individual.}
    \label{fig:routine_1}
\end{figure}{}

\begin{figure}
    \centering
    \tikzset{every picture/.style={line width=0.75pt}} %set default line width to 0.75pt        

\begin{tikzpicture}[x=0.75pt,y=0.75pt,yscale=-1,xscale=1]

%Shape: Rectangle [id:dp974259444763385] 
\draw  [fill={rgb, 255:red, 126; green, 211; blue, 33 }  ,fill opacity=0.39 ][line width=0.75]  (39,89.17) -- (184,89.17) -- (184,114.17) -- (39,114.17) -- cycle ;
%Shape: Rectangle [id:dp1457414964556758] 
\draw  [fill={rgb, 255:red, 126; green, 211; blue, 33 }  ,fill opacity=0.39 ][line width=0.75]  (39,165.25) -- (184,165.25) -- (184,190.25) -- (39,190.25) -- cycle ;
%Shape: Rectangle [id:dp8480688434708062] 
\draw  [fill={rgb, 255:red, 126; green, 211; blue, 33 }  ,fill opacity=0.39 ][line width=0.75]  (39,241.33) -- (184,241.33) -- (184,266.33) -- (39,266.33) -- cycle ;
%Shape: Rectangle [id:dp9823103326517663] 
\draw  [fill={rgb, 255:red, 74; green, 144; blue, 226 }  ,fill opacity=0.24 ][line width=0.75]  (236,127.21) -- (381,127.21) -- (381,152.21) -- (236,152.21) -- cycle ;
%Straight Lines [id:da22512986954348035] 
\draw    (452,89) -- (505,89) ;
\draw [shift={(508,89)}, rotate = 180] [fill={rgb, 255:red, 0; green, 0; blue, 0 }  ][line width=0.08]  [draw opacity=0] (8.93,-4.29) -- (0,0) -- (8.93,4.29) -- cycle    ;
%Shape: Rectangle [id:dp10096827047725765] 
\draw  [fill={rgb, 255:red, 74; green, 144; blue, 226 }  ,fill opacity=0.24 ][line width=0.75]  (362,192.29) -- (507,192.29) -- (507,217.29) -- (362,217.29) -- cycle ;
%Curve Lines [id:da6487326705019393] 
\draw    (184,178.33) .. controls (221.29,178.29) and (172.78,140.86) .. (232.53,139.69) ;
\draw [shift={(235.33,139.67)}, rotate = 539.96] [fill={rgb, 255:red, 0; green, 0; blue, 0 }  ][line width=0.08]  [draw opacity=0] (8.93,-4.29) -- (0,0) -- (8.93,4.29) -- cycle    ;
%Curve Lines [id:da7976095260803209] 
\draw    (381.4,139.6) .. controls (431.49,86.04) and (310.66,37.77) .. (309.01,124.53) ;
\draw [shift={(309,127.2)}, rotate = 269.49] [fill={rgb, 255:red, 0; green, 0; blue, 0 }  ][line width=0.08]  [draw opacity=0] (8.93,-4.29) -- (0,0) -- (8.93,4.29) -- cycle    ;
%Curve Lines [id:da2389580348633935] 
\draw    (381.4,140) .. controls (441.77,139.22) and (434.14,144.19) .. (433.81,188.82) ;
\draw [shift={(433.8,191.6)}, rotate = 270] [fill={rgb, 255:red, 0; green, 0; blue, 0 }  ][line width=0.08]  [draw opacity=0] (8.93,-4.29) -- (0,0) -- (8.93,4.29) -- cycle    ;
%Curve Lines [id:da7182631436364411] 
\draw    (184.6,254.4) .. controls (276.54,254.8) and (252.05,204.9) .. (359.77,204.8) ;
\draw [shift={(361.4,204.8)}, rotate = 180.21] [fill={rgb, 255:red, 0; green, 0; blue, 0 }  ][line width=0.08]  [draw opacity=0] (8.93,-4.29) -- (0,0) -- (8.93,4.29) -- cycle    ;
%Curve Lines [id:da6336276610279128] 
\draw    (184,101) .. controls (221.29,101.05) and (172.78,138.47) .. (232.53,139.64) ;
\draw [shift={(235.33,139.67)}, rotate = 180.04] [fill={rgb, 255:red, 0; green, 0; blue, 0 }  ][line width=0.08]  [draw opacity=0] (8.93,-4.29) -- (0,0) -- (8.93,4.29) -- cycle    ;

% Text Node
\draw (457,74) node [anchor=north west][inner sep=0.75pt]  [font=\small] [align=left] {if fails};
% Text Node
\draw  [draw opacity=0]  (384,193.34) -- (485,193.34) -- (485,216.34) -- (384,216.34) -- cycle  ;
\draw (434.5,197.34) node [anchor=north] [inner sep=0.75pt]  [font=\small] [align=left] {{\small Return-Individual}};
% Text Node
\draw  [draw opacity=0]  (70,242.38) -- (153,242.38) -- (153,265.38) -- (70,265.38) -- cycle  ;
\draw (111.5,246.38) node [anchor=north] [inner sep=0.75pt]  [font=\small] [align=left] {{\small Return-Social}};
% Text Node
\draw  [draw opacity=0]  (49,90.22) -- (174,90.22) -- (174,113.22) -- (49,113.22) -- cycle  ;
\draw (111.5,94.22) node [anchor=north] [inner sep=0.75pt]  [font=\small] [align=left] {{\small Exploration-Individual}};
% Text Node
\draw  [draw opacity=0]  (58,166.3) -- (165,166.3) -- (165,189.3) -- (58,189.3) -- cycle  ;
\draw (111.5,170.3) node [anchor=north] [inner sep=0.75pt]  [font=\small] [align=left] {{\small Exploration-Social}};
% Text Node
\draw  [draw opacity=0]  (246,128.26) -- (371,128.26) -- (371,151.26) -- (246,151.26) -- cycle  ;
\draw (308.5,132.26) node [anchor=north] [inner sep=0.75pt]  [font=\small] [align=left] {{\small Exploration-Individual}};
% Text Node
\draw (41,117.17) node [anchor=north west][inner sep=0.75pt]   [align=left] {{\footnotesize \textit{i = 0}}};
% Text Node
\draw (41,193.25) node [anchor=north west][inner sep=0.75pt]   [align=left] {{\footnotesize \textit{i = 0}}};
% Text Node
\draw (238,155.21) node [anchor=north west][inner sep=0.75pt]   [align=left] {{\footnotesize \textit{i++}}};
% Text Node
\draw (256,60.21) node [anchor=north west][inner sep=0.75pt]   [align=left] {{\footnotesize \textit{if i }$\displaystyle \leq $\textit{ n\_max}}\\ \ \ {\footnotesize \textit{pick} $\displaystyle \Delta t$}};
% Text Node
\draw (439,153.21) node [anchor=north west][inner sep=0.75pt]   [align=left] {{\footnotesize \textit{if i }$\displaystyle  >$\textit{ n\_max}}};
\end{tikzpicture}
    \caption[The action correction routine for handling cases where the agent can not reach any of the possible locations]{In the action correction routine designed for handling cases where the agent can not reach any of the possible locations, if the individual was performing an exploration action a new waiting time, greater than the current one, is picked; if it still can not reach any location the latter procedure is repeated at most $n_{max}$ times. If the agent was performing a Return-Social action or after $n_{max}$ increments, the action is corrected with a Return-Individual; Return-Individual can always be performed since the agent can return in its current location $r_j$ from the moment that $dist(r_j,r_j)=0$.}
    \label{fig:routine_2}
\end{figure}{}

\begin{figure}
    \centering
    \tikzset{every picture/.style={line width=0.75pt}} %set default line width to 0.75pt        

\begin{tikzpicture}[x=0.75pt,y=0.75pt,yscale=-1,xscale=0.96]

%Shape: Rectangle [id:dp9834864376310755] 
\draw  [fill={rgb, 255:red, 126; green, 211; blue, 33 }  ,fill opacity=0.39 ][line width=0.75]  (5,25.33) -- (150,25.33) -- (150,50.33) -- (5,50.33) -- cycle ;
%Shape: Rectangle [id:dp7477956579662924] 
\draw  [fill={rgb, 255:red, 126; green, 211; blue, 33 }  ,fill opacity=0.39 ][line width=0.75]  (5,252.25) -- (150,252.25) -- (150,277.25) -- (5,277.25) -- cycle ;
%Shape: Rectangle [id:dp06197678997697065] 
\draw  [fill={rgb, 255:red, 126; green, 211; blue, 33 }  ,fill opacity=0.39 ][line width=0.75]  (57,202.17) -- (202,202.17) -- (202,227.17) -- (57,227.17) -- cycle ;
%Shape: Rectangle [id:dp5471962864909669] 
\draw  [fill={rgb, 255:red, 126; green, 211; blue, 33 }  ,fill opacity=0.39 ][line width=0.75]  (57,75.17) -- (202,75.17) -- (202,100.17) -- (57,100.17) -- cycle ;
%Shape: Rectangle [id:dp4894812552173363] 
\draw  [fill={rgb, 255:red, 74; green, 144; blue, 226 }  ,fill opacity=0.24 ][line width=0.75]  (242,75.17) -- (387,75.17) -- (387,100.17) -- (242,100.17) -- cycle ;
%Shape: Rectangle [id:dp08769918989632797] 
\draw  [fill={rgb, 255:red, 74; green, 144; blue, 226 }  ,fill opacity=0.24 ][line width=0.75]  (242,202.17) -- (387,202.17) -- (387,227.17) -- (242,227.17) -- cycle ;
%Straight Lines [id:da0223920854812919] 
\draw    (202.51,88.2) -- (238,88.2) ;
\draw [shift={(241,88.2)}, rotate = 180] [fill={rgb, 255:red, 0; green, 0; blue, 0 }  ][line width=0.08]  [draw opacity=0] (8.93,-4.29) -- (0,0) -- (8.93,4.29) -- cycle    ;
%Straight Lines [id:da4573560195403087] 
\draw    (202.2,214.4) -- (237.69,214.4) ;
\draw [shift={(240.69,214.4)}, rotate = 180] [fill={rgb, 255:red, 0; green, 0; blue, 0 }  ][line width=0.08]  [draw opacity=0] (8.93,-4.29) -- (0,0) -- (8.93,4.29) -- cycle    ;
%Curve Lines [id:da008506612234683963] 
\draw    (24.33,252.17) .. controls (23.75,209.87) and (23.42,213.91) .. (52.24,214.46) ;
\draw [shift={(55,214.5)}, rotate = 180.54] [fill={rgb, 255:red, 0; green, 0; blue, 0 }  ][line width=0.08]  [draw opacity=0] (8.93,-4.29) -- (0,0) -- (8.93,4.29) -- cycle    ;
%Curve Lines [id:da18109780648993323] 
\draw    (24.33,50.33) .. controls (23.75,92.63) and (23.42,88.59) .. (52.24,88.04) ;
\draw [shift={(55,88)}, rotate = 539.46] [fill={rgb, 255:red, 0; green, 0; blue, 0 }  ][line width=0.08]  [draw opacity=0] (8.93,-4.29) -- (0,0) -- (8.93,4.29) -- cycle    ;
%Shape: Rectangle [id:dp6222145563005488] 
\draw  [fill={rgb, 255:red, 74; green, 144; blue, 226 }  ,fill opacity=0.24 ][line width=0.75]  (347,138.17) -- (492,138.17) -- (492,163.17) -- (347,163.17) -- cycle ;
%Curve Lines [id:da3793029739915291] 
\draw    (387.57,87.71) .. controls (426.21,86.33) and (419.3,91.68) .. (419.76,135.28) ;
\draw [shift={(419.8,138)}, rotate = 269.01] [fill={rgb, 255:red, 0; green, 0; blue, 0 }  ][line width=0.08]  [draw opacity=0] (8.93,-4.29) -- (0,0) -- (8.93,4.29) -- cycle    ;
%Curve Lines [id:da2704480948483904] 
\draw    (387.57,214.81) .. controls (426.21,216.2) and (419.3,210.85) .. (419.76,167.24) ;
\draw [shift={(419.8,164.53)}, rotate = 450.99] [fill={rgb, 255:red, 0; green, 0; blue, 0 }  ][line width=0.08]  [draw opacity=0] (8.93,-4.29) -- (0,0) -- (8.93,4.29) -- cycle    ;
%Straight Lines [id:da9968386004676839] 
\draw    (429,35) -- (482,35) ;
\draw [shift={(485,35)}, rotate = 180] [fill={rgb, 255:red, 0; green, 0; blue, 0 }  ][line width=0.08]  [draw opacity=0] (8.93,-4.29) -- (0,0) -- (8.93,4.29) -- cycle    ;

% Text Node
\draw  [draw opacity=0]  (36,26.38) -- (119,26.38) -- (119,49.38) -- (36,49.38) -- cycle  ;
\draw (77.5,30.38) node [anchor=north] [inner sep=0.75pt]  [font=\small] [align=left] {{\small Return-Social}};
% Text Node
\draw  [draw opacity=0]  (24,253.3) -- (131,253.3) -- (131,276.3) -- (24,276.3) -- cycle  ;
\draw (77.5,257.3) node [anchor=north] [inner sep=0.75pt]  [font=\small] [align=left] {{\small Exploration-Social}};
% Text Node
\draw  [draw opacity=0]  (67,203.22) -- (192,203.22) -- (192,226.22) -- (67,226.22) -- cycle  ;
\draw (129.5,207.22) node [anchor=north] [inner sep=0.75pt]  [font=\small] [align=left] {{\small Exploration-Individual}};
% Text Node
\draw  [draw opacity=0]  (79,76.22) -- (180,76.22) -- (180,99.22) -- (79,99.22) -- cycle  ;
\draw (129.5,80.22) node [anchor=north] [inner sep=0.75pt]  [font=\small] [align=left] {{\small Return-Individual}};
% Text Node
\draw  [draw opacity=0]  (252,76.22) -- (377,76.22) -- (377,99.22) -- (252,99.22) -- cycle  ;
\draw (314.5,80.22) node [anchor=north] [inner sep=0.75pt]  [font=\small] [align=left] {{\small Exploration-Individual}};
% Text Node
\draw  [draw opacity=0]  (264,203.22) -- (365,203.22) -- (365,226.22) -- (264,226.22) -- cycle  ;
\draw (314.5,207.22) node [anchor=north] [inner sep=0.75pt]  [font=\small] [align=left] {{\small Return-Individual}};
% Text Node
\draw  [draw opacity=0]  (373.5,139.22) -- (465.5,139.22) -- (465.5,162.22) -- (373.5,162.22) -- cycle  ;
\draw (419.5,143.22) node [anchor=north] [inner sep=0.75pt]  [font=\small] [align=left] {{\small Home Location}};
% Text Node
\draw (434,20) node [anchor=north west][inner sep=0.75pt]  [font=\small] [align=left] {if fails};
\end{tikzpicture}
    \caption[The action correction routine for guarantee the injective mapping of the abstract location visited during a run]{In the action correction routine, in order to guarantee the injective mapping of the abstract location visited during a run, if the agent was executing a social choice the action is corrected in an individual one, preserving the mechanism selected. If an individual action can not be performed the complementary one is executed, if it fails then the agent returns home.}
    \label{fig:routine_3}
\end{figure}{}

%%%%%%%%%%%%%%%%%%%%%%%%%%%%%%%%%%%%%%%%%%%%%%
%%                 RESULTS                  %%
%%%%%%%%%%%%%%%%%%%%%%%%%%%%%%%%%%%%%%%%%%%%%%
\section{Results}\label{chapter:results}
In this chapter, we show the results of the experiments that simulate the mobility of 1,001 agents in the urban area of New York City for an observation period of three months. We compare the synthetic trajectories with the trajectories of the real individuals moving in the same city for the same number of months. We use a set of well-known mobility measures to assess the similarity between the two sets of trajectories.

\subsection{Mobility Measures}\label{sect:measures}
We can classify the socio-mobility measures along the spatial, temporal, and social dimensions \cite{opp_net}.
All the measures, except the social one, are computed through the 
\textit{scikit-mobility}\footnote{\href{https://github.com/scikit-mobility}{https://github.com/scikit-mobility}} library \cite{scikit_mobility}.

\subsection*{Jump Length}
A key factor in modeling human mobility is the distance an individual travels in an amount of time. Given a trajectory, the jump length $\Delta r$ is the geographical distance between two consecutive locations visited by an individual $u$ \cite{gonzalez2008understanding, pappalardo2013understanding}:
\begin{equation}
\Delta r = dist(r_i, r_{i+1})
\end{equation}

\noindent where $r_i$ and $r_{i+1}$ are two consecutive spatial points in the trajectory of $u$ and $dist$ is the distance on the spherical earth between two points.

\subsection*{Radius of Gyration}
The radius of gyration $r_g(u)$ describes the typical distance traveled by an individual during the period of observation.
It characterizes the spatial spread of the locations visited by the individual $u$ from the locations' center of mass $r_{cm}$ \cite{gonzalez2008understanding, ret_exp_pappalardo}.

\begin{equation}
r_g(u) = \sqrt{\frac{1}{N}\sum_{i=1}^Ndist(r_i(u), r_{cm}(u))^2}
\end{equation}
\noindent where $N$ is the number of locations in the trajectory of the individual $u$ and the center of mass $r_{cm} = \frac{1}{N}\sum_{i=1}^N r_i$.

\subsection*{Visits per Location}
A useful measure to understand how individuals move in a physical space is the number of visits per location.
This quantity describes the relevance of a location, namely the attractiveness at a collective level.

\subsection*{Location Frequency}
Humans exhibit a strong tendency to return to locations they visited before \cite{gonzalez2008understanding}. The location frequency $f(r_i)$ measures the probability of visiting a location $r_i$: 
\begin{equation}\label{equation:loc_freq}
f(r_i) = \frac{n(r_i)}{n_u}
\end{equation}
\noindent where $n(r_i)$ is the number of visits to location $r_i$ and $n_u$ is the total number of points in the trajectory of the individual $u$.\\
One method to describe the importance of a location for an individual $u$ is the concept of location's rank; a location $r_i$ has rank $k$ if it is the \textit{k-th} most visited location by an individual $u$.\\

\subsection*{Waiting Time}
The waiting time $\Delta_t$ is defined as the elapsed time between two consecutive points in the mobility trajectory of an individual $u$, or equivalently as the time spent in a location:
\begin{equation}
\Delta_t = t_{i+1}-t_i
\end{equation}

\subsection*{Uncorrelated Entropy}
The uncorrelated entropy gives an estimation of the \textit{predictability} of the movements of an individual $u$ \cite{song_limits}:
\begin{equation}
E_{unc}(u) = - \sum_{i=1}^{N_u} p_u(i)\:log_2(p_u(i))
\end{equation}

\noindent where $N_u$ is the number of distinct locations visited by $u$ and $p_u(i)$ is the historical probability that location $i$ was visited by user $u$.

\subsection*{Activity per Hour}
The movements of individuals are not distributed uniformly during the hours of the day.
Humans' actions follow a circadian rhythm \cite{ditras,opp_net}; people tend to be stationary during the night hours while they preferentially move at specific times of the day, for example, to reach the workplace or return home.
To measure this distinctiveness of human mobility, we compute the number of movements made by the individuals at every hour of the day.

\subsection*{Mobility Similarity}
Several studies demonstrate the correlation between human mobility and sociality \cite{geosim,link_prediction,cho,chao,datasetFS}; the movements of friends are more similar than those of strangers, mainly because we are more likely to visit a location if a social contact explored that location before. Furthermore, individuals with a similar visitation pattern are more likely to establish a social link.
We define the mobility similarity $mob_{sim}$ between two individuals $u_i,u_j$ as the cosine-similarity of their location vectors $lv_i, lv_j$.
\begin{equation}
mob_{sim}(u_i,u_j) = \frac{lv_i \cdot lv_j}{\lVert lv_i \rVert \lVert lv_j \rVert}
\end{equation}

\subsection{Statistical similarity}\label{sect:scores}
We quantify the statistical similarity between the distributions of the human mobility measures of the generated and the real trajectories using five metrics: \cite{timegeo,ditras}.
\begin{itemize}
\item \textbf{RMSE}: The Root Mean Square Error (RMSE) between a ground truth distribution $p$ and a synthetic distribution $q$ is defined as:
\begin{equation}
    \mbox{RMSE}(p,q) = \sqrt{\frac{\sum_{i=1}^n(p_i-q_i)^2}{n}}
\end{equation}
where $q_i \in q$, $p_i \in p$ and the number of observations in both the distributions is $n$.
\item \textbf{Kullback–Leibler divergence}: The Kullback–Leibler divergence (KL) between a ground truth distribution $p$ and a synthetic distribution $q$ quantifies how much information is lost when $q$ is used to approximate $p$.
\begin{equation}
    \mbox{KL}(p\parallel q)=\sum _{i=1} ^ n p_i \log\left({\frac {p_i}{q_i}}\right)
\end{equation}{}
\item \textbf{Hellinger distance}: The Hellinger distance (H) measure the distance between two distributions $p$ and $q$.
\begin{equation}
    \mbox{H}(p,q) = {\frac {1}{\sqrt {2}}}\;{\sqrt {\sum _{i=1}^{n}({\sqrt {p_{i}}}-{\sqrt {q_{i}}})^{2}}}
\end{equation}

\item \textbf{Pearson's correlation coefficient}: The Pearson's correlation coefficient (r) is a measure of the linear relationship between two set of observations $p$ and $q$.
\begin{equation}
    r_{pq}={\frac {\sum _{i=1}^{n}(p_{i}-{\bar {p}})(q_{i}-{\bar {q}})}{{\sqrt {\sum _{i=1}^{n}(p_{i}-{\bar {p}})^{2}}}{\sqrt {\sum _{i=1}^{n}(q_{i}-{\bar {q}})^{2}}}}}
\end{equation}
where $\bar {p}$ and $\bar {q}$ are the mean values of $p$ and $q$ respectively.
\item \textbf{Spearman's Rank correlation coefficient}: The Spearman's Rank correlation coefficient ($\rho$) measure the monotonic relationships (linear or non-linear) while Pearson's correlation measure only linear relationships.
\begin{equation}
    \rho_{pq}= 1 - 6 \frac{\sum_{i=1}^n (r_k(p_i)-r_k(q_i))^2}{n(n^2-1)}
\end{equation}
where $r_k(p_i)$ is the \textit{rank} of value $p_i$ in the sorted list $(p_1,...,p_n)$, analogously $r_k(q_i)$.
\end{itemize}{}

\subsection{Dataset}\label{sect:dataset}

We compare the trajectories generated by STS-EPR with real trajectories obtained from an LBSN (Location-Based Social Network) data set collected by Yang et al. \cite{datasetFS}. 
The data set contains a set of global-scale check-ins gathered from Foursquare over 22 months (from April 2012 to January 2014). 
A check-in describes a user's real-time position with its social contacts.
In Foursquare, the check-ins made by a user are not publicly available; despite this, many users share their check-ins on Twitter to make them public. The authors of the dataset collected the Foursquare check-ins from Twitter by searching the Foursquare hashtag \cite{datasetFS}.
The dataset is associated with a lookup dataset for the locations, and with a snapshot of the social network obtained from Twitter, antecedent at the collection period.

The LBSN dataset $D_{FS}$ \cite{datasetFS}, contains 90,048,627 check-ins made by 2,733,324 users all around the globe.
The attributes of $D_{FS}$ are an anonymized user identifier, an identifier of the location where the user made the check-in, the UTC (Coordinated Universal Time) when the check-in occurred, and the location's timezone offset (Table \ref{tab:records}).
A lookup dataset $D_{loc}$ associates the location's identifier with the respective coordinates and other information.
LBSN datasets allow the reconstruction of the mobility of an individual considering the check-ins as points in the individual's trajectories.

\begin{table}[h]
\scriptsize
\centering
\begin{tabular}{ccccl}
\multicolumn{1}{l}{(a)} & \multicolumn{1}{l}{} & \multicolumn{1}{l}{}      & \multicolumn{1}{l}{} &                              \\ \hline
user\_id                & location\_id            & UTC time                  & timezone             &                              \\ \hline
$\vdots$                & $\vdots$             & $\vdots$                  & $\vdots$             &                              \\
268846                  & 42872fd9b60caeb      & Tue Apr 03 18:27:37 2012  & -240                 &                              \\
377500                  & 3c38c65be1b8c04      & Tue Apr 03 18:27:38 2012  & -240                 &                              \\
248657                  & 1855f964a520be3      & Tue Apr 03 18:27:38  2012 & -240                 &                              \\
$\vdots$                & $\vdots$             & $\vdots$                  & $\vdots$             &                              \\
\multicolumn{1}{l}{}    & \multicolumn{1}{l}{} & \multicolumn{1}{l}{}      & \multicolumn{1}{l}{} &                              \\
\multicolumn{1}{l}{(b)} & \multicolumn{1}{l}{} & \multicolumn{1}{l}{}      & \multicolumn{1}{l}{} &                              \\ \hline
location\_id               & latitude             & longitude                 & category             & \multicolumn{1}{c}{cc}       \\ \hline
$\vdots$                & $\vdots$             & $\vdots$                  & $\vdots$             & \multicolumn{1}{c}{$\vdots$} \\
42872fd9b60caeb         & 41.660393            & -83.615227                & College Cafeteria    & \multicolumn{1}{c}{US}       \\
6200f964a520ee3         & 40.722206            & -73.981720                & Theater              & \multicolumn{1}{c}{US}       \\
9cadf964a521fe3         & 44.972814            & -93.235313                & Student Center       & \multicolumn{1}{c}{US}       \\
$\vdots$                & $\vdots$             & $\vdots$                  & $\vdots$             & \multicolumn{1}{c}{$\vdots$}
\end{tabular}
\caption[An example of records for the dataset $D_{FS}$ and the lookup dataset $D_{loc}$]{An example of records for the dataset $D_{FS}$ (a) and the lookup dataset $D_{loc}$ (b), In $D_{loc}$ the \texttt{location\_id} is associated with the coordinates, the category and the country code.}
\label{tab:records}
\end{table}

We create the mobility dataset $D_{NYC} \subset D_{FS}$ relative at the area of New York City performing a join between $D_{FS}$ and $D_{loc}$ on the attribute \texttt{location\_id}, obtaining all the displacements made by the individuals in New York City, associated with the relative coordinates. 
Before performing the join, we apply some filters to obtain only the check-ins in NYC and to reduce the number of records in $D_{FS}$ and $D_{loc}$, avoiding a computationally expensive operation.
The resulting dataset $D_{FS\_loc}$ is composed of 925,289 check-ins relative to 80,146 users. After converting the UTC in the time of New York City, we remove all the users not included in the snapshot of the social graph $G$ scraped from Twitter.
Of the 80,146 users only 8,452 appear in $G$ ($10.5\%$).
In the next filter operation, we take only the check-ins of the 8,452 users performed during a period of three months, from April 2012 to July 2012, in this period of observation the check-ins made by the filtered users are 80,032.
Then, we substitute the fast check-ins, defined as a set of check-ins such that the time difference between them is less or equal than $t=7s$, with a single check-in where the coordinates and timestamp are the averages of the respective attributes for the \textit{fast check-ins}.
Then, we select only the users with mobility (at least two check-ins) and the users who appear in at least one edge with another of the filtered users.
After the latter filtering operations, the users left are 1,780. We removed the users not in the main component of the social graph $G$ (considering only the edges between the 1,780 users).
The final dataset, $D_{NYC}$, contains 37,489 check-ins made by 1,001 connected users during an observation period of three months (April 2012 to July 2012).

\begin{figure}[!h]
\centering
    \subfigure[]{\label{fig:dist_dist}
\includegraphics[width=0.45\textwidth]{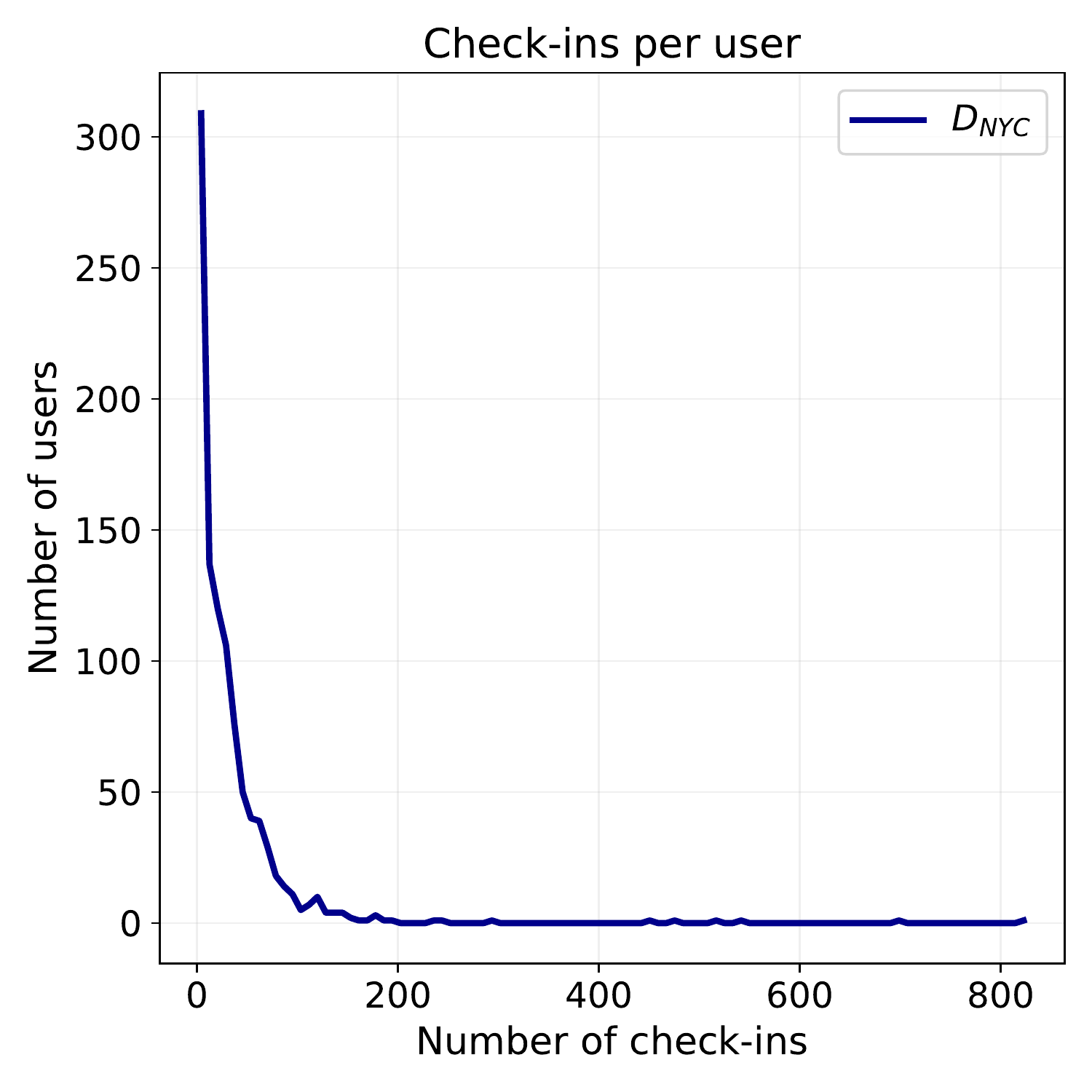}}
\hspace{2mm}
\subfigure[]{\label{fig:dist_unc}
\includegraphics[width=0.45\textwidth]{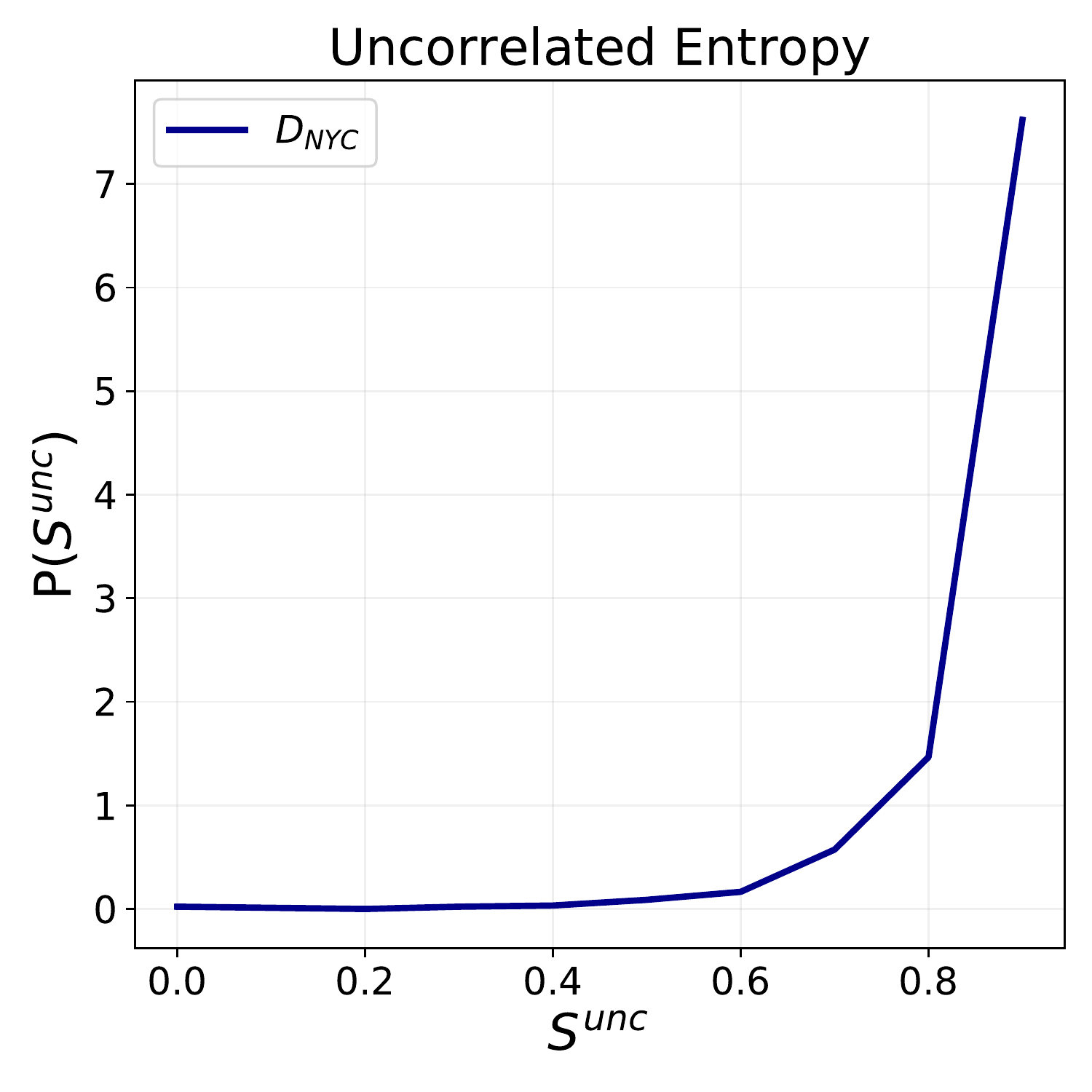}}
\subfigure[]{\label{fig:dist_jl}
\includegraphics[width=0.45\textwidth]{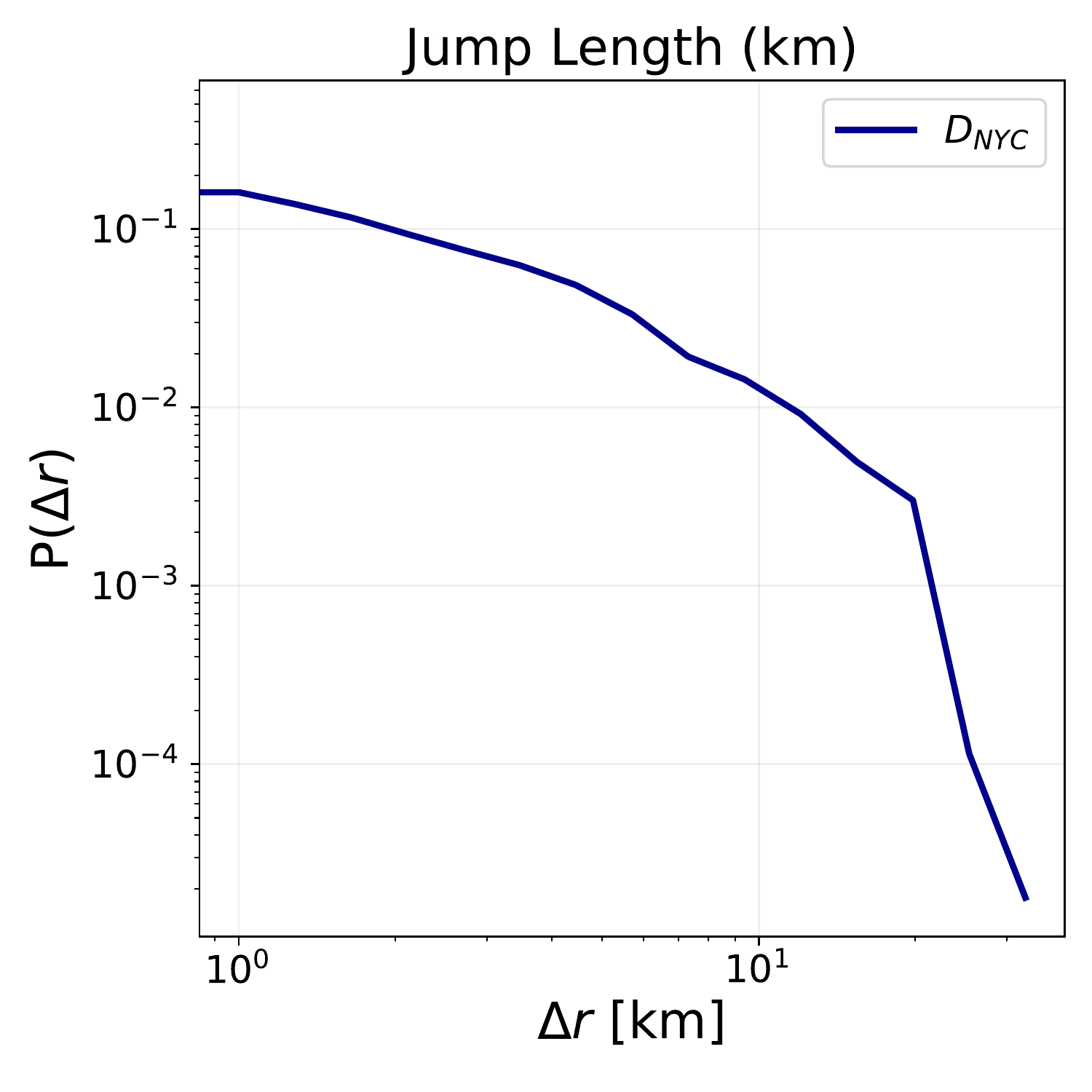}}
\hspace{2mm}
    \subfigure[]{\label{fig:dist_rog}
\includegraphics[width=0.45\textwidth]{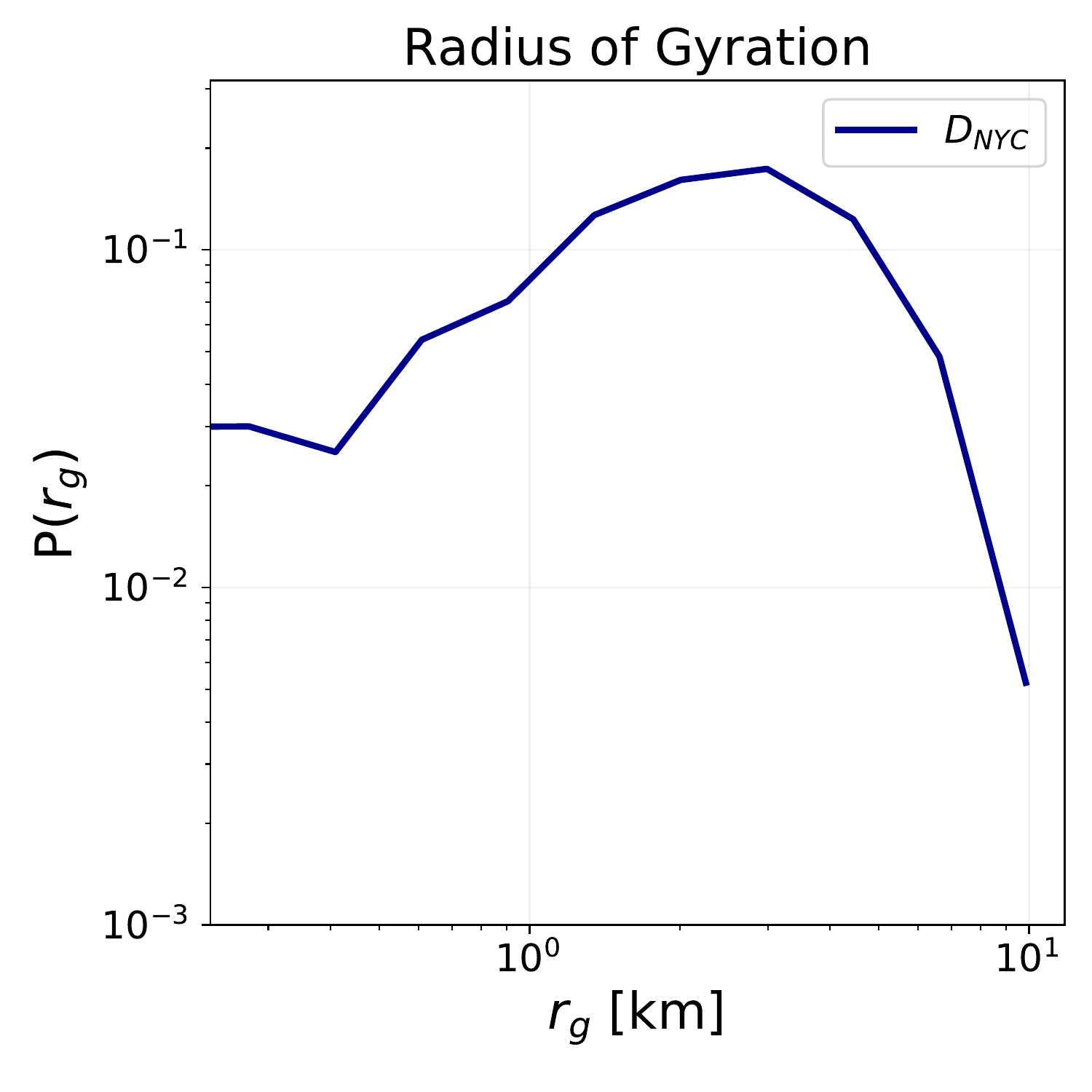}}
 \subfigure[]{
\includegraphics[width=0.45\textwidth]{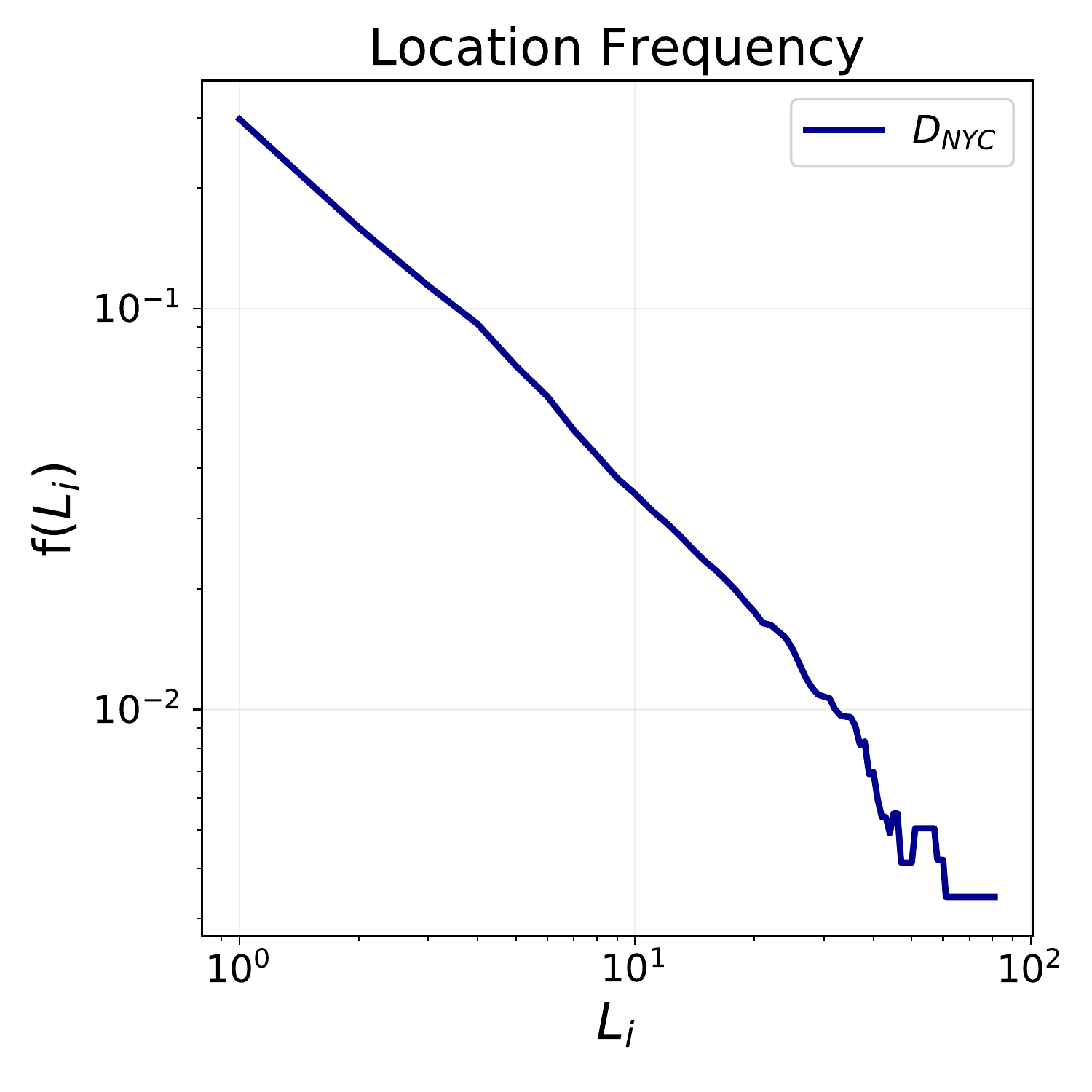}}
\hspace{2mm}
\subfigure[]{\label{fig:dist_visits}
\includegraphics[width=0.45\textwidth]{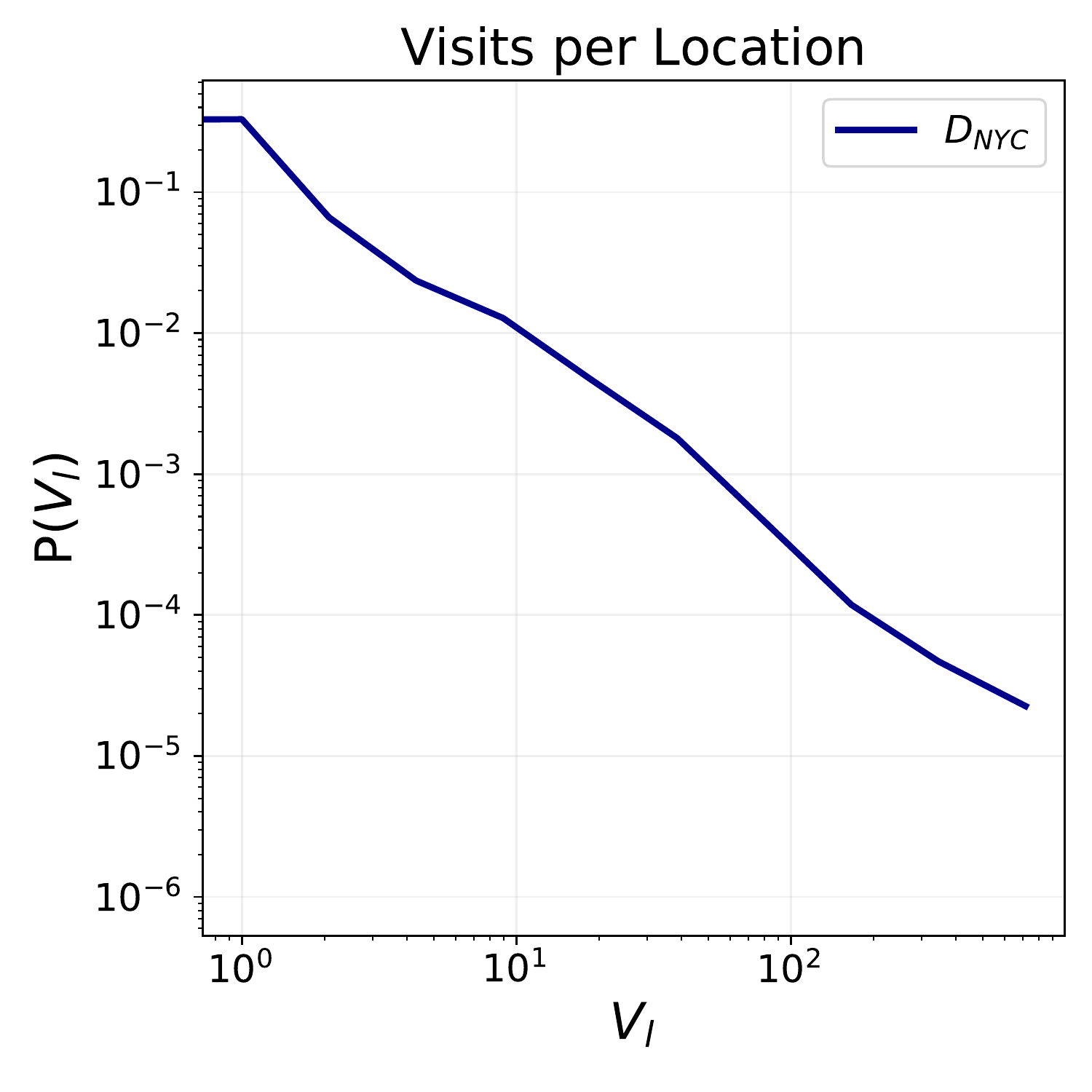}}
\caption{The distributions of the number of check-ins per user (a), uncorrelated entropy (b), jump length and (c) radius of gyration (d), for the filtered data set $D_{\mbox{\tiny NYC}}$. The distributions of the location frequency (e) and visits per location (f) for the dataset $D_{\mbox{\tiny NYC}}$ computed with a weighted squared tessellation with size 1000 meters.}
\label{fig:dist}
\end{figure}

We analyzed the probability density functions (PDF) of the measures presented in Section \ref{sect:measures}, to check whether or not the obtained trajectories $\in D_{NYC}$ present the significant analytical proprieties of individuals' displacement.
The distribution of the number of check-ins per user is heavy-tailed (Figure \ref{fig:dist}). 
This behavior is typical of the LBSN datasets \cite{datasetFS}. The movements of the individuals in New York City are not highly predictable, as attested by the uncorrelated entropy measure (Figure \ref{fig:dist}).

The jump length confirms the tendency of individuals to move at small rather than long distances \cite{brockmann,gonzalez2008understanding}, as we can see from Figure \ref{fig:dist_jl} individuals in the area of NYC move rarely at distances greater than $\approx22\mbox{km}$. The distribution of the radius of gyration (Figure \ref{fig:dist_rog}) shows that the typical spatial spread of individuals' displacements is likely to be included between 1 and 7 $km$.

\begin{figure}
    \centering
    \includegraphics[width=0.95\textwidth]{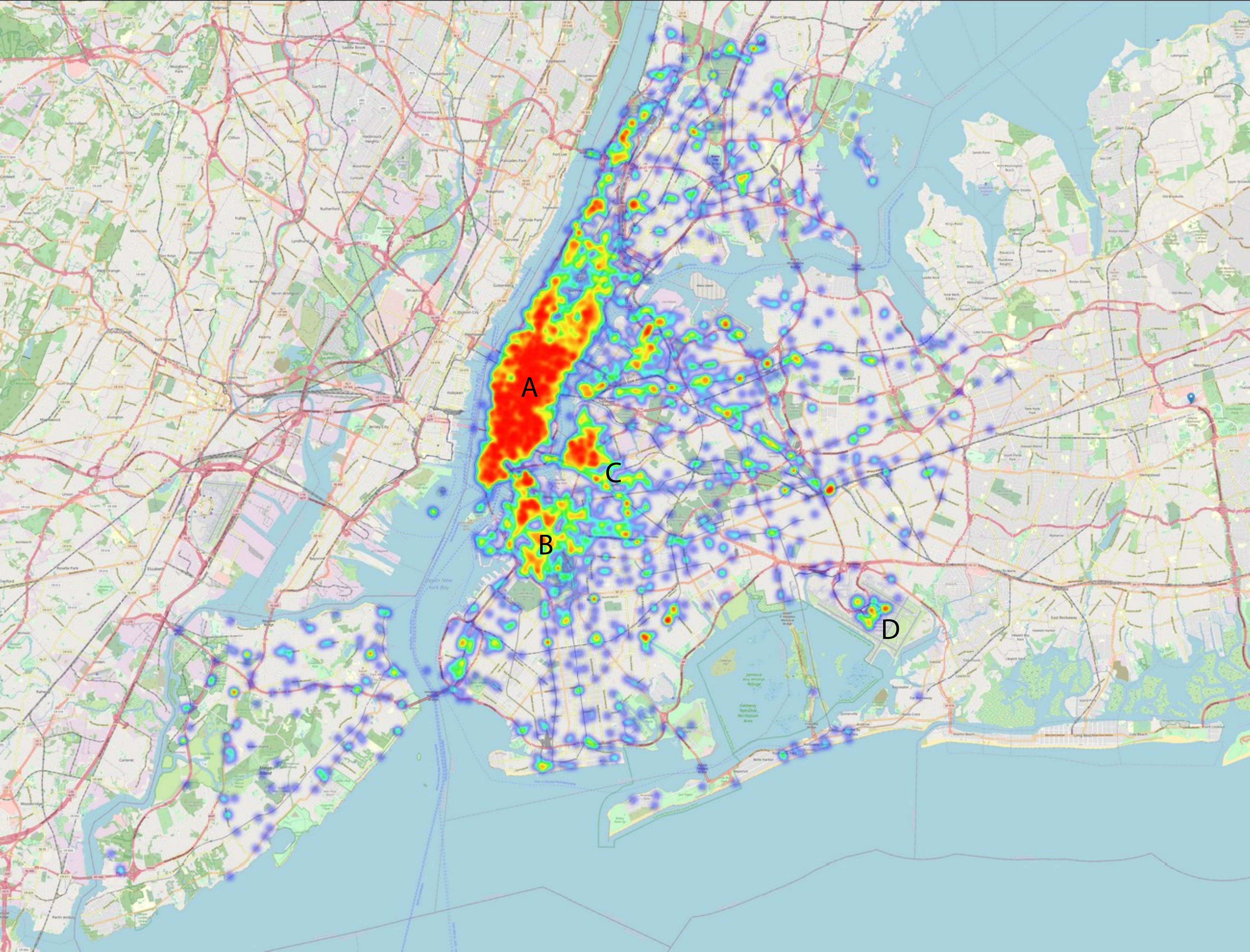}
    \caption[The heatmap relative at the mobility in New York City]{The heatmap relative at the 37,489 check-ins made by 1,001 individuals during an observation period of three months (April 2012 to July 2012) in New York City.
    There is a high concentration of check-ins in the borough of Manhattan (A) and in its surroundings (upper part of Brooklyn (B) and of Queens (C)); this high concentration of check-ins in that area can be explained mainly because Manhattan is the most densely populated of the five boroughs of New York City; another reason is that Manhattan is the touristic center of New York City, it contains attractive locations such as Times Square, Central Park, the Empire State Build, Statue of Liberty, Wall Street, One World Trade Center, and many others.
    Another area of dense check-ins is the one that is associated to the JFK airport (D).
    The distribution of the check-ins in the physical space can be considered as a continuous and non-aggregated form of relevance, from the moment that the relevance of a location is computed as the number of check-ins made in that location.}
    \label{fig:heatmap_checkins_nyc}
\end{figure}

The probability for an individual to visit a location of rank $i$ (Figure \ref{fig:dist_dist}), namely the location frequency, follows a distinctive distribution of this measure, the Zipf law \cite{gonzalez2008understanding}. The number of visits per each location, which correspond to the relevance of a tessellation, results in a power-law distribution (Figure \ref{fig:dist_visits}); most locations have a few visits while only rare locations receive
a significant number of visits \cite{ditras}. 
Figure \ref{fig:heatmap_checkins_nyc} shows a heatmap of the check-ins $\in D_{NYC}$.

{}

\begin{figure}[!h]
\centering
    \subfigure[]{
\includegraphics[width=0.45\textwidth]{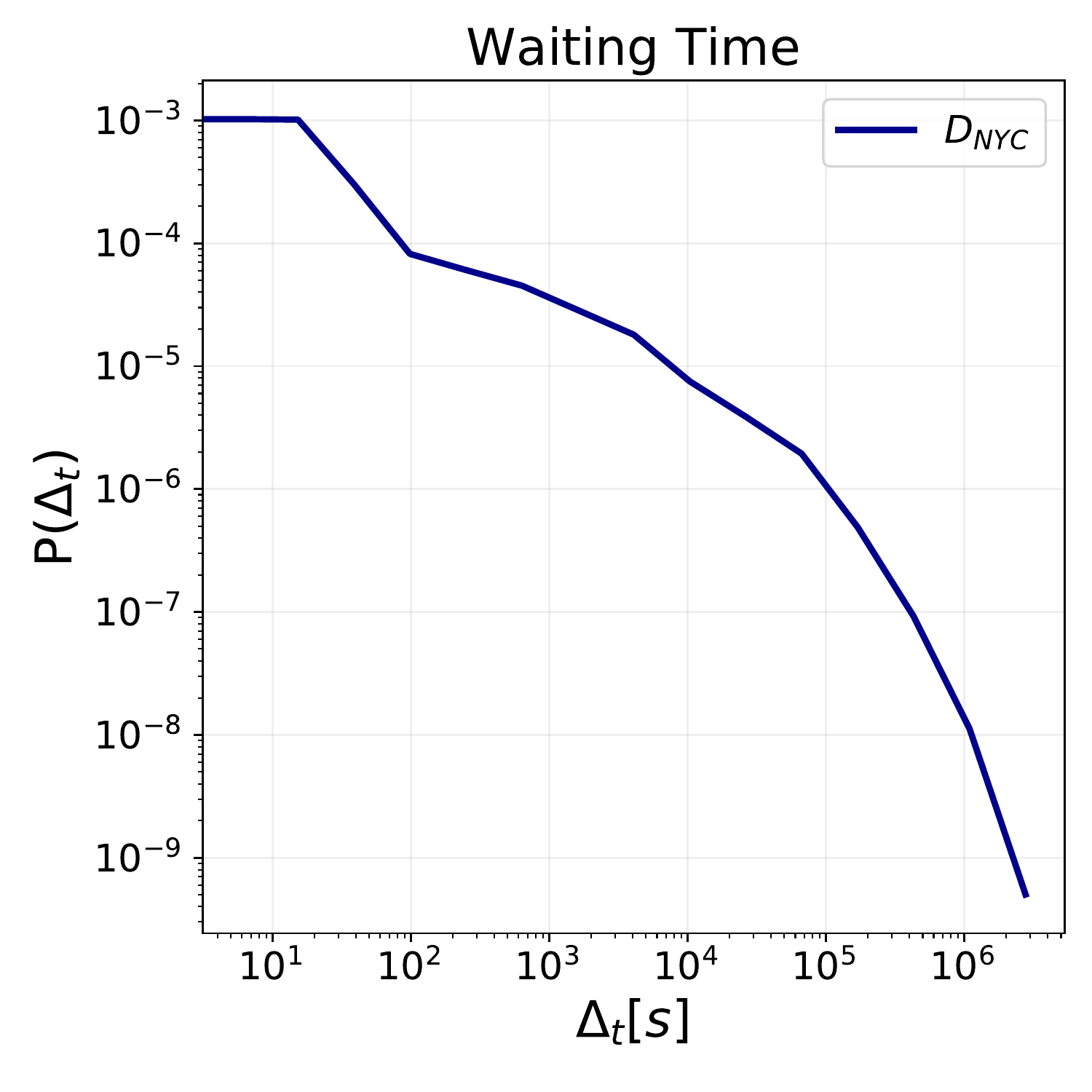}}
\hspace{2mm}
    \subfigure[]{
\includegraphics[width=0.45\textwidth]{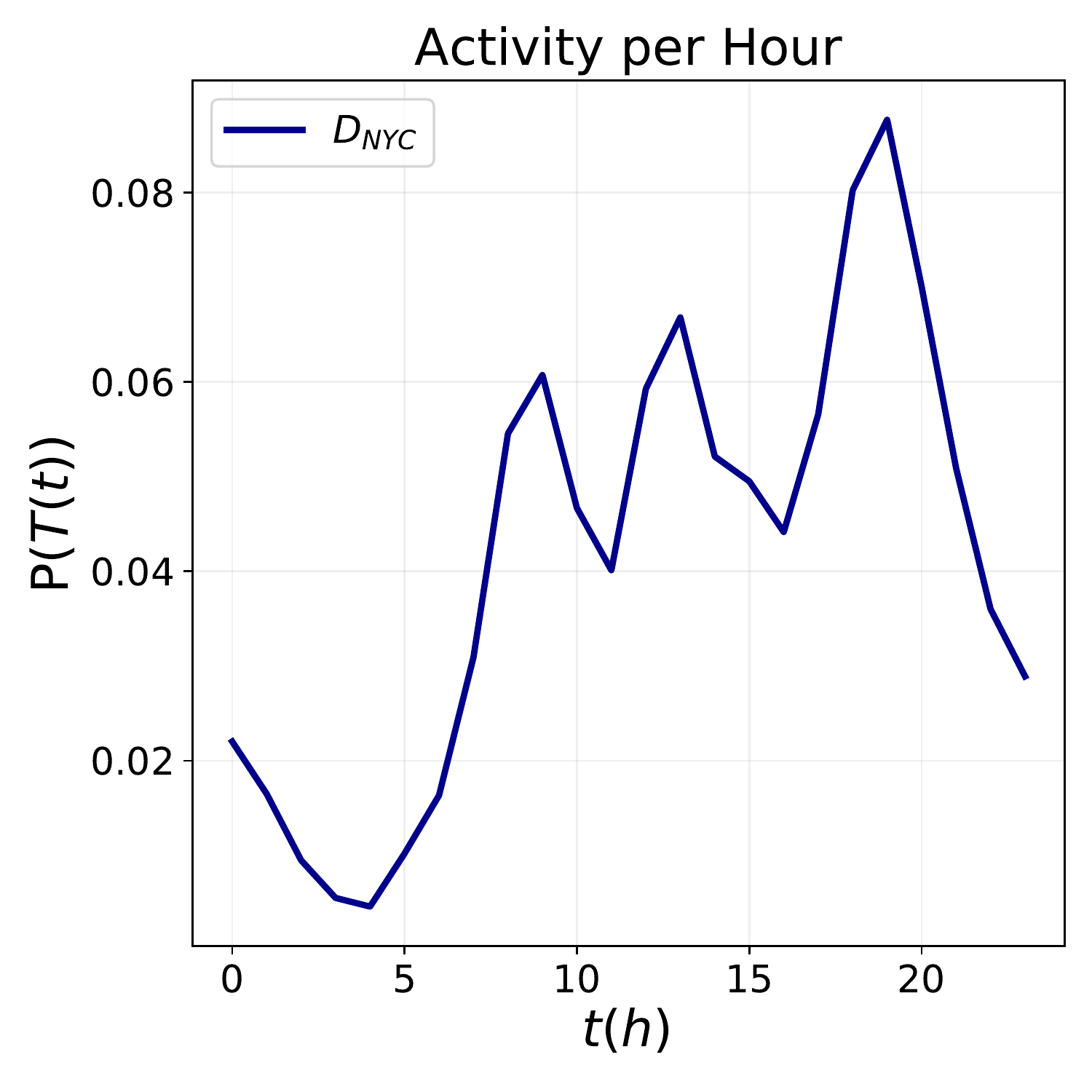}}
\subfigure[]{\label{fig:mob_sims_nyc}
\includegraphics[width=0.45\textwidth]{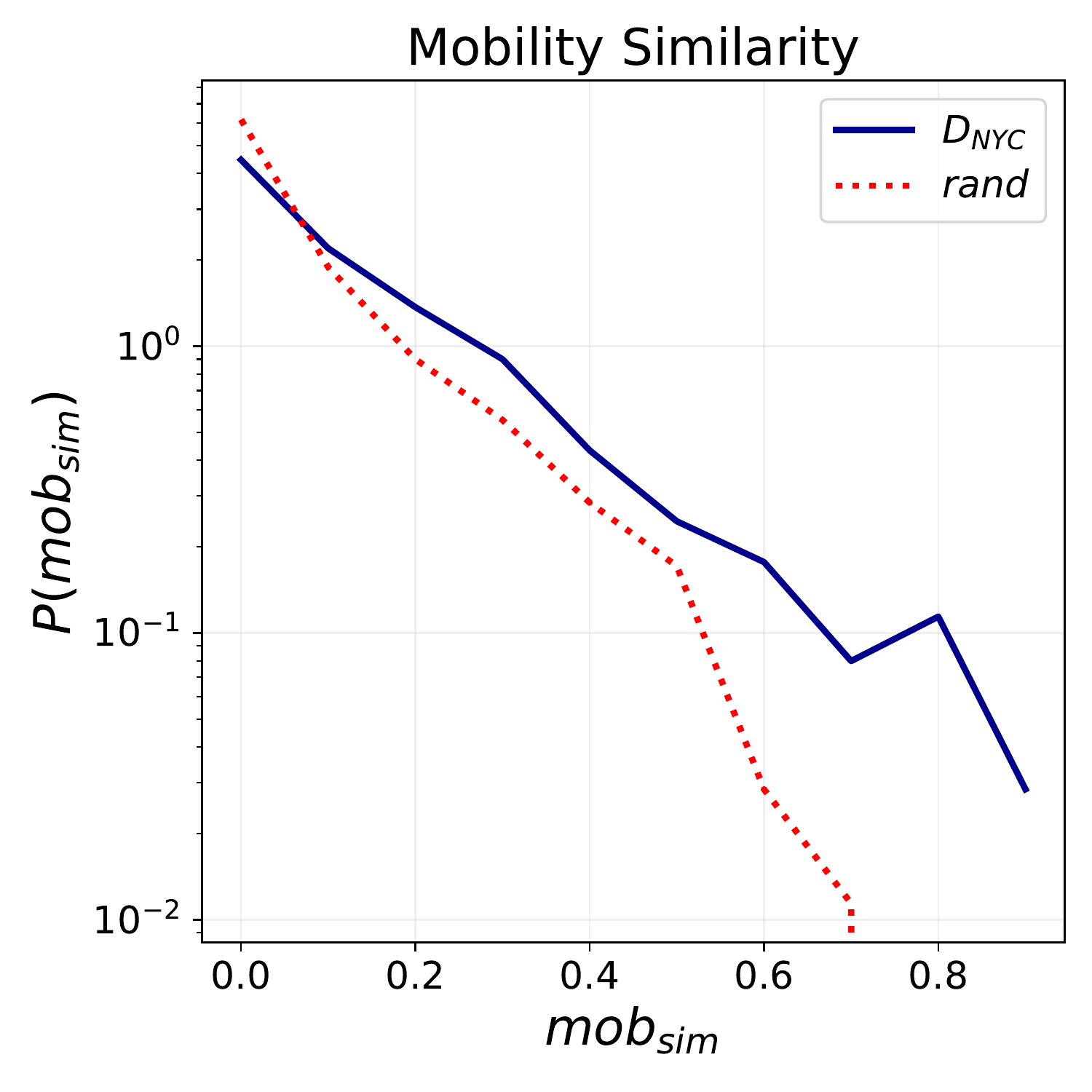}}
\hspace{2mm}
    \subfigure[]{
\includegraphics[width=0.45\textwidth]{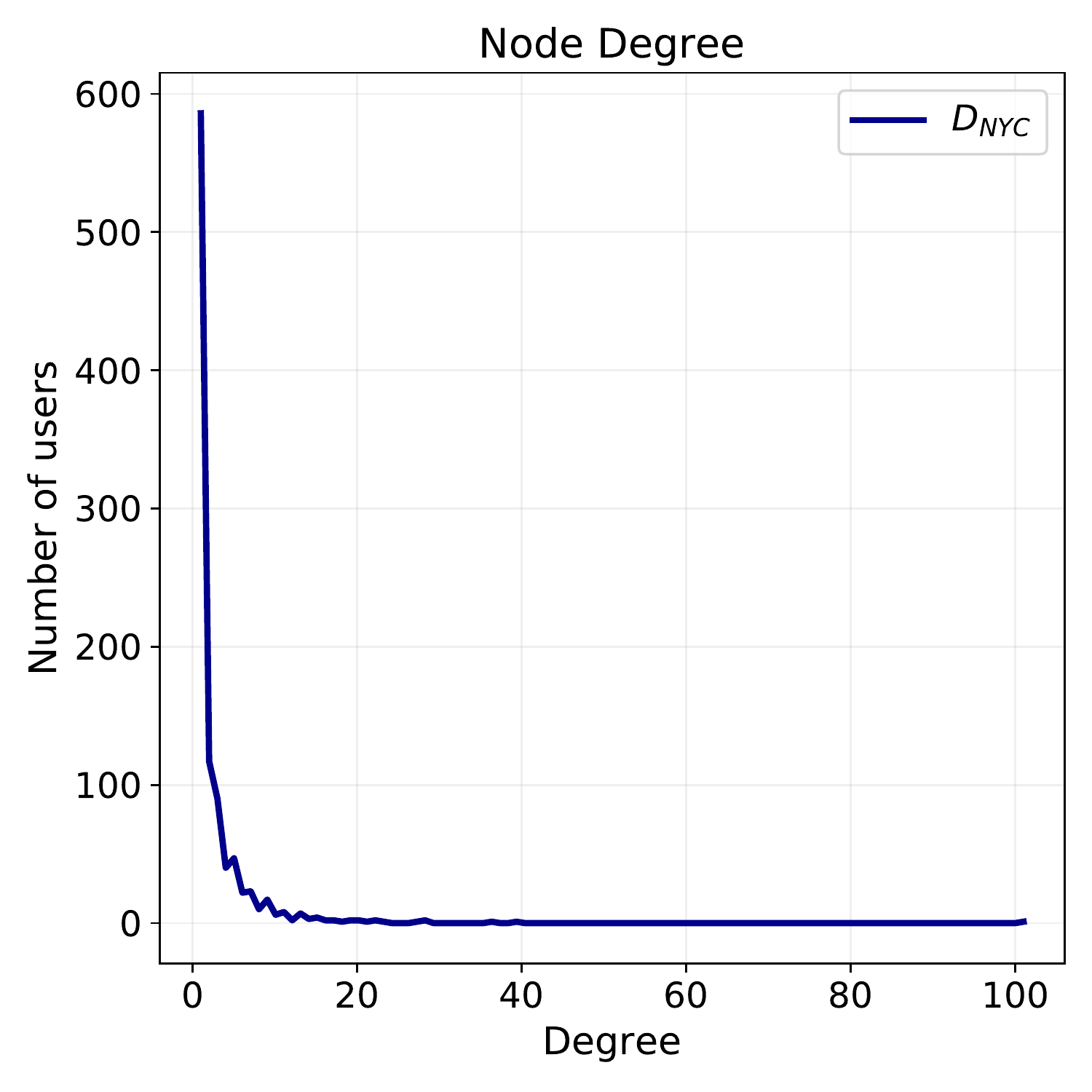}}
\caption{The distributions of waiting time (a) and activity per hour (b) for $D_{\mbox{\tiny NYC}}$.
In (b), note how the circadian rhythm of the individuals presents three peaks at hours corresponding to the following activities: reach the workplace (8-9 am), work lunch break (12am-1pm) and return home (6-7pm). 
(c) Probability density function of the mobility similarity, computed among the pair of users connected in the social graph $G_{\mbox{\tiny NYC}}$ (blue solid line) and for random pairs of users not connected in the social graph (red dotted line). (d) Distribution of the node degree for the social graph $G_{\mbox{\tiny NYC}}$.}
\label{fig:dist_tmp}
\end{figure}

The distribution of the time spent in a location, for the 1,001 individuals, during the three months follows a power law \cite{ditras,opp_net} (Figure \ref{fig:dist_tmp}).
Humans' actions follow a circadian rhythm \cite{ditras,opp_net}: the activity per hour measures, depicts the non-uniform distribution of the movements of individuals during the hours of a day (Figure \ref{fig:dist_tmp}).

We compute the mobility similarity for the users connected in $G$ and for a random graph with the same number of nodes and edges. 
As Figure \ref{fig:mob_sims_nyc} shows, the mobility similarity within users connected in the social graph is generally higher than the ones of random pairs of users. This result confirms the correlation between human mobility and sociality: the movements of friends are more similar than those of strangers \cite{geosim,link_prediction,cho,chao}.

\subsection{Social Graph}
\begin{figure}[!ht]
\centering
\includegraphics[width=0.81\textwidth]{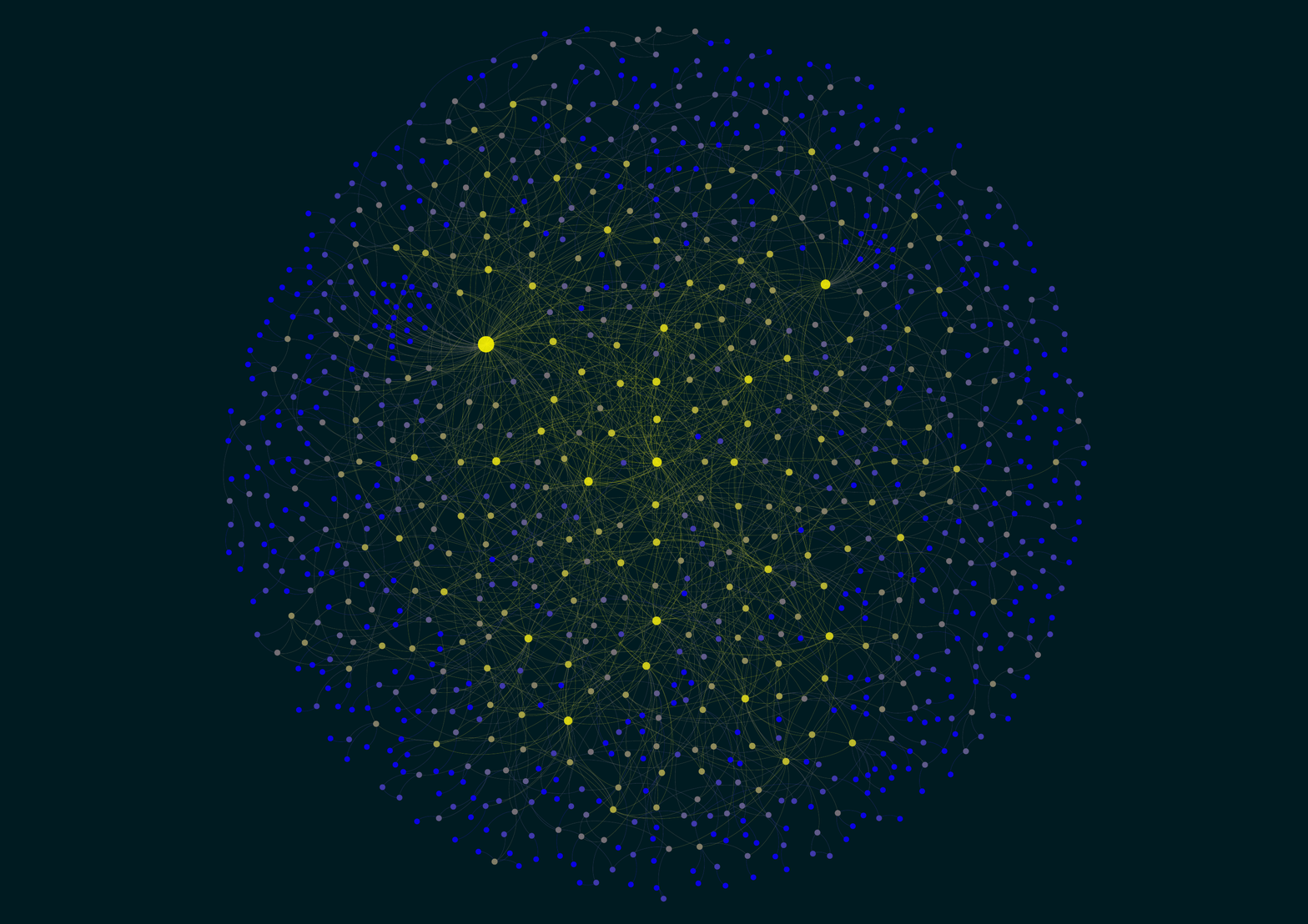}
\caption[A visualization of the social graph $G_{NYC}$]{A visualization of the social graph $G_{\mbox{\tiny NYC}}$. The size of a node is proportional to the degree, as well as the color that varies from purple to yellow.}
\label{fig:social_graph}
\end{figure}

The social graph $G$, a snapshot relative to March 2012 of the social connections among a subset of the users in $D_{FS}$, obtained from Twitter, is composed of 114,324 nodes and 363,704 edges.
During the data preprocessing, we filter $G$ several times, obtaining a new graph $G_{NYC}$ (Figure \ref{fig:social_graph}); it is an undirected and connected graph, with 1,001 nodes which represent the users and 1,755 edges which represent the social connections between users. 
The graph's statistical properties follow the well-know significant properties of social graphs; the node degree distribution follows a power-law, and the average path length is $5.154$ ($\approx 6$ in social graphs according to Karamshuk et al. \cite{opp_net}). The density and average node degree are respectively $4\cdot 10^{-3}$ and $3.506$.

\subsection{Weighted Spatial Tessellation}\label{sect:tess}
To partition the area of New York City into a discrete number of relevant and \textit{non-overlapped} locations, we used a weighted spatial \textit{squared} tessellation $L$.
The first consideration is that some of the locations are in the water area of New York City and, consequently, they are unreachable for the agents in our models. 
We exclude these locations obtaining a new tessellation $L_{\mbox{\tiny LAND}} \subseteq L$ since we consider only displacements within land locations.
We compute the relevance $w_i$ of each location $r_i \in L_{\mbox{\tiny LAND}}$ as the total number of check-ins in $D_{NYC}$ made in that location by the individuals; we assigned a default relevance of $0.1$ at the locations without any associated check-in.
We compute another subset of locations, $L_{\mbox{\tiny REL}} \subseteq L_{\mbox{\tiny LAND}}$ defined as follows:
\begin{equation}
    L_{\mbox{\tiny REL}} = \{r_i\in L_{\mbox{\tiny LAND}}\;|\; w_i \geq 1\}
\end{equation}
We build the tessellations $L_{\mbox{\tiny LAND}}$ and $L_{\mbox{\tiny REL}}$ for different levels of granularity, we select the side $s$, in meters, of the squared tiles from the set $\{250,500,750,1000,2000\}$; a tessellation with locations of side $s$ is referred as $L_{\mbox{\tiny LAND}}(s)$ or $L_{\mbox{\tiny REL}}(s)$.
Table \ref{tab:tiles} shows the number of locations in each tessellation.

\begin{table}[h]
    \centering
    \begin{tabular}{c|ccc}
    tile size & $|L|$ & $|L_{\mbox{\tiny LAND}}|$ & $|L_{\mbox{\tiny REL}}|$ \\ \hline
    250 m     & 34,408 & 23,740        & 2,893        \\
    500 m     & 8,745  & 6,285         & 1,604        \\
    750 m     & 3,951  & 2,918         & 1,082        \\
    1000 m    & 2,256  & 1,709         & 800         \\
    2000 m    & 596   & 475          & 333        
    \end{tabular}
    \caption[The number of locations for each tile size for the tessellation $L$, $L_{\mbox{\tiny LAND}}$ and $L_{\mbox{\tiny REL}}$]{The number of locations for each tile size for the tessellation $L$, $L_{\mbox{\tiny LAND}}$ and $L_{\mbox{\tiny REL}}$.}
    \label{tab:tiles}
\end{table}{}

\subsection{Experimental settings}\label{sect:exp_setting}
We generate the synthetic trajectories simulating for three months the displacements of 1,001 individuals connected in the graph $G_{\mbox{\tiny NYC}}$, moving in the urban area of New York City, represented through the tessellation $L_{\mbox{\tiny REL}}$ presented in Section \ref{sect:tess}.
During the experiments, we compare the trajectories generated by STS-EPR with the ones generated from GeoSim and two extensions of it: $\mbox{GeoSim}_d$ and $\mbox{GeoSim}_{gravity}$. 
In the first proposed extension, $\mbox{GeoSim}_d$, we introduce into GeoSim a mechanism that takes into account the distance from the current location and the location to explore.
In the second extension, $\mbox{GeoSim}_{gravity}$, we take into account the relevance of a location together with the distance from the current location using a gravity-law.
More information and technical details on the proposed extensions can be found in the Appendix.
Each of the proposed extensions is instantiated with the additional features RSL and the action correction phase\footnote{We tried to include also the other additional features during the experiments, namely reachable locations and social choice by degree, but they did not improve the performance of the generative models.}.
The Markov model of STS-EPR, relative to the mobility diary generator, is trained on the displacements of the individuals included in the dataset $D_{\mbox{\tiny NYC}}$.
For each model we use the weighted spatial tessellations $L_{\mbox{\tiny REL}}$ for different levels of granularity, we select the side $s$ in meters from the set $\{250,500,750,1000,2000\}$.
In the experimental phase, for each model and for each tessellation we make five executions, to collect the mean and the standard deviation for each mobility measure, resulting from the comparison of the synthetic trajectories with the trajectories of real individuals.

\begin{figure}[!h]
\centering
    \subfigure[]{
\includegraphics[width=0.45\textwidth]{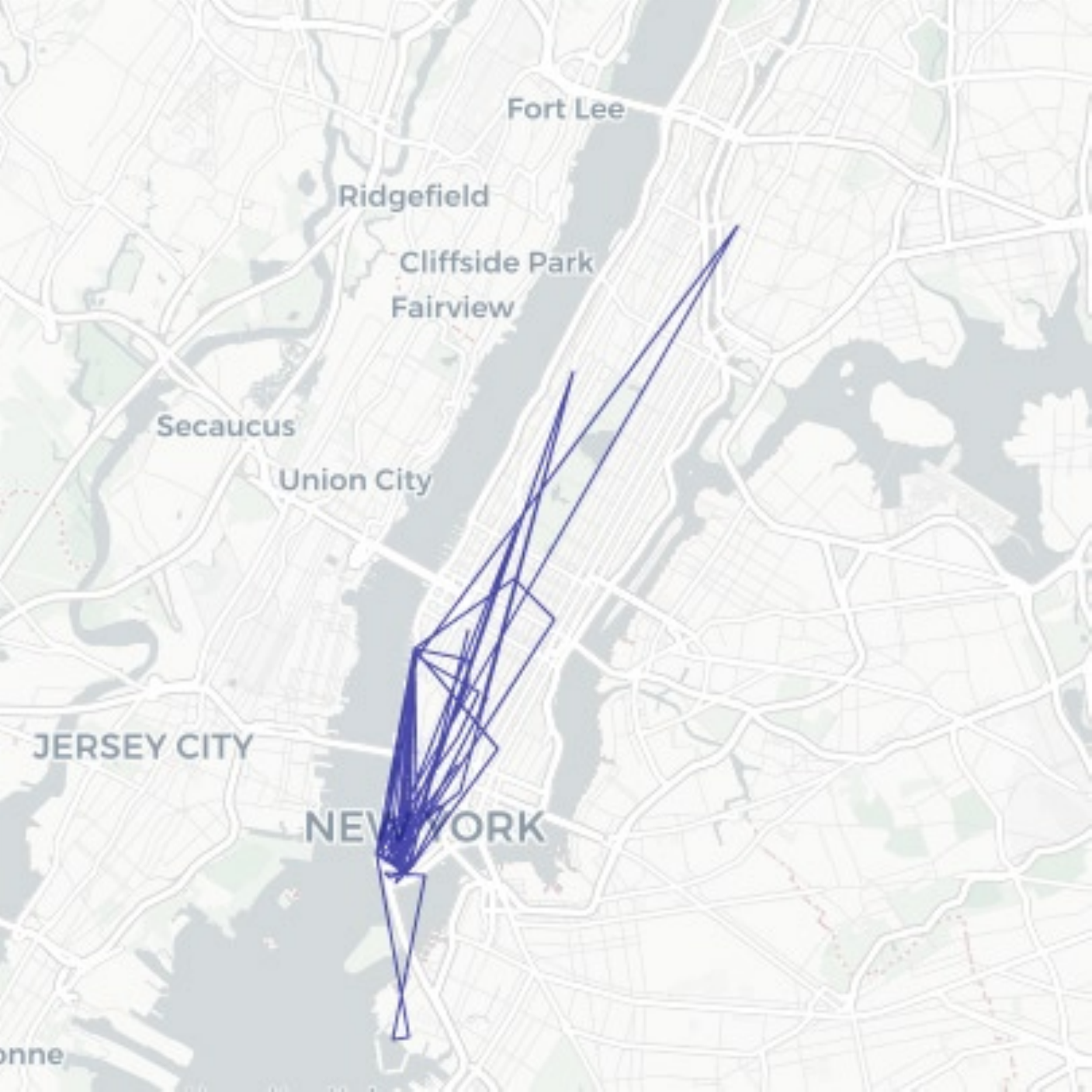}}
\hspace{2mm}
    \subfigure[]{
\includegraphics[width=0.45\textwidth]{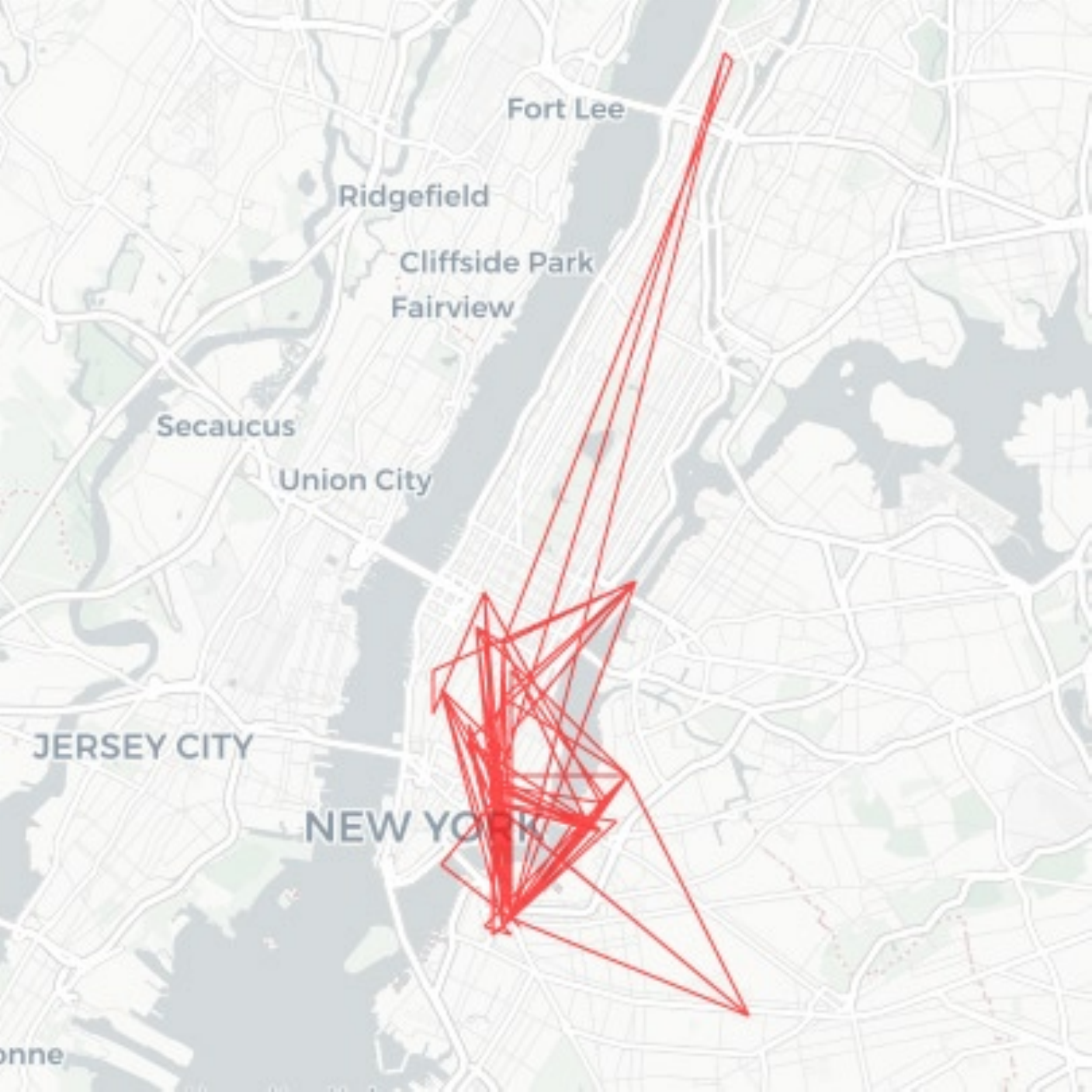}}
\caption{A spatial representation of the trajectory of a real individual (a) and a synthetic individual (b); the latter is generated using the STS-EPR with the weighted spatial tessellation $L_{\mbox{\tiny REL}}(250)$.
Figures generated with \textit{scikit-mobility} \cite{scikit_mobility}.
}
\label{fig:real_syn_traj}
\end{figure}

\begin{figure}[!h]
\centering
    \subfigure[]{
\includegraphics[width=0.45\textwidth]{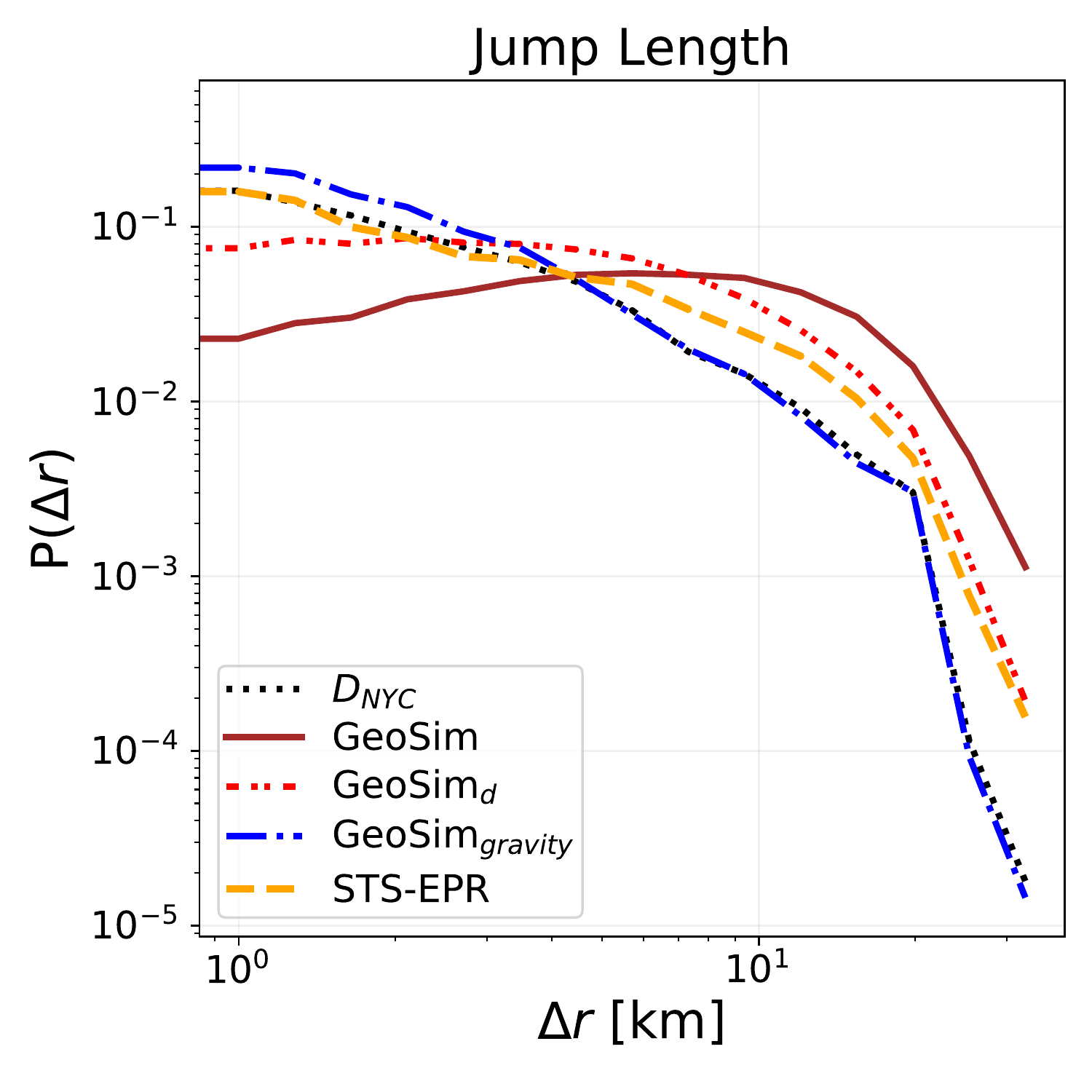}}
\hspace{2mm}
    \subfigure[]{
\includegraphics[width=0.45\textwidth]{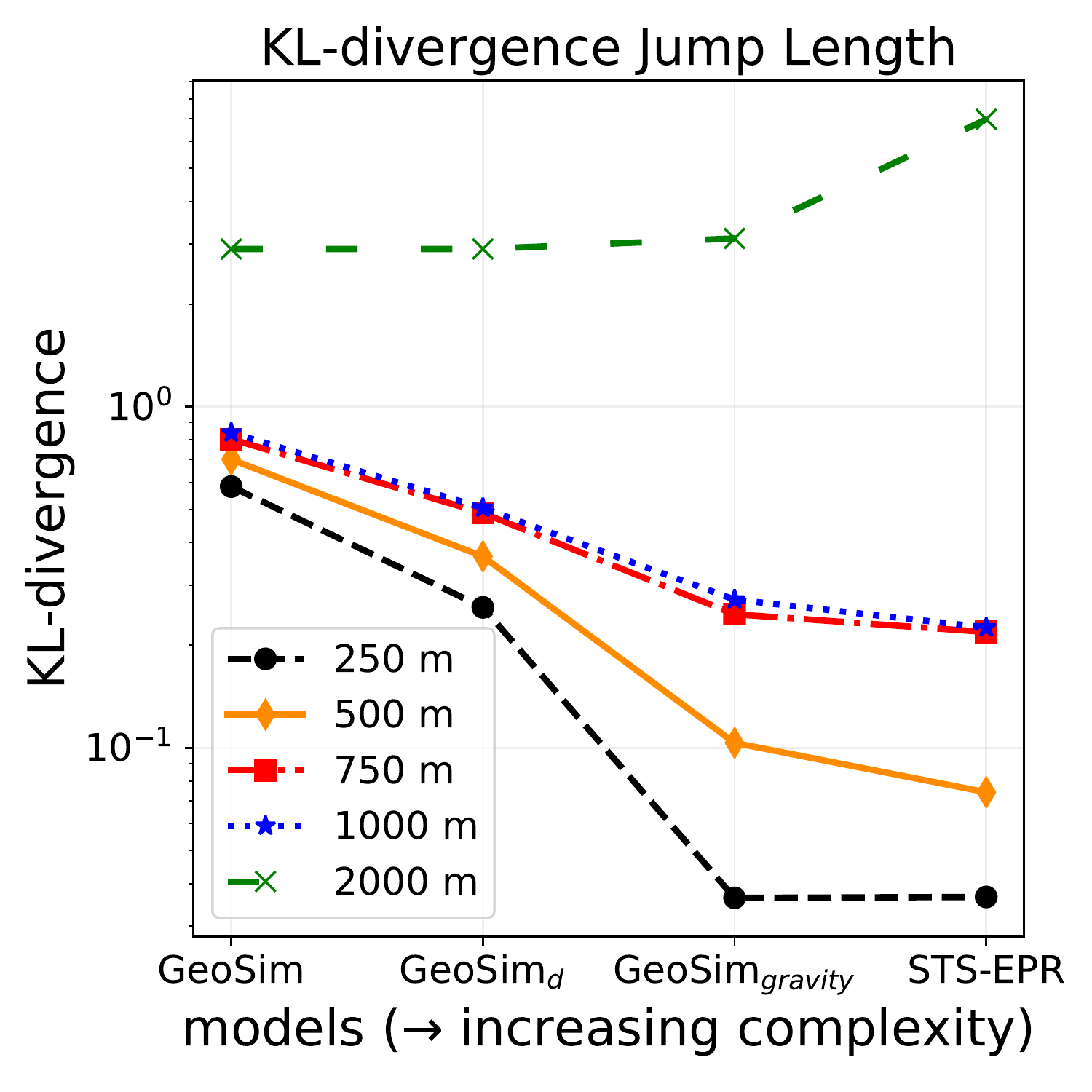}}
 \subfigure[]{
\includegraphics[width=0.45\textwidth]{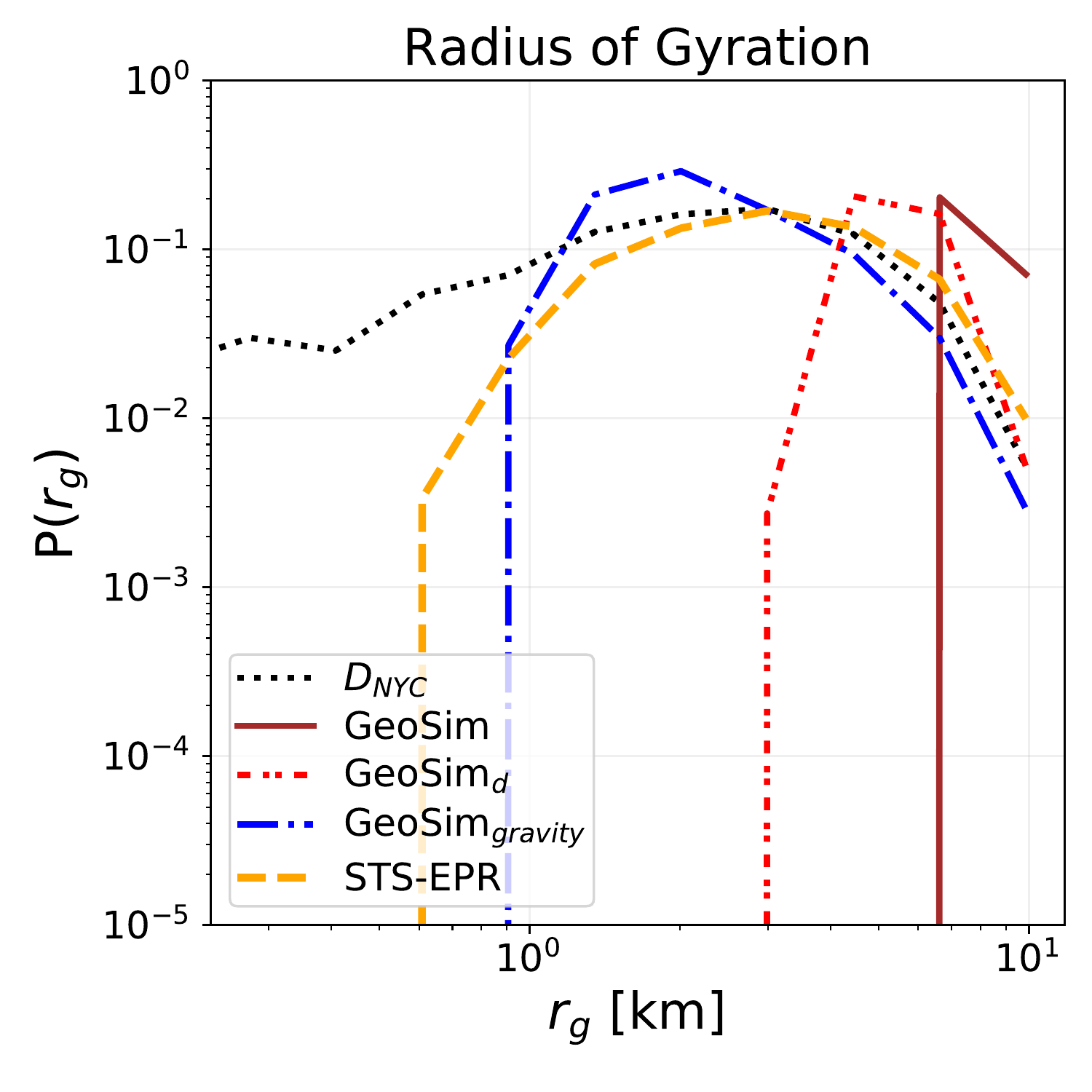}}
\hspace{2mm}
    \subfigure[]{
\includegraphics[width=0.45\textwidth]{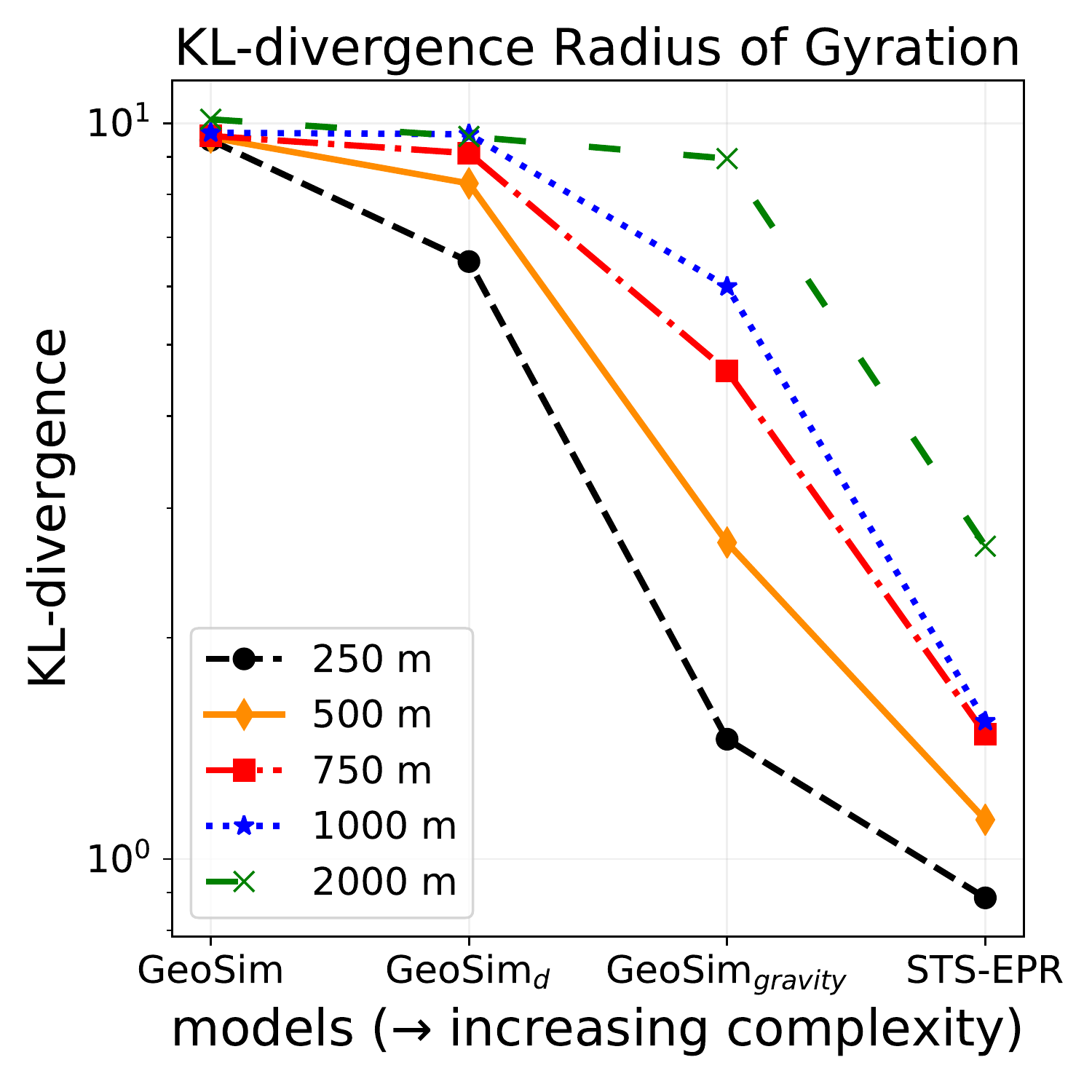}}
\caption{The probability density function of the jump length (a) for the real trajectories and for each of the proposed spatial mechanisms computed with the tessellation $L_{\mbox{\tiny REL}}(250)$.
The Kullback-Leibler divergence (b) for each model and for each granularity of the weighted spatial tessellation. 
None of the presented models can reproduce the radius of gyration for small values (c). The Kullback-Leibler divergence for each model and for each granularity of the weighted spatial tessellation (d).}
\label{fig:jl_results}
\end{figure}

The experiments show that the role of both the model mechanisms and the tessellation granularity is crucial to produce realistic trajectories.
The probability density function of the spatial measures computed over the generated trajectories, shapes according to the mechanism used in the generative model.
For what concerns the jump length (Figure \ref{fig:jl_results}), in GeoSim, no mechanism takes into account the spatial distance between locations, consequently, the model can not even replicate correctly the monotonicity of the distribution.
In $\mbox{GeoSim}_d$, with the introduction of a mechanism that models the spatial distance between locations, the probability density function of the jump length follows the power-law behavior; the tendency of the individuals to move at small rather than long distances is preserved.
However, taking into account only the distance is not sufficient, $\mbox{GeoSim}_d$ underestimates small-distance trips and overestimates long-distance trips.
With the introduction of the gravity model, $\mbox{GeoSim}_{gravity}$ generates trajectories that reproduce the distribution of distances accurately, although slightly overestimating small displacements.
To reproduce the jump length distribution accurately, as shown in Figure \ref{fig:jl_results}, the granularity of the tessellation plays a crucial role together with the mechanism used in the generative model.
With a fine-grained weighted spatial tessellation, the model generates more realistic trajectories (Figure \ref{fig:real_syn_traj} shows a real and a synthetic trajectory).

\begin{figure}[!h]
\centering
    \subfigure[]{\label{fig:cinss}
\includegraphics[width=0.45\textwidth]{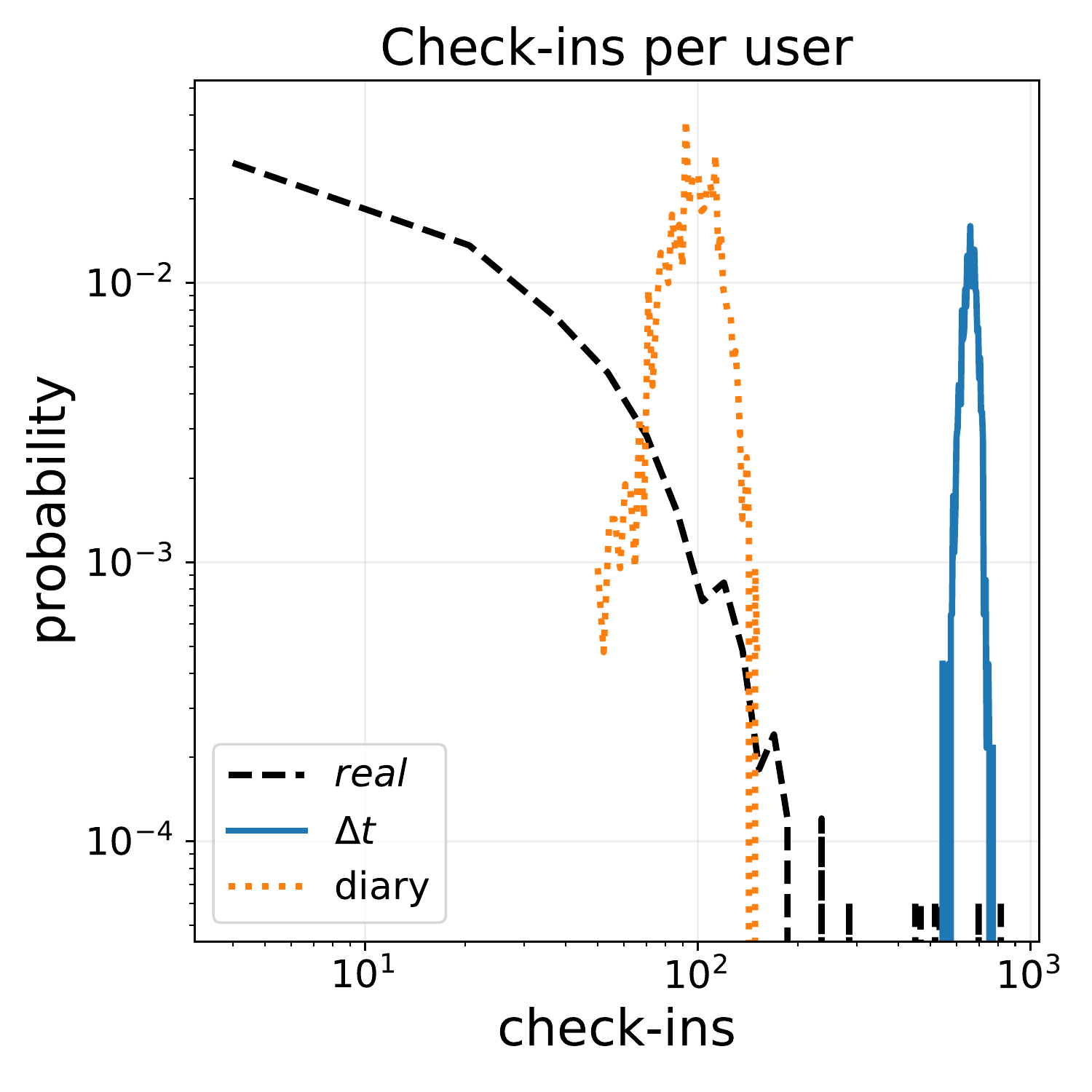}}
\hspace{2mm}
    \subfigure[]{\label{fig:unc_ent}
\includegraphics[width=0.45\textwidth]{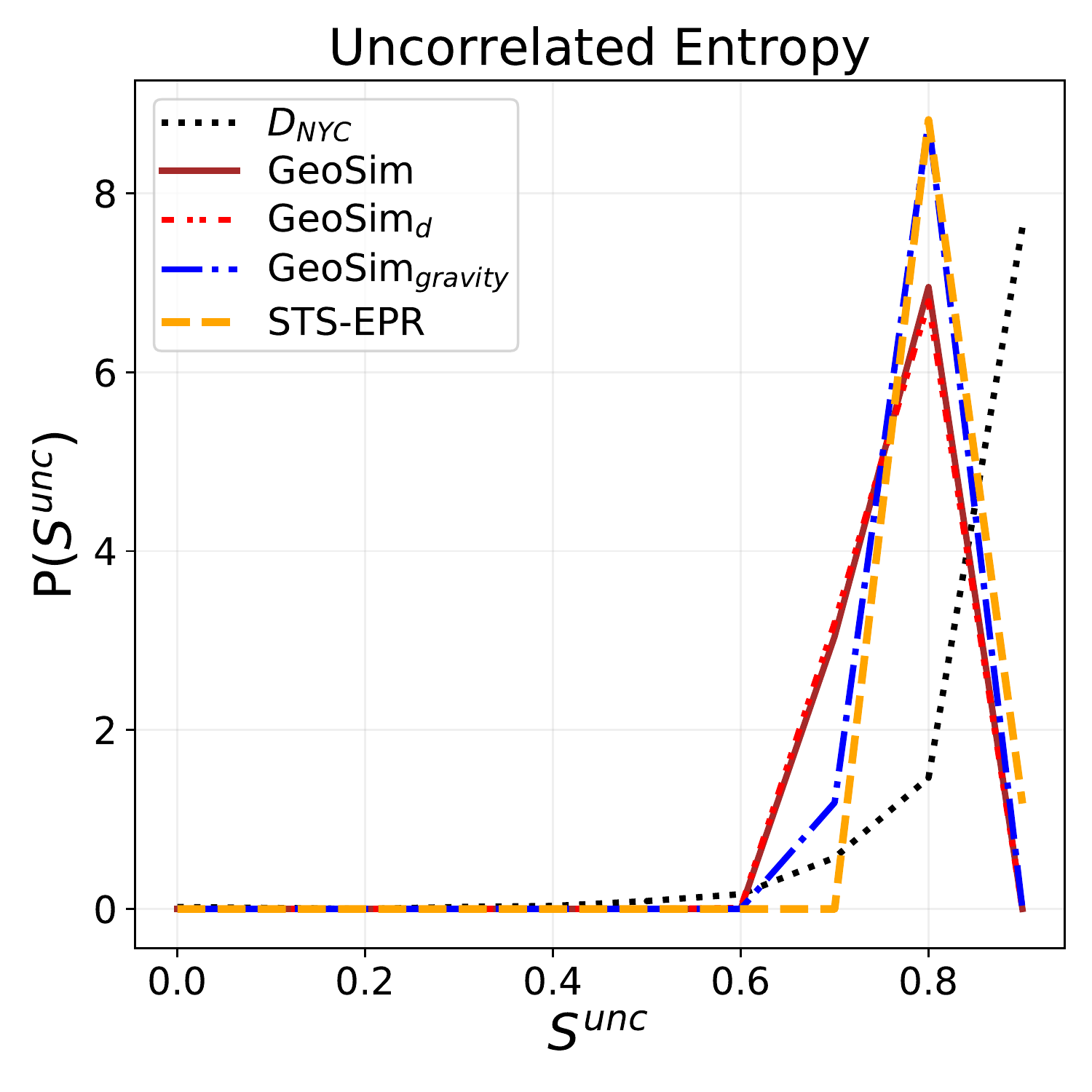}}
\caption[Number of check-ins for real and synthetic individuals]{The probability density function of the number of check-ins made by the real individuals and for models that pick the waiting time from the empirical distribution of Song et al. \cite{song_limits} and for the model that use the diary generator (a).
The probability density function of the uncorrelated entropy (b).}
\end{figure}

The considerations made for the jump length also hold for the radius of gyration; the typical spatial spread of the agents is mechanism and tessellation dependent (Figure \ref{fig:jl_results}).
With every introduction of a more sophisticated mechanism the models generate more realistic synthetic trajectories.
STS-EPR produces the most accurate trajectories, outperforming $\mbox{GeoSim}_{gravity}$ in terms of Kullback-Leibler divergence despite both use the gravity law. This can be due to the fact that the number of check-ins per user using the diary generator is similar to the real ones, while the other models overestimate this measure, as shown in Figure \ref{fig:cinss}.
None of the proposed models are able to reproduce correctly the radius of gyration for values smaller than $1 km$.

\begin{figure}[!h]
\centering
    \subfigure[]{\label{fig:loc_freqs}
\includegraphics[width=0.45\textwidth]{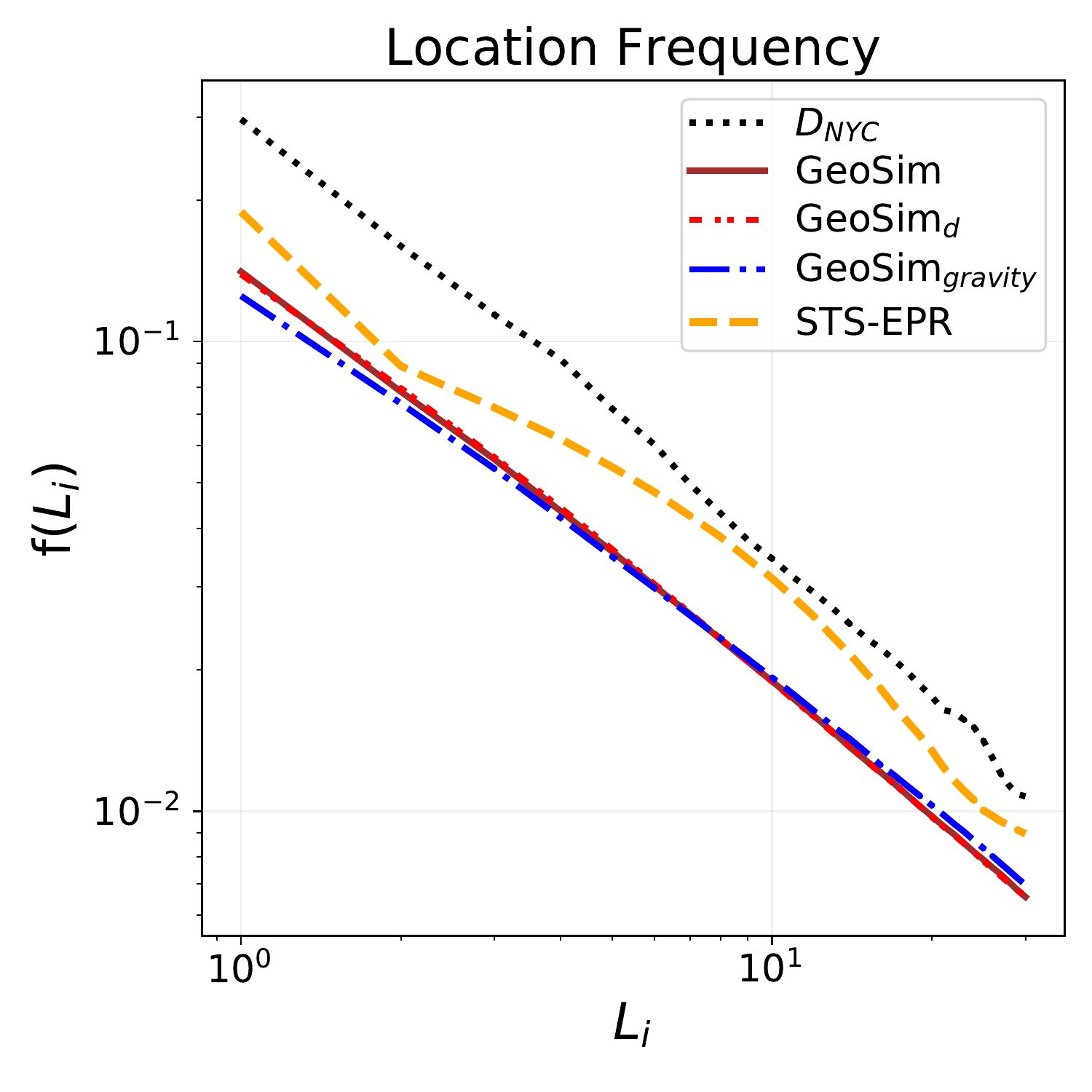}}
\hspace{2mm}
    \subfigure[]{\label{fig:loc_freqs_tex}
\includegraphics[width=0.45\textwidth]{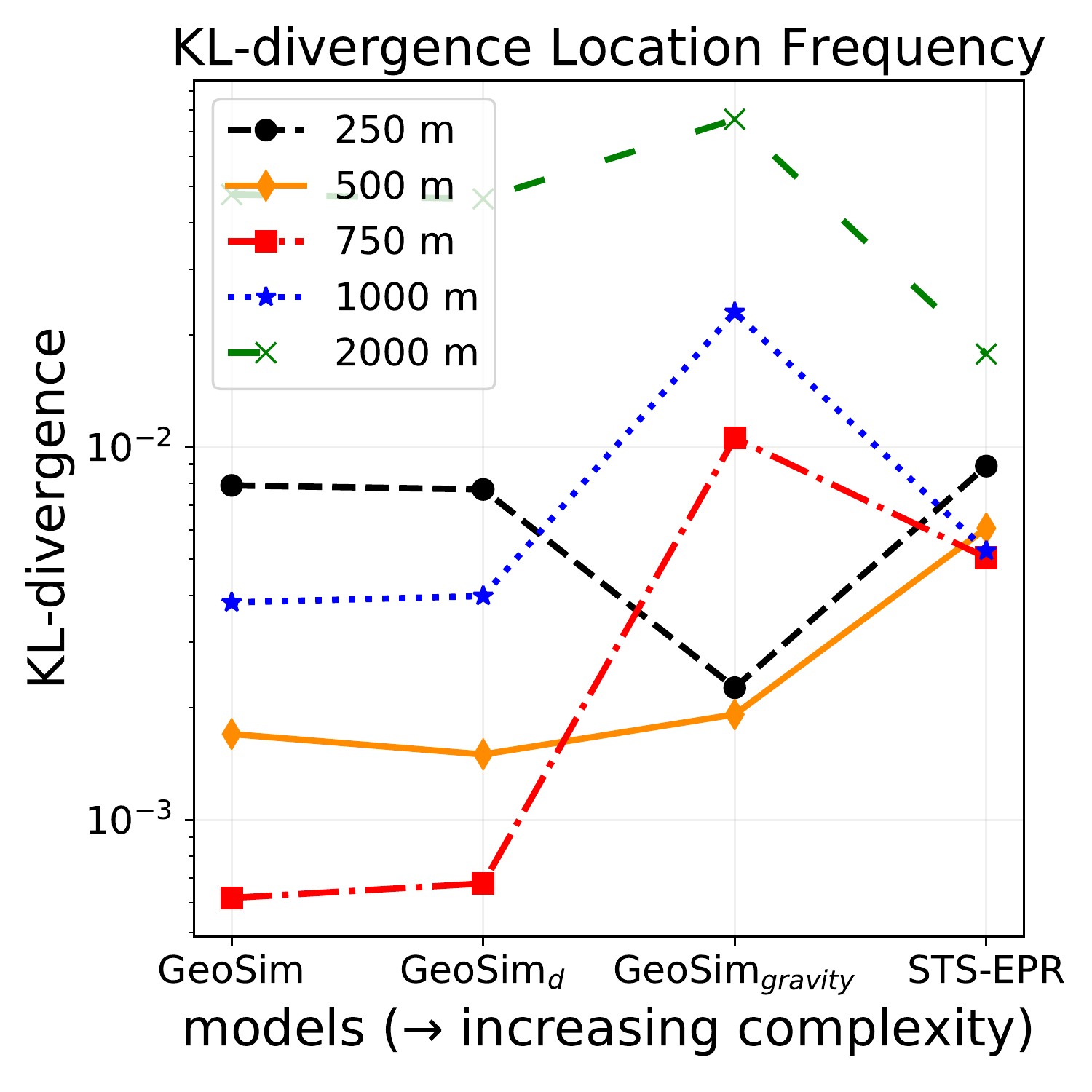}}
    \subfigure[]{\label{fig:visits}
\includegraphics[width=0.45\textwidth]{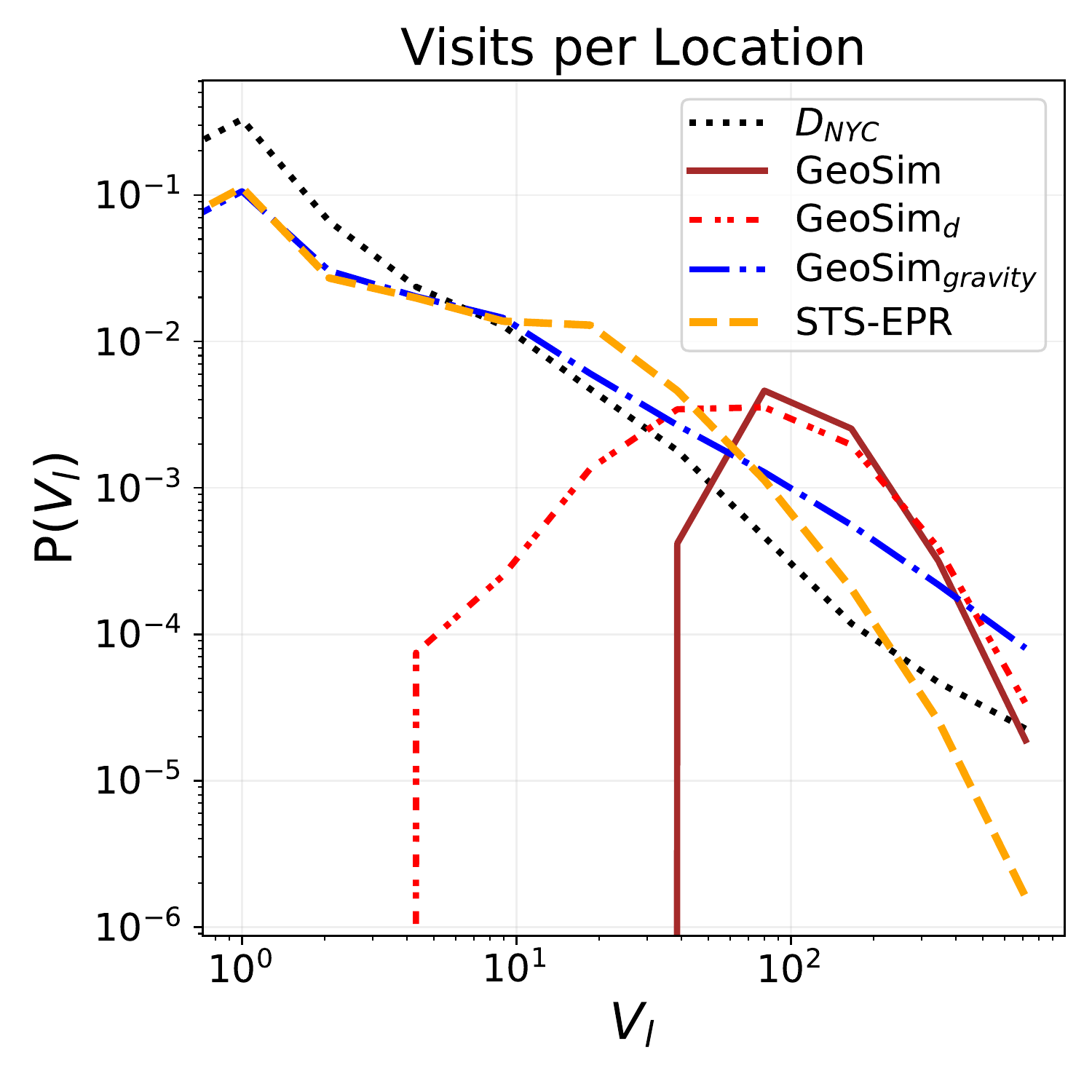}}
\hspace{2mm}
    \subfigure[]{\label{fig:conf_visits}
\includegraphics[width=0.45\textwidth]{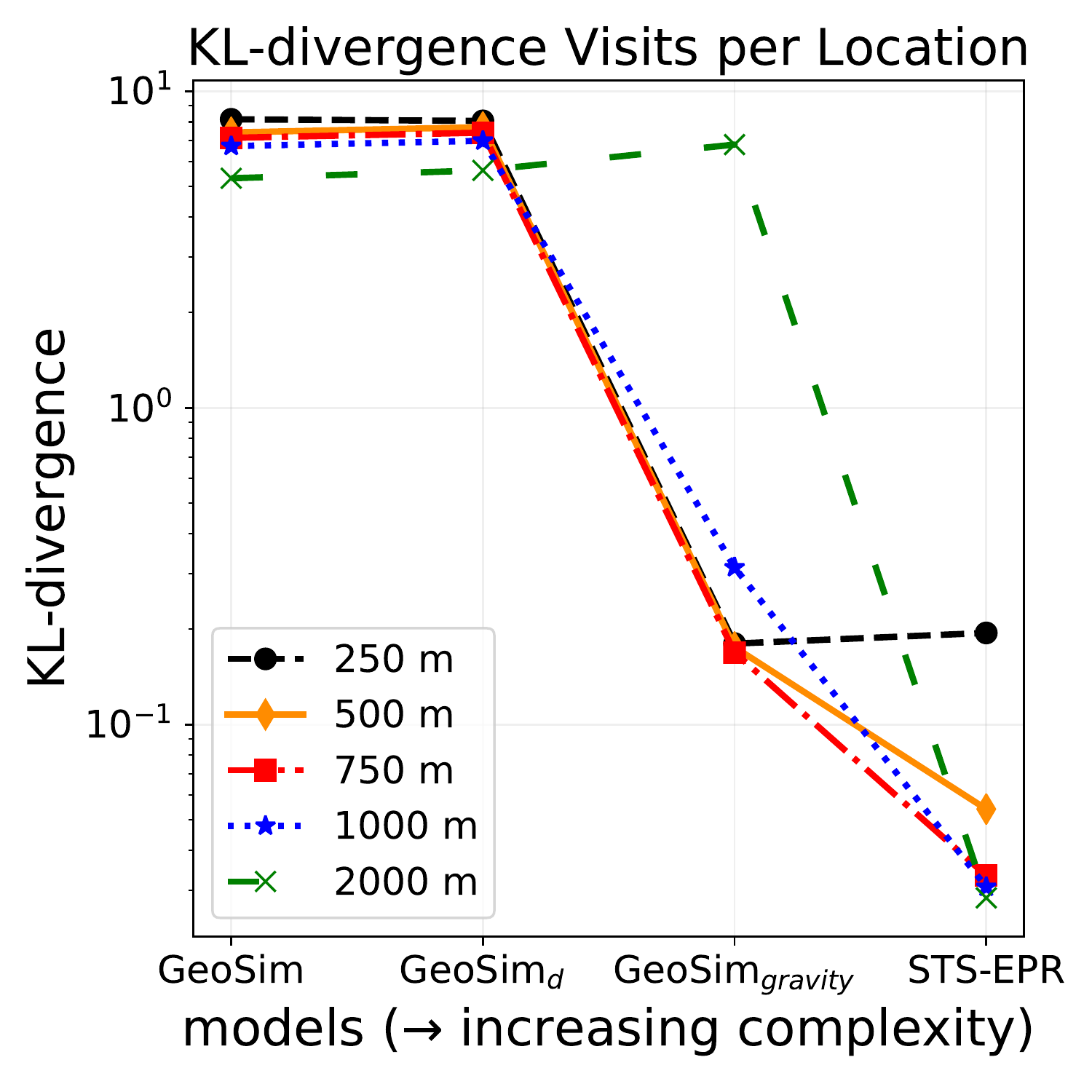}}
\caption[The probability density function of the top 20 locations for real and synthetic trajectories for the area of New York City]{The probability density function of the top 20 locations visited by an individual (a) and the Kullback-Leibler divergence for each model and tessellation granularity; in this case a fine granularity does not ensure better result.
The models without the gravity law mechanism overestimate the number of visits per location (c). The tessellation does not play a crucial role except for the STS-EPR model.}
\end{figure}

The location frequency distribution of the real data is better reproduced by STS-EPR (Figure \ref{fig:loc_freqs}); STS-EPR underestimates the location frequency of the top ten locations visited by the individuals.
For this measure the use of the fine-grained tessellation does not ensure better results; all the weighted spatial tessellation with size $\leq 1000\;m$ produce good result in terms of KL-divergence (order of $10^{-3}$ for different models) as shown by the plot in Figure \ref{fig:loc_freqs_tex}.

The distribution of the visits per location, equivalent at the relevance of each location, for the generated trajectories changes according to the individual exploration mechanism used in the model. The introduction of more sophisticated mechanisms produces trajectories with a more realistic number of visits per location.
Both GeoSim and $\mbox{GeoSim}_d$ underestimate the number of locations with less than 30 and 50 visits respectively (Figure \ref{fig:visits}).
With the introduction of the gravity law and the concept of relevance, $\mbox{GeoSim}_{gravity}$ and STS-EPR replicates accurately the power-law behavior of the number of visits per location.
The choice of the spatial partition is crucial for the number of visits; for the baseline model as well as $\mbox{GeoSim}_{d}$ and $\mbox{GeoSim}_{gravity}$ the tessellation used does not play a crucial role: all the tessellation with granularity $\leq 1000 m$ produce similar results.
In STS-EPR the results are better with a tessellation $> 250 m$ because the agents perform a small number of displacement. Consequently, an aggregated number of visits considering larger locations are more similar to the real one.

\begin{figure}[!h]
\centering
    \subfigure[]{\label{fig:wt}
\includegraphics[width=0.47\textwidth]{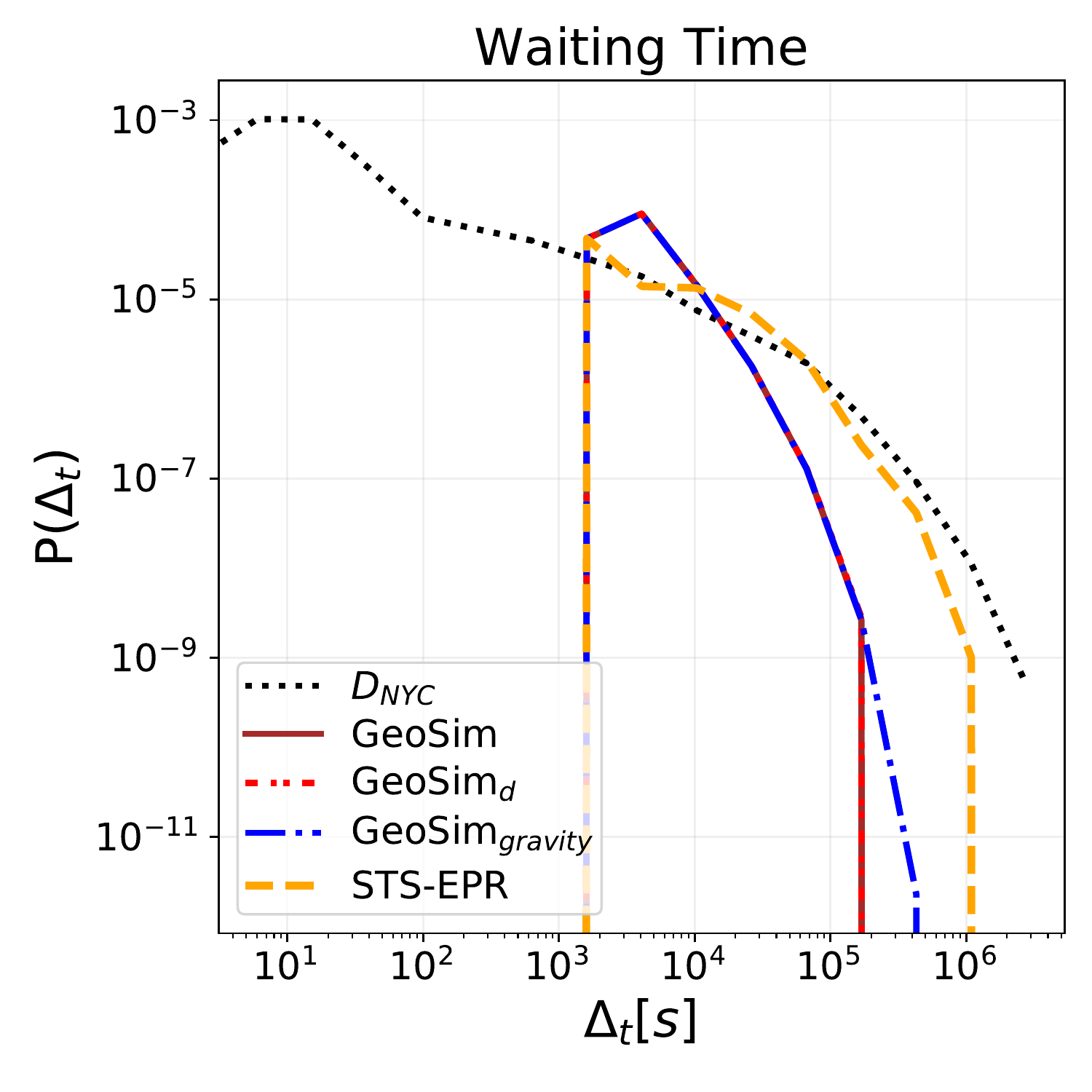}}
\hspace{2mm}
    \subfigure[]{\label{fig:wt_1h}
\includegraphics[width=0.45\textwidth]{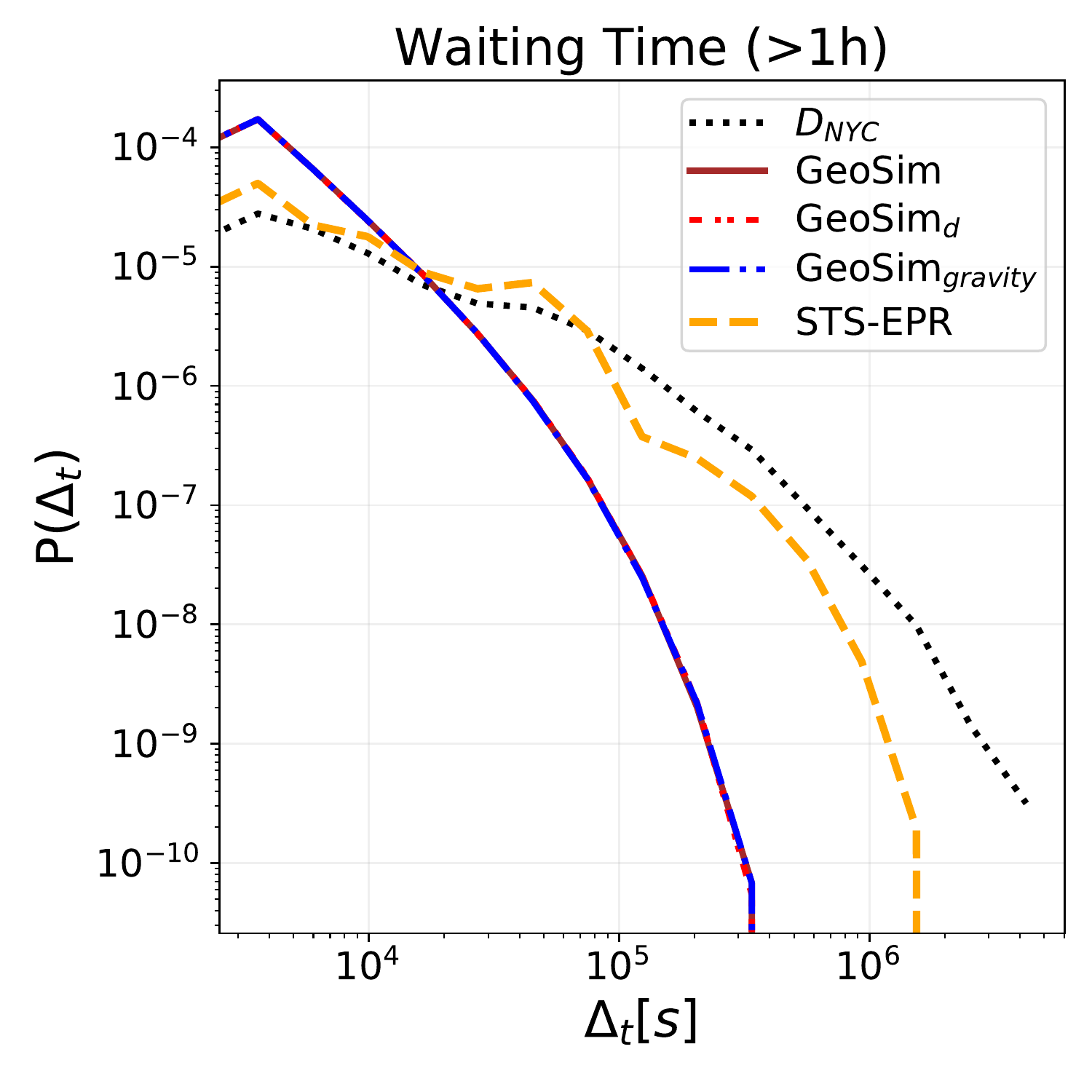}}
    \subfigure[]{\label{fig:wt_remap}
\includegraphics[width=0.45\textwidth]{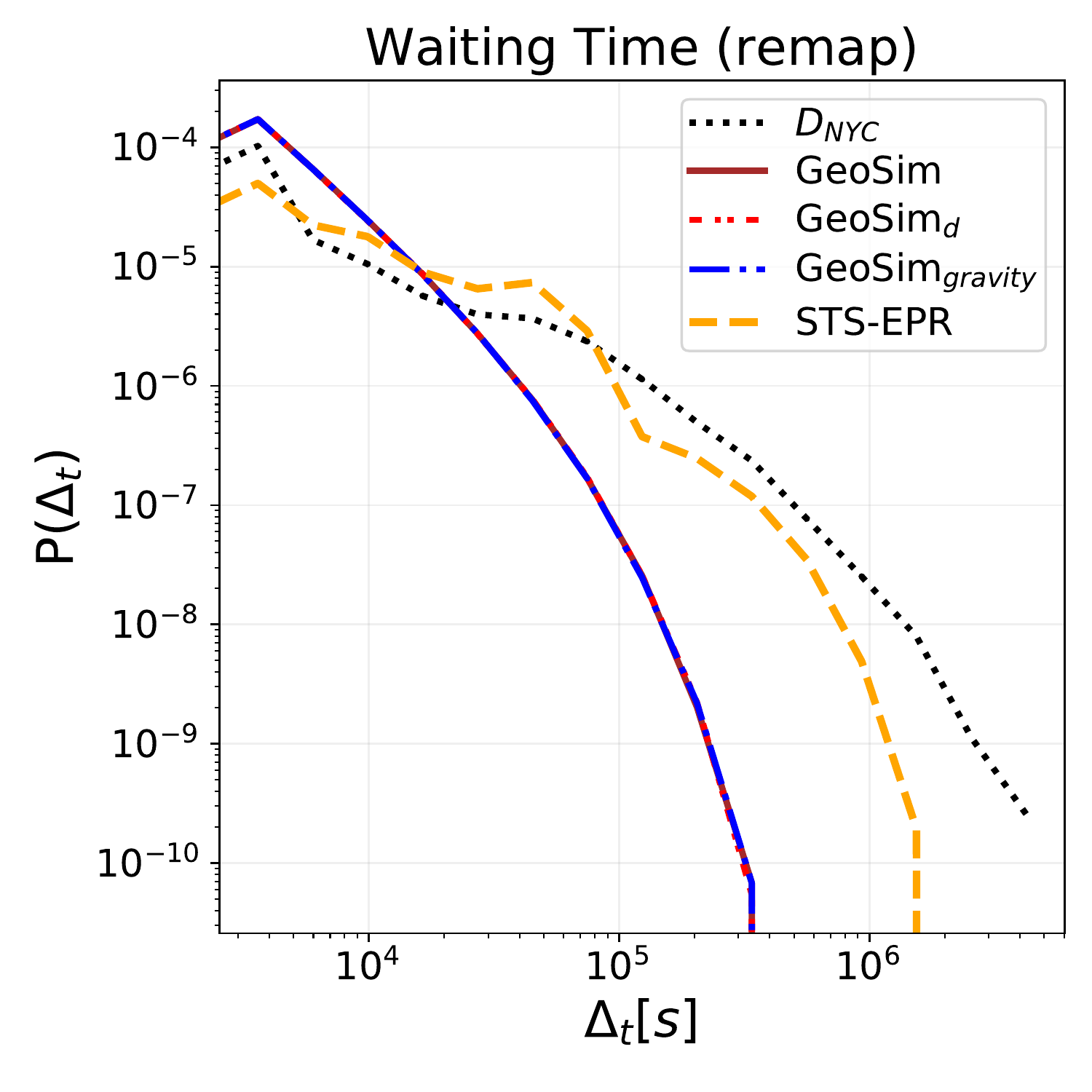}}
\caption[The three options considered for dealing with waiting time values under one hour]{The three options considered for dealing with waiting time values under one hour; in the first case we compare the distributions with all the values (a), in the second options we cut the real distribution (b) and finally we re-map all the values $<1h$ in the value $1h$.}
\end{figure}

The time spent in a location, namely the waiting time, has a minimum temporal resolution of one hour for the proposed model, since $min_{wt}=1$. Since in the real distribution of the waiting time, there are values $<1h$, to compare the synthetic and the real distributions we consider three cases: (i) we compare the distributions as we do with the other measures (Figure \ref{fig:wt}); (ii) we cut from the real distribution the waiting times $<1h$ (Figure \ref{fig:wt_1h}); and (iii) we map all the values $<1h$ in $1h$, preserving the number of points in the distribution (Figure \ref{fig:wt_remap}).
All the models that assign the waiting time using the empirical distribution computed by Song et al. \cite{im_model_song} behave in the same way, while the model with use the diary generator is able to reproduce more accurately the characteristic waiting time of the set of individuals in New York City.

\begin{figure}[!h]
\centering
    \subfigure[]{\label{fig:act_h}
\includegraphics[width=0.45\textwidth]{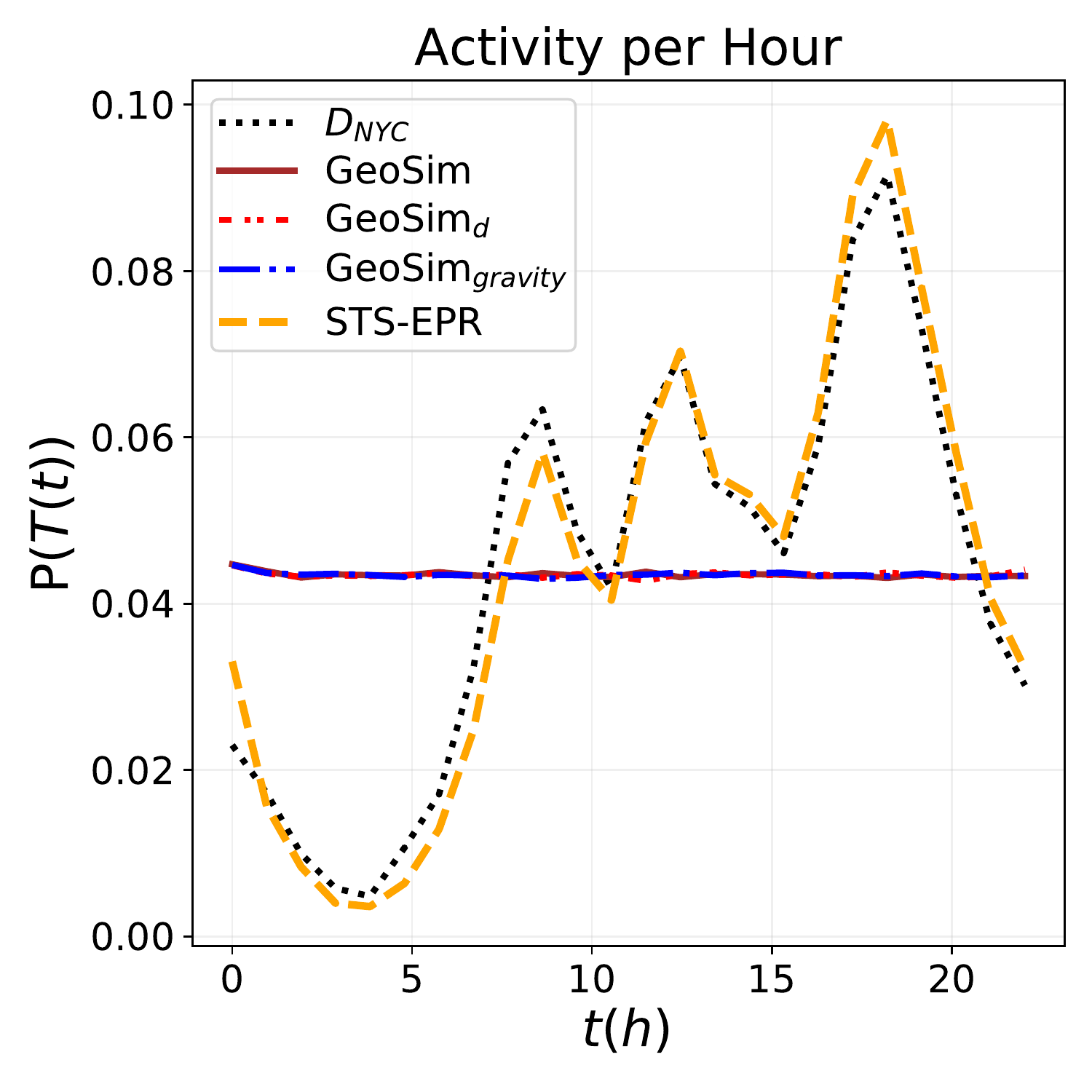}}
\hspace{2mm}
    \subfigure[]{\label{fig:tex_act_h}
\includegraphics[width=0.45\textwidth]{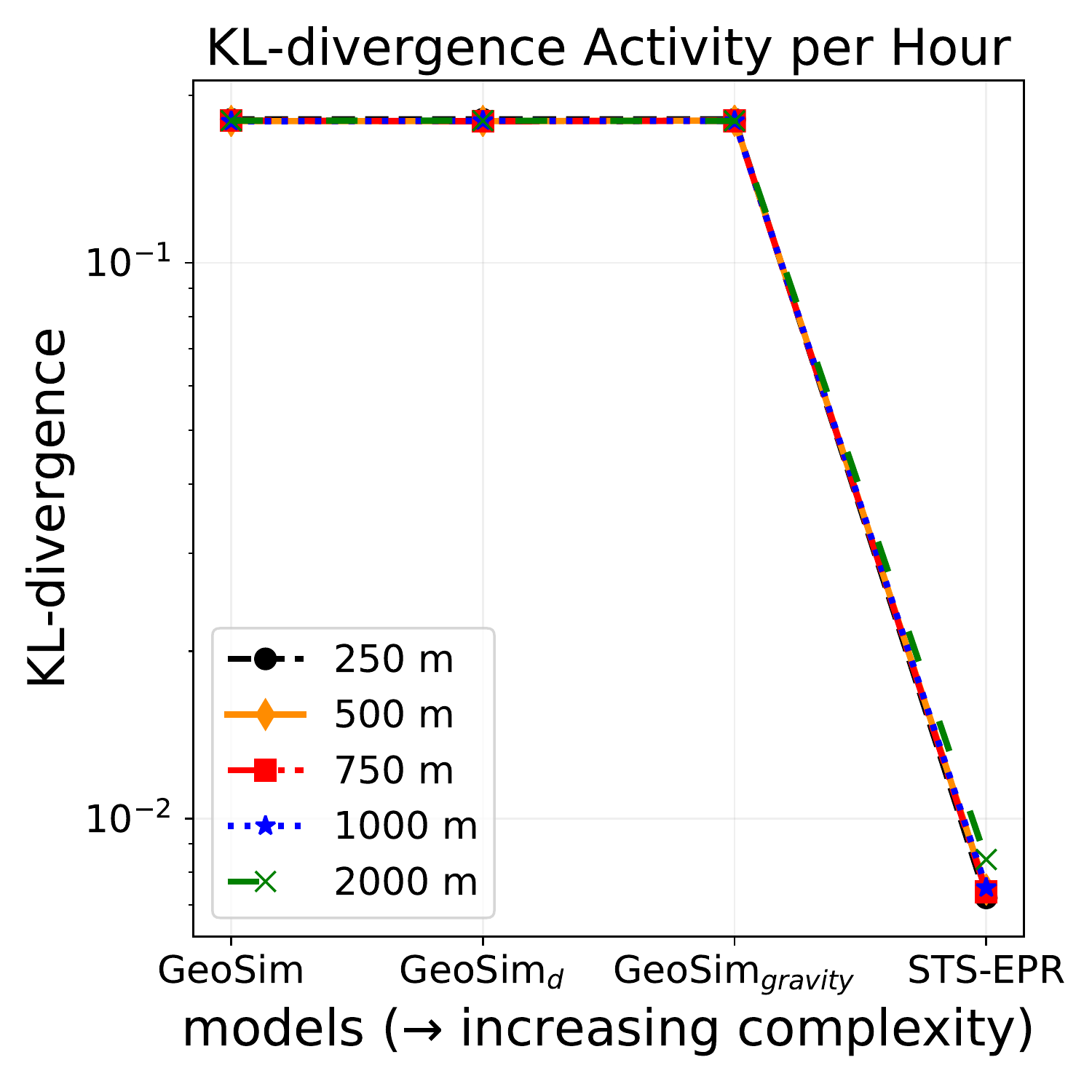}}
\caption[The circadian rhythm of the individuals in the dataset $D_{NYC}$]{The only model able to reproduce the circadian rhythm of the individuals in the dataset $D_{NYC}$ is $\mbox{GeoSim}_{gravity}$(a); in the other models the probability of moving at a certain hour is uniform. The choice of the tessellation does not influence this temporal measure (b).}
\end{figure}

The number of trips made at each hour of the day, namely the activity per hour, depicts the tendency of individuals to move at a certain hour of the day.
This measure is affected only by the time of the movements of individuals and neither by the mechanism on which they choose the next location to explore nor by the spatial tessellation used during the experiments.
As we can see in Figure \ref{fig:tex_act_h}, the models produce the same results with every tessellation, and the only model able to reproduce the circadian rhythm of the individuals is STS-EPR. This is caused by the fact that STS-EPR takes into account the preference of individuals to move at specific times, while GeoSim, $\mbox{GeoSim}_{d}$ and $\mbox{GeoSim}_{gravity}$ takes into account only the waiting times, without considering the hour of the day nor the preference of an individual.

The behavior of the synthetic agents is slightly more predictable than the real counterparts; this can be caused by the fact that in the presented models, a fraction $\alpha$ (the social factor) of the displacements of an agent are based on the previous movements of its social contacts; this can increase the predictability of the movements of the agents.
The model that reproduces in a more similar way this characteristic is STS-EPR.

\begin{figure}[!h]
\centering
    \subfigure[][]{
\includegraphics[width=0.45\textwidth]{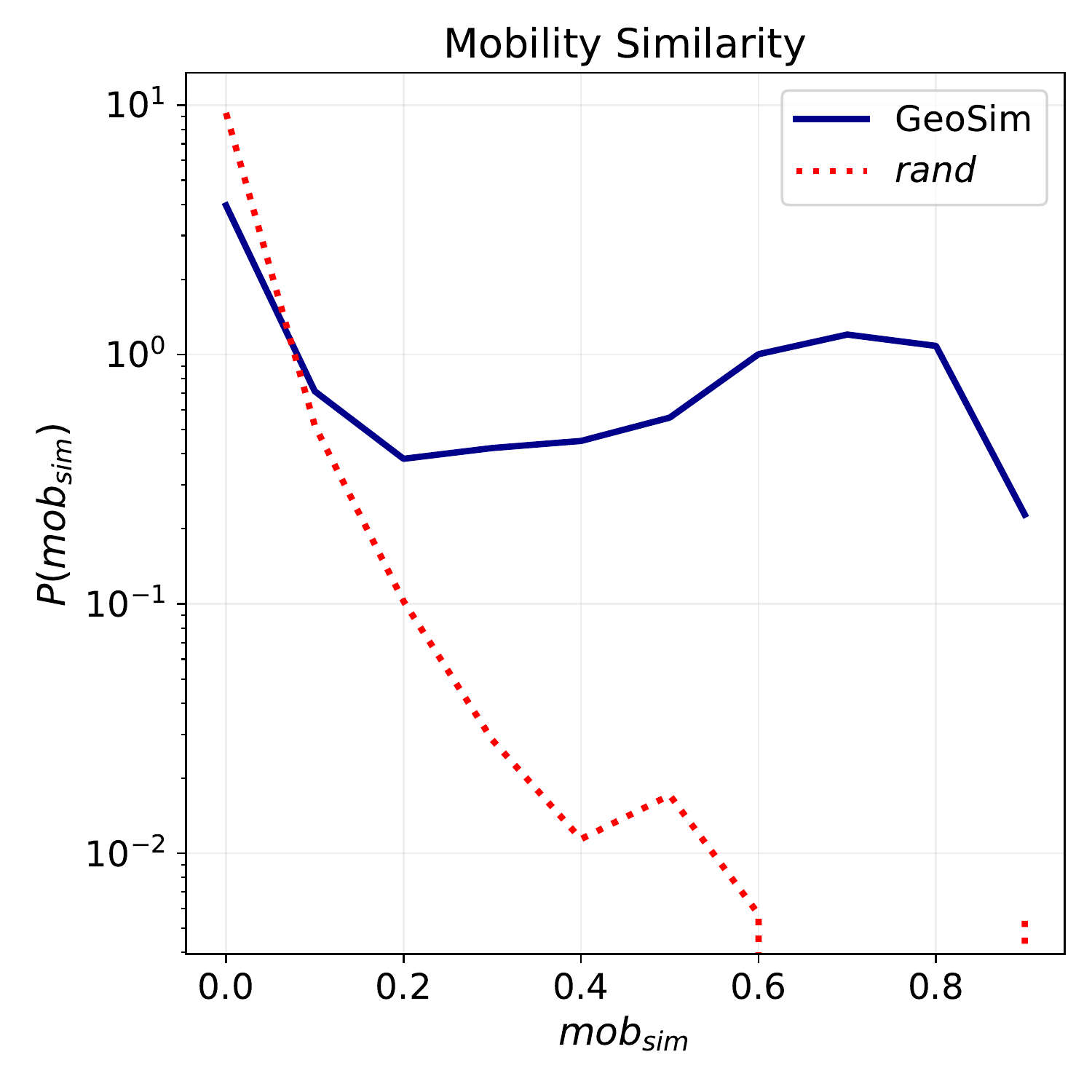}}
\hspace{2mm}
    \subfigure[][]{
\includegraphics[width=0.45\textwidth]{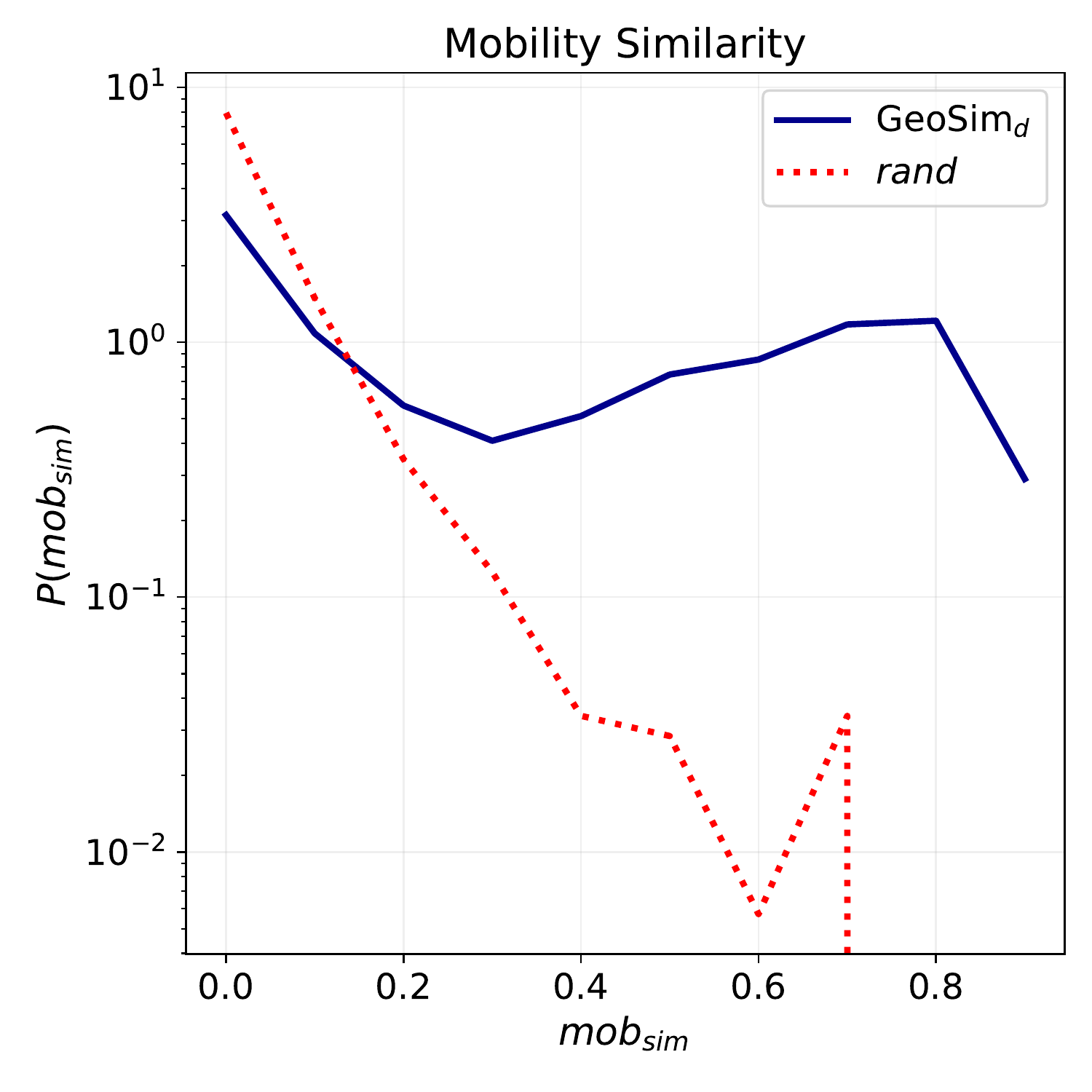}}
    \subfigure[][]{
\includegraphics[width=0.45\textwidth]{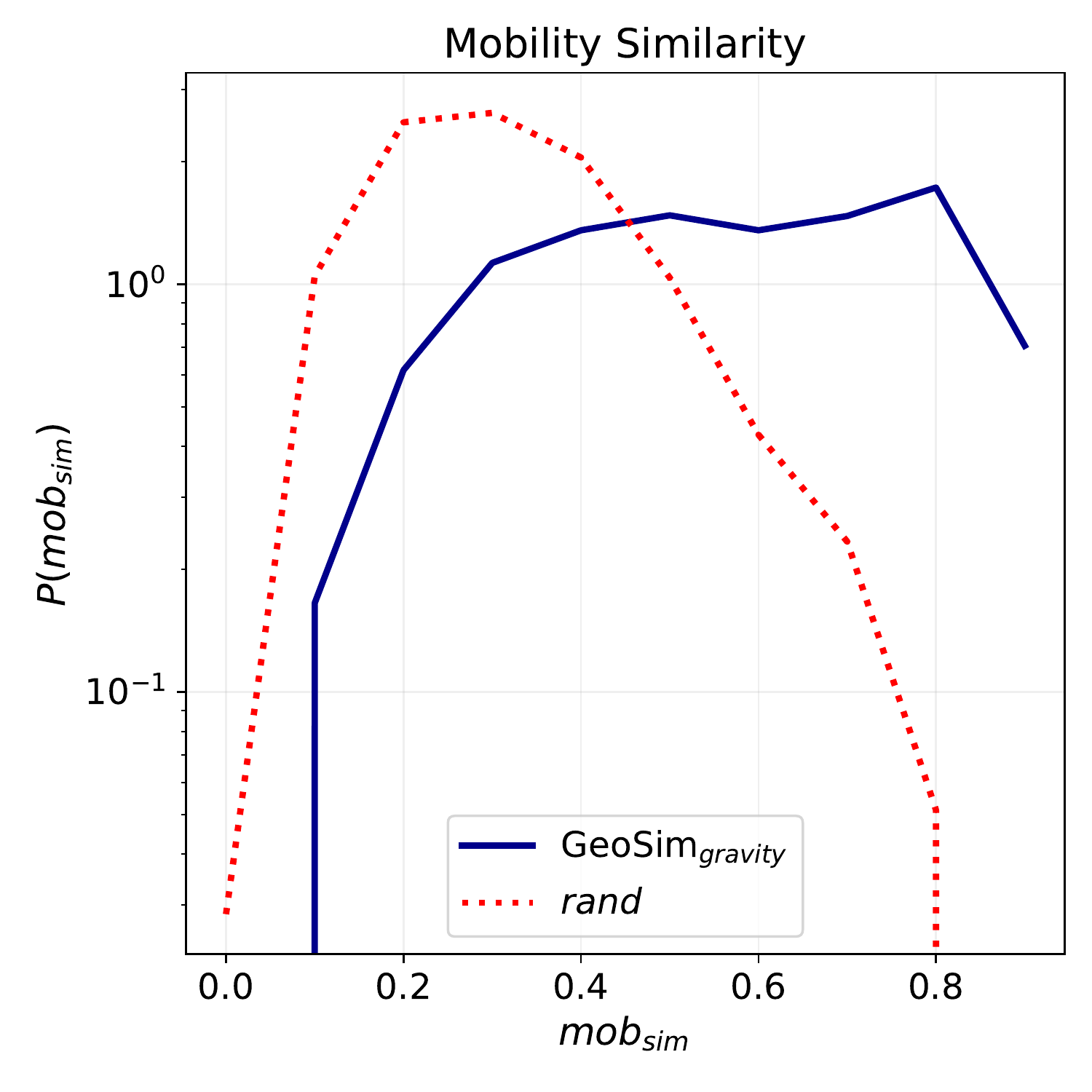}}
\hspace{2mm}
    \subfigure[][]{
\includegraphics[width=0.45\textwidth]{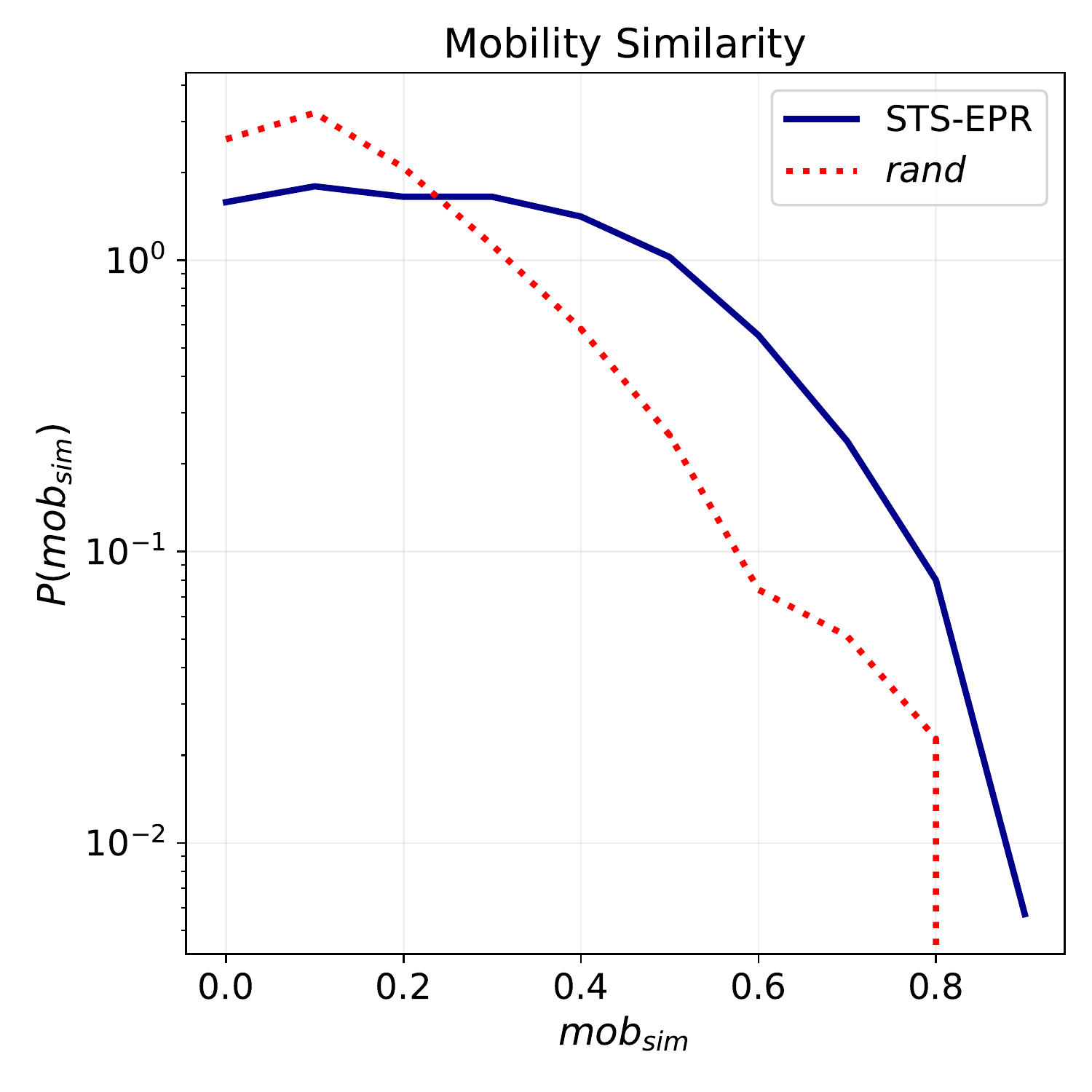}}
\caption[The distributions of the mobility similarity for the synthetic trajectories generated by different models]{The shape of the generated mobility similarity changes according to the model used; all the generative models preserve the correlation between human mobility and sociality, the movements of friends are more similar than those of strangers}
\label{fig:mob_02}
\end{figure}

\begin{figure}[!h]
\centering
    \subfigure[]{\label{fig:mob_sim_all}
\includegraphics[width=0.45\textwidth]{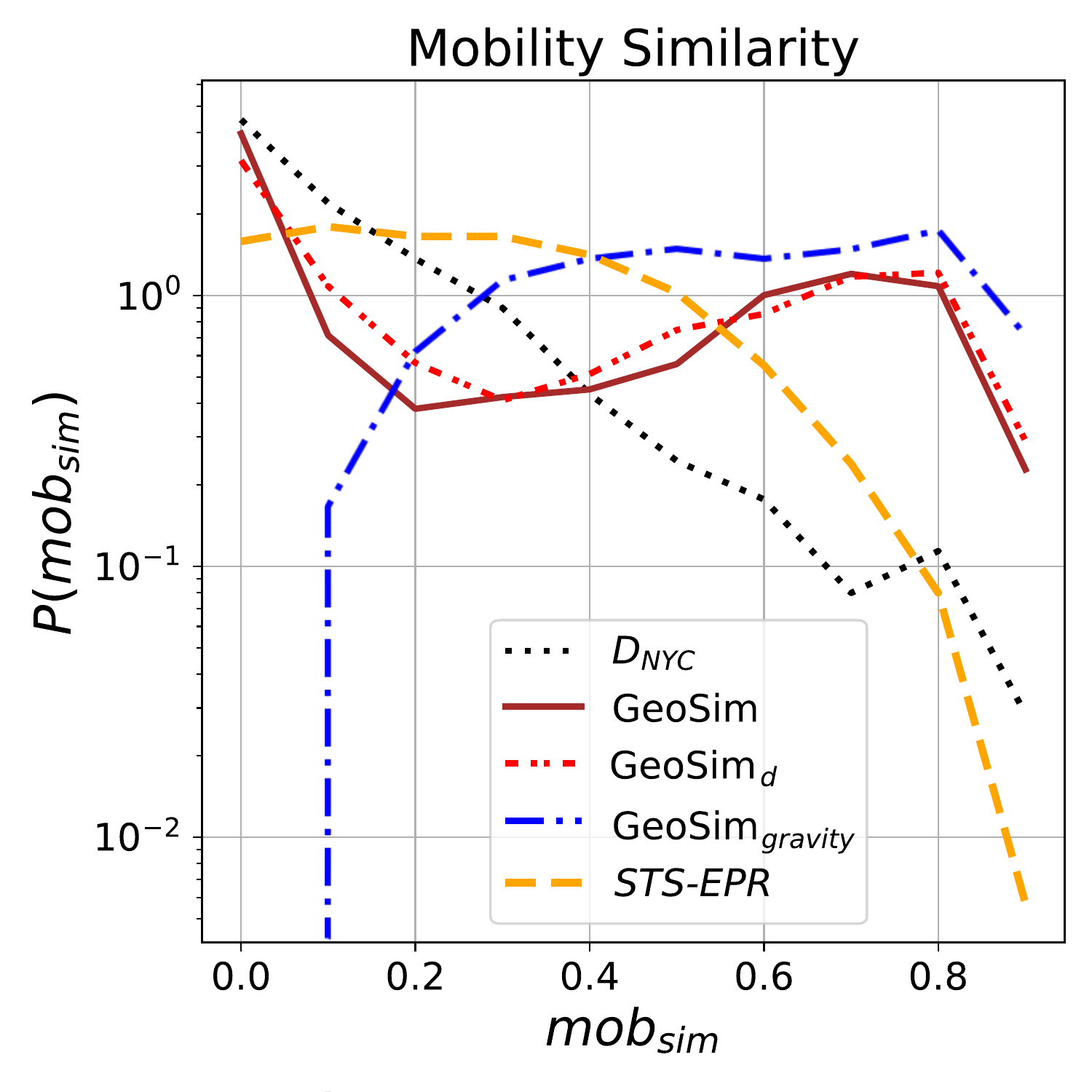}}
\hspace{2mm}
    \subfigure[]{\label{fig:conf_mob_sim}
\includegraphics[width=0.45\textwidth]{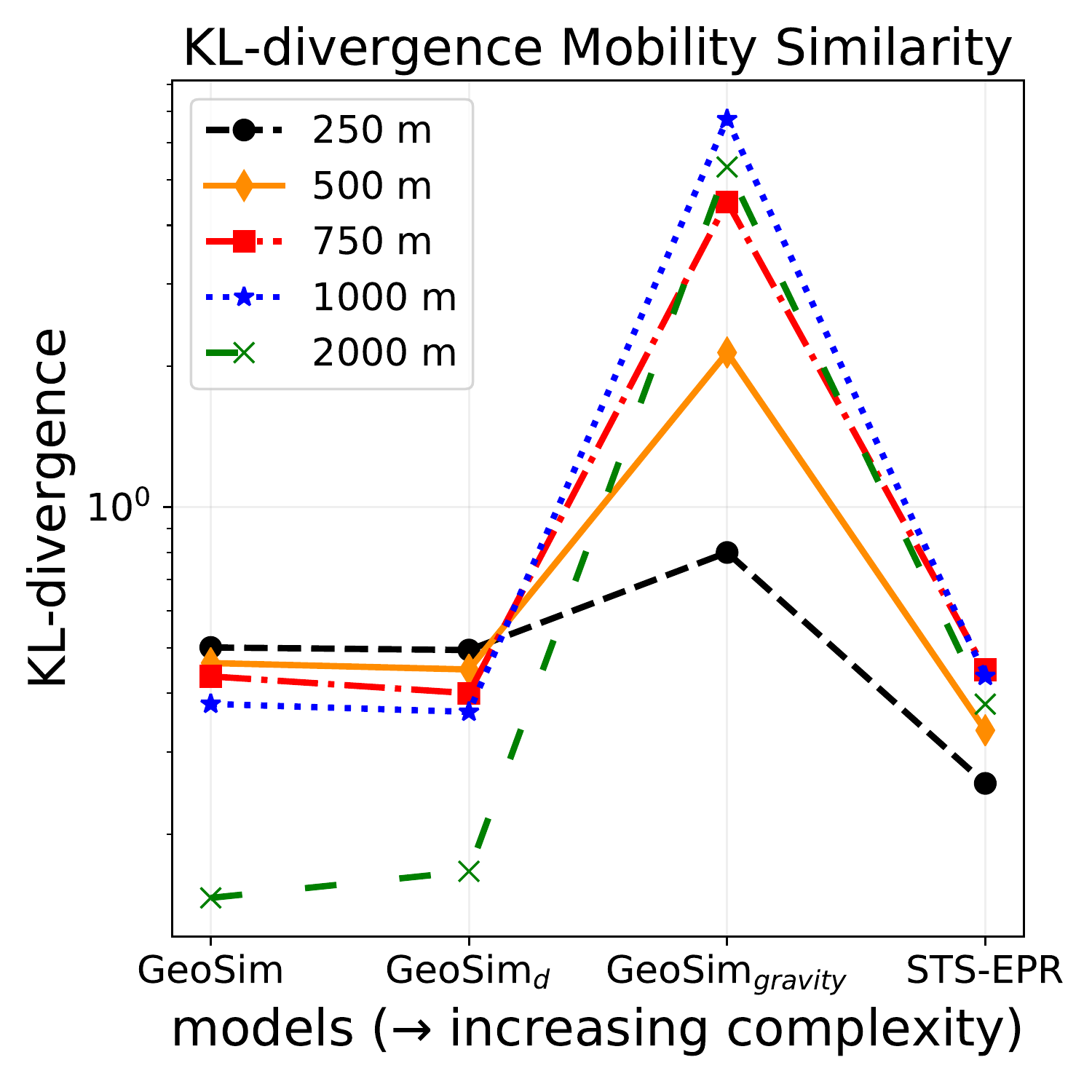}}
\caption[The probability density function of the mobility similarity for real and synthetic trajectories]{The probability density function of the mobility similarity (a); generally a fine grained tessellation produces better results, in GeoSim the good score with the tessellation $L_{\mbox{\tiny REL}}(2000)$ can be caused by the smaller dimension of the location vectors assigned at an agent, and consequently more likely to be similar at the one of its social contacts.}
\end{figure}

The probability density function of the mobility similarity changes according to the used model; as we can see from Figure \ref{fig:mob_02}, the baseline model and $\mbox{GeoSim}_{d}$ produces similar shapes: the users connected in the graph $G_{\mbox{\tiny NYC}}$ have higher mobility similarity than random pairs of non connected users.
The correlation between mobility and sociality also holds for the trajectories produced by $\mbox{GeoSim}_{gravity}$ and STS-EPR; the model that reproduces the mobility similarity more accurately is STS-EPR (Figure \ref{fig:mob_02}(d)).
Except for the models that does not use the gravity law, the fine-grained tessellations produce better results (Figure \ref{fig:conf_mob_sim}).

Our experiments reveal three main results. First, the proposed extension STS-EPR produces synthetic trajectories having in general the best fit to the trajectories in the dataset $D_{\mbox{\tiny NYC}}$.
Second result is that the choice of the spatial mechanism and the temporal mechanism (empirical distribution or diary generator) used for picking the waiting time is important in order to reproduce accurately some properties of the mobility trajectories; in general the gravity-law is the crucial mechanism for reproducing correctly almost all the measure.
Last, the choice of the granularity of the weighted spatial tessellation depends on which measure we consider: for spatial measures (jump length and radius of gyration) as well as for the location frequency and mobility similarity a fine tessellation produces better results. In contrast, for the other measures a larger tessellation produces better trajectories.

\subsection{Summary of Results}\label{sect:table_results}
For each mobility measure we report the table of the scores obtained through the experiments, referred at the tessellation which gives the best overall result for that specific measure in terms of the five scores used to quantify the similarity between the real and the synthetic distributions. 
For each score, we report mean and standard deviation, the best values for each score are reported in bold.

\paragraph{Jump Length.}
The following table summarizes the scores for the jump length, using a tessellation $L_{\mbox{\tiny REL}}(250)$.
The best models are the ones that use the gravity-law mechanism in the choice of the next location to explore.

\begin{table}[H]
\footnotesize
\renewcommand*{\arraystretch}{1.2}
\begin{tabular}{@{}lcccc@{}}
\toprule
                          & GeoSim             & $\mbox{GeoSim}_{d}$                      & $\mbox{GeoSim}_{gravity}$            & STS-EPR       \\ \midrule
\multicolumn{1}{l|}{RMSE} & 0.1044$ \pm$0.0002 & 0.0821$ \pm$0.0004 & \textbf{0.0364$ \pm$0.0006}          & 0.0371$ \pm$0.001  \\
\multicolumn{1}{l|}{KL}   & 0.5838$ \pm$0.0052 & 0.2583$ \pm$0.004           & \textbf{0.0363$ \pm$0.0014} & 0.0366$ \pm$0.0016 \\
\multicolumn{1}{l|}{Hellinger} & 0.4298$ \pm$0.0009 & 0.2783$ \pm$0.0024 & 0.1149$ \pm$0.0032    & \textbf{0.1114$ \pm$0.004}  \\
\multicolumn{1}{l|}{spearman}  & 0.3412$ \pm$0.0213 & 0.9594$ \pm$0.0183 & \textbf{1.0$ \pm$0.0} & \textbf{1.0$ \pm$0.0}       \\
\multicolumn{1}{l|}{pearson}   & 0.2934$ \pm$0.0132 & 0.7201$ \pm$0.0048 & 0.9366$ \pm$0.0018    & \textbf{0.9787$ \pm$0.0031} \\ \bottomrule
\end{tabular}
\caption[Scores of the jump length measure]{Scores of the jump length measure.}
\label{tab:res_jl}
\end{table}

\paragraph{Radius of Gyration.}
The scores reflect the fact that all the proposed models are not able to reproduce small radii correctly (Figure \ref{fig:jl_results}). The most accurate model is STS-EPR, the latter is the only model able to reduce in a significant way the Kullback-Leibler divergence and the Hellinger-distance score.

\begin{table}[H]
\small
\renewcommand*{\arraystretch}{1.2}
\begin{tabular}{@{}lcccc@{}}
\toprule
                              & GeoSim           & $\mbox{GeoSim}_{d}$           & $\mbox{GeoSim}_{gravity}$    & STS-EPR               \\ \midrule
\multicolumn{1}{l|}{RMSE}      & 0.1064$\pm$0.0004  & 0.0957$\pm$0.0003 & 0.0627$\pm$0.0065 & \textbf{0.0345$\pm$0.0044} \\
\multicolumn{1}{l|}{KL}       & 9.481$\pm$0.411  & 6.4871$\pm$0.0229 & 1.4549$\pm$0.0442 & \textbf{0.8855$\pm$0.2103} \\
\multicolumn{1}{l|}{Hellinger} & 0.6507$\pm$0.0031  & 0.5645$\pm$0.0011 & 0.2977$\pm$0.0136 & \textbf{0.2446$\pm$0.0112} \\
\multicolumn{1}{l|}{spearman} & -0.3256$\pm$0.05 & 0.0842$\pm$0.0    & 0.8281$\pm$0.0    & \textbf{0.8299$\pm$0.0208} \\
\multicolumn{1}{l|}{pearson}   & -0.2666$\pm$0.0057 & 0.1402$\pm$0.0122 & 0.8613$\pm$0.017  & \textbf{0.8826$\pm$0.0361} \\ \bottomrule
\end{tabular}
\caption[Scores of the radius of gyration measure]{Scores of the radius of gyration measure; the considered spatial tessellation is $L_{\mbox{\tiny REL}}(250)$.}
\end{table}

\paragraph{Location Frequency.}
For this measure the tessellation considered is $L_{\mbox{\tiny REL}}(250)$; all the models are good in terms of Kullback-Leibler divergence, the only model able to reduce the Hellinger-distance is STS-EPR.
\begin{table}[H]
\small
\renewcommand*{\arraystretch}{1.2}
\begin{tabular}{@{}lcccc@{}}
\toprule
                          & GeoSim            & $\mbox{GeoSim}_{d}$            & $\mbox{GeoSim}_{gravity}$             & STS-EPR               \\ \midrule
\multicolumn{1}{l|}{RMSE} & 0.0254$\pm$0.0002 & 0.0256$\pm$0.0002 & 0.0279$\pm$0.0001          & \textbf{0.0139$\pm$0.0002} \\
\multicolumn{1}{l|}{KL}   & 0.0079$\pm$0.0005 & 0.0077$\pm$0.0005 & \textbf{0.0023$\pm$0.0001} & 0.0089$\pm$0.0001          \\
\multicolumn{1}{l|}{Hellinger} & 0.219$\pm$0.0005     & 0.2195$\pm$0.0003          & 0.2232$\pm$0.0002    & \textbf{0.1175$\pm$0.0005} \\
\multicolumn{1}{l|}{spearman}  & \textbf{1.0$\pm$0.0} & \textbf{1.0$\pm$0.0}       & \textbf{1.0$\pm$0.0} & \textbf{1.0$\pm$0.0}       \\
\multicolumn{1}{l|}{pearson}   & 0.9991$\pm$0.0002    & \textbf{0.9993$\pm$0.0002} & 0.9992$\pm$0.0003    & 0.9943$\pm$0.0003          \\ \bottomrule
\end{tabular}
\caption[Scores for the location frequency measure]{Scores for the location frequency measure, the spatial tessellation considered is $L_{\mbox{\tiny REL}}(250)$.}
\end{table}

\paragraph{Visits per Location.}
From the scores (Table \ref{tab:score_vpl}) emerges the overestimation of both GeoSim and $\mbox{GeoSim}_{d}$ (Kullback-Leibler of 7.7095 and 6.9719 respectively) in the number of visits per location.
The introduction of the concept of relevance and the mechanism of the gravity-law give the best scores.
\begin{table}[H]
\small
\renewcommand*{\arraystretch}{1.2}
\begin{tabular}{@{}lcccc@{}}
\toprule
                               & GeoSim             & $\mbox{GeoSim}_{d}$             & $\mbox{GeoSim}_{gravity}$       & STS-EPR               \\ \midrule
\multicolumn{1}{l|}{RMSE}      & 0.1013$\pm$0.0     & 0.1013$\pm$0.0     & 0.0869$\pm$0.0008    & \textbf{0.0464$\pm$0.0025} \\
\multicolumn{1}{l|}{KL}        & 6.7095$\pm$0.0067  & 6.9719$\pm$0.0137  & 0.3135$\pm$0.024     & \textbf{0.0307$\pm$0.0052} \\
\multicolumn{1}{l|}{Hellinger} & 0.4674$\pm$0.0     & 0.4674$\pm$0.0     & 0.2696$\pm$0.0047    & \textbf{0.1187$\pm$0.0055} \\
\multicolumn{1}{l|}{spearman}  & -0.5267$\pm$0.0286 & -0.5965$\pm$0.0196 & \textbf{1.0$\pm$0.0} & 0.9982$\pm$0.0036          \\
\multicolumn{1}{l|}{pearson} & -0.1955$\pm$0.0011 & -0.2276$\pm$0.0066 & 0.8856$\pm$0.0166 & \textbf{0.9976$\pm$0.0014} \\ \bottomrule
\end{tabular}
\caption[Scores for the visits per location measure]{Scores for the visits per location measure, the spatial tessellation considered is $L_{\mbox{\tiny REL}}(1000)$.}
\label{tab:score_vpl}
\end{table}

\paragraph{Waiting Time(s).}
From the three tables below we can see how the scores changes according to the distribution of the waiting time considered; considering also values $1<h$ (Table \ref{tab:wt_score1}) no model is able to reproduce correctly the real distribution, cutting from the real distribution all the waiting times $1<h$ (Table \ref{tab:wt_score2}) or remapping to an hour (Table \ref{tab:wt_score3}) produce better scores; the model that best fits this measure is STS-EPR.

\begin{table}[H]
\small
\renewcommand*{\arraystretch}{1.2}
\begin{tabular}{@{}lcccc@{}}
\toprule
                               & GeoSim            & $\mbox{GeoSim}_{d}$            & $\mbox{GeoSim}_{gravity}$    & STS-EPR               \\ \midrule
\multicolumn{1}{l|}{RMSE}    & \textbf{0.0004$\pm$0.0} & \textbf{0.0004$\pm$0.0}     & \textbf{0.0004$\pm$0.0} & \textbf{0.0004$\pm$0.0} \\
\multicolumn{1}{l|}{KL}        & 3.6461$\pm$0.0008 & 3.6405$\pm$0.0008 & 3.642$\pm$0.0025  & \textbf{2.9627$\pm$0.0057} \\
\multicolumn{1}{l|}{Hellinger} & 0.0344$\pm$0.0    & 0.0344$\pm$0.0    & 0.0344$\pm$0.0    & \textbf{0.0341$\pm$0.0}    \\
\multicolumn{1}{l|}{spearman}  & 0.1588$\pm$0.0    & 0.2658$\pm$0.0    & 0.2444$\pm$0.0428 & \textbf{0.322$\pm$0.0655}  \\
\multicolumn{1}{l|}{pearson} & -0.1667$\pm$0.0001      & \textbf{-0.1653$\pm$0.0001} & -0.1655$\pm$0.0006      & -0.1779$\pm$0.0022      \\ \bottomrule
\end{tabular}
\caption[Scores for the waiting time distribution]{Scores for the waiting time distribution.}
\label{tab:wt_score1}
\end{table}

\begin{table}[H]
\small
\renewcommand*{\arraystretch}{1.2}
\begin{tabular}{@{}lcccc@{}}
\toprule
                               & GeoSim               & $\mbox{GeoSim}_{d}$              & $\mbox{GeoSim}_{gravity}$      & STS-EPR               \\ \midrule
\multicolumn{1}{l|}{RMSE}      & \textbf{0.0$\pm$0.0} & \textbf{0.0$\pm$0.0} & \textbf{0.0$\pm$0.0} & \textbf{0.0$\pm$0.0}       \\
\multicolumn{1}{l|}{KL}        & 0.5619$\pm$0.0011    & 0.5609$\pm$0.0003    & 0.5644$\pm$0.0027    & \textbf{0.0423$\pm$0.0031} \\
\multicolumn{1}{l|}{Hellinger} & 0.0065$\pm$0.0       & 0.0065$\pm$0.0       & 0.0065$\pm$0.0       & \textbf{0.0014$\pm$0.0001} \\
\multicolumn{1}{l|}{spearman}  & 0.5179$\pm$0.0       & 0.5179$\pm$0.0       & 0.5254$\pm$0.0149    & \textbf{0.8851$\pm$0.0281} \\
\multicolumn{1}{l|}{pearson}   & 0.9154$\pm$0.0001    & 0.9154$\pm$0.0002    & 0.9153$\pm$0.0003    & \textbf{0.9769$\pm$0.0009} \\ \bottomrule
\end{tabular}
\caption[Scores for the waiting time distribution where are considered only values $\geq 1$ hour]{Scores for the waiting time distribution where are considered only values $\geq 1$ hour.}
\label{tab:wt_score2}
\end{table}

\begin{table}[H]
\small
\renewcommand*{\arraystretch}{1.2}
\begin{tabular}{@{}lcccc@{}}
\toprule
                               & GeoSim            & $\mbox{GeoSim}_{d}$                     & $\mbox{GeoSim}_{gravity}$    & STS-EPR               \\ \midrule
\multicolumn{1}{l|}{RMSE}    & \textbf{0.0$\pm$0.0} & \textbf{0.0$\pm$0.0} & \textbf{0.0$\pm$0.0}       & \textbf{0.0$\pm$0.0} \\
\multicolumn{1}{l|}{KL}        & 0.1976$\pm$0.0006 & \textbf{0.1972$\pm$0.0004} & 0.198$\pm$0.0007  & 0.2045$\pm$0.0051          \\
\multicolumn{1}{l|}{Hellinger} & 0.0043$\pm$0.0    & 0.0043$\pm$0.0             & 0.0042$\pm$0.0    & \textbf{0.0029$\pm$0.0}    \\
\multicolumn{1}{l|}{spearman}  & 0.5179$\pm$0.0    & 0.5179$\pm$0.0             & 0.5254$\pm$0.0149 & \textbf{0.8851$\pm$0.0281} \\
\multicolumn{1}{l|}{pearson} & 0.975$\pm$0.0001     & 0.9749$\pm$0.0002    & \textbf{0.9751$\pm$0.0001} & 0.9178$\pm$0.0015    \\ \bottomrule
\end{tabular}
\caption[Scores for the waiting time distribution where the values $\geq 1$ hour are considered as 1 hour]{Scores for the waiting time distribution where the values $\geq 1$ hour are considered as 1 hour.}
\label{tab:wt_score3}
\end{table}

\paragraph{Activity per Hour.}
The three models that use the empirical distribution of Song et al. \cite{song_limits} in the waiting time choice behave the same.
The only model able to reproduce the circadian rhythm is STS-EPR.

\begin{table}[H]
\small
\renewcommand*{\arraystretch}{1.2}
\begin{tabular}{@{}lcccc@{}}
\toprule
                          & GeoSim            & $\mbox{GeoSim}_{d}$            & $\mbox{GeoSim}_{gravity}$  & STS-EPR               \\ \midrule
\multicolumn{1}{l|}{RMSE} & 0.0234$\pm$0.0001 & 0.0234$\pm$0.0    & 0.0234$\pm$0.0  & \textbf{0.0047$\pm$0.0001} \\
\multicolumn{1}{l|}{KL}   & 0.1797$\pm$0.0007 & 0.1801$\pm$0.0004 & 0.18$\pm$0.0006 & \textbf{0.0075$\pm$0.0003} \\
\multicolumn{1}{l|}{Hellinger} & 0.2268$\pm$0.0004  & 0.227$\pm$0.0002   & 0.2269$\pm$0.0004  & \textbf{0.0431$\pm$0.0009} \\
\multicolumn{1}{l|}{spearman}  & -0.1078$\pm$0.2374 & -0.2607$\pm$0.1109 & -0.2214$\pm$0.1681 & \textbf{0.9786$\pm$0.0016} \\
\multicolumn{1}{l|}{pearson}   & -0.2008$\pm$0.1717 & -0.2962$\pm$0.0873 & -0.2531$\pm$0.1352 & \textbf{0.9834$\pm$0.0006} \\ \bottomrule
\end{tabular}
\caption[Scores for the activity per hour measure]{Scores for the activity per hour measure.}
\end{table}

\paragraph{Uncorrelated Entropy.}
The best tessellation for this measure is $L_{\mbox{\tiny REL}}(2000)$; all the models are not able to reproduce in an accurate way the distribution of the uncorrelated entropy of the individuals in New York City.
\begin{table}[H]
\small
\renewcommand*{\arraystretch}{1.2}
\begin{tabular}{@{}lcccc@{}}
\toprule
 & GeoSim & $\mbox{GeoSim}_{d}$ & $\mbox{GeoSim}_{gravity}$ & STS-EPR \\ \midrule
\multicolumn{1}{l|}{RMSE}      & 2.0942$\pm$0.0405 & \textbf{2.0842$\pm$0.0207} & 2.4114$\pm$0.0141 & 2.2734$\pm$0.0129          \\
\multicolumn{1}{l|}{KL}        & 7.3433$\pm$0.0099 & 7.3404$\pm$0.0048          & 7.4983$\pm$0.0147 & \textbf{3.4449$\pm$0.2474} \\
\multicolumn{1}{l|}{Hellinger} & 1.7999$\pm$0.0085 & 1.7974$\pm$0.0042          & 1.9059$\pm$0.0078 & \textbf{1.7716$\pm$0.0242} \\
\multicolumn{1}{l|}{spearman}  & 0.5276$\pm$0.0    & 0.5276$\pm$0.0             & 0.5276$\pm$0.0    & \textbf{0.768$\pm$0.0}     \\
\multicolumn{1}{l|}{pearson}   & 0.5544$\pm$0.0021 & \textbf{0.555$\pm$0.0011}  & 0.5357$\pm$0.0009 & 0.5456$\pm$0.0014          \\ \bottomrule
\end{tabular}
\caption[Scores for the uncorrelated entropy measure]{Scores for the uncorrelated entropy measure.}
\end{table}

\paragraph{Mobility Similarity.}
The best model for what concern the distribution of the mobility similarity with respect the social graph $G$ is STS-EPR.
In this case the use of the gravity-law with the waiting times chosen from the empirical distribution of Song et al. \cite{song_limits} gives the worst result; using a diary generator and the gravity law like in STS-EPR is the best choice according to the results presented in Table \ref{tab:scores_mobism}.
\begin{table}[H]
\small
\renewcommand*{\arraystretch}{1.2}
\begin{tabular}{@{}lcccc@{}}
\toprule
                               & GeoSim            & $\mbox{GeoSim}_{d}$           & $\mbox{GeoSim}_{gravity}$    & STS-EPR               \\ \midrule
\multicolumn{1}{l|}{RMSE}      & 1.3609$\pm$0.0236 & 1.3677$\pm$0.021  & 1.904$\pm$0.0478  & \textbf{0.9007$\pm$0.0318} \\
\multicolumn{1}{l|}{KL}        & 0.5013$\pm$0.0085 & 0.495$\pm$0.007   & 0.7998$\pm$0.0403 & \textbf{0.2568$\pm$0.0238} \\
\multicolumn{1}{l|}{Hellinger} & 1.2843$\pm$0.0091 & 1.2743$\pm$0.0092 & 1.5087$\pm$0.0284 & \textbf{0.7958$\pm$0.018}  \\
\multicolumn{1}{l|}{spearman}  & 0.1345$\pm$0.0225 & 0.1879$\pm$0.0153 & 0.4594$\pm$0.0145 & \textbf{0.9222$\pm$0.0102} \\
\multicolumn{1}{l|}{pearson}   & 0.9432$\pm$0.0041 & 0.9517$\pm$0.0033 & 0.8638$\pm$0.0311 & \textbf{0.9785$\pm$0.0028} \\ \bottomrule
\end{tabular}
\caption[Scores for the mobility similarity measure]{The scores referred to the measure mobility similarity using a tessellation $L_{\mbox{\tiny REL}}(250)$.}
\label{tab:scores_mobism}
\end{table}

\subsection{Modeling ability in new scenarios}\label{sect:london}

\begin{figure}
    \centering
    \includegraphics[width=0.95\textwidth]{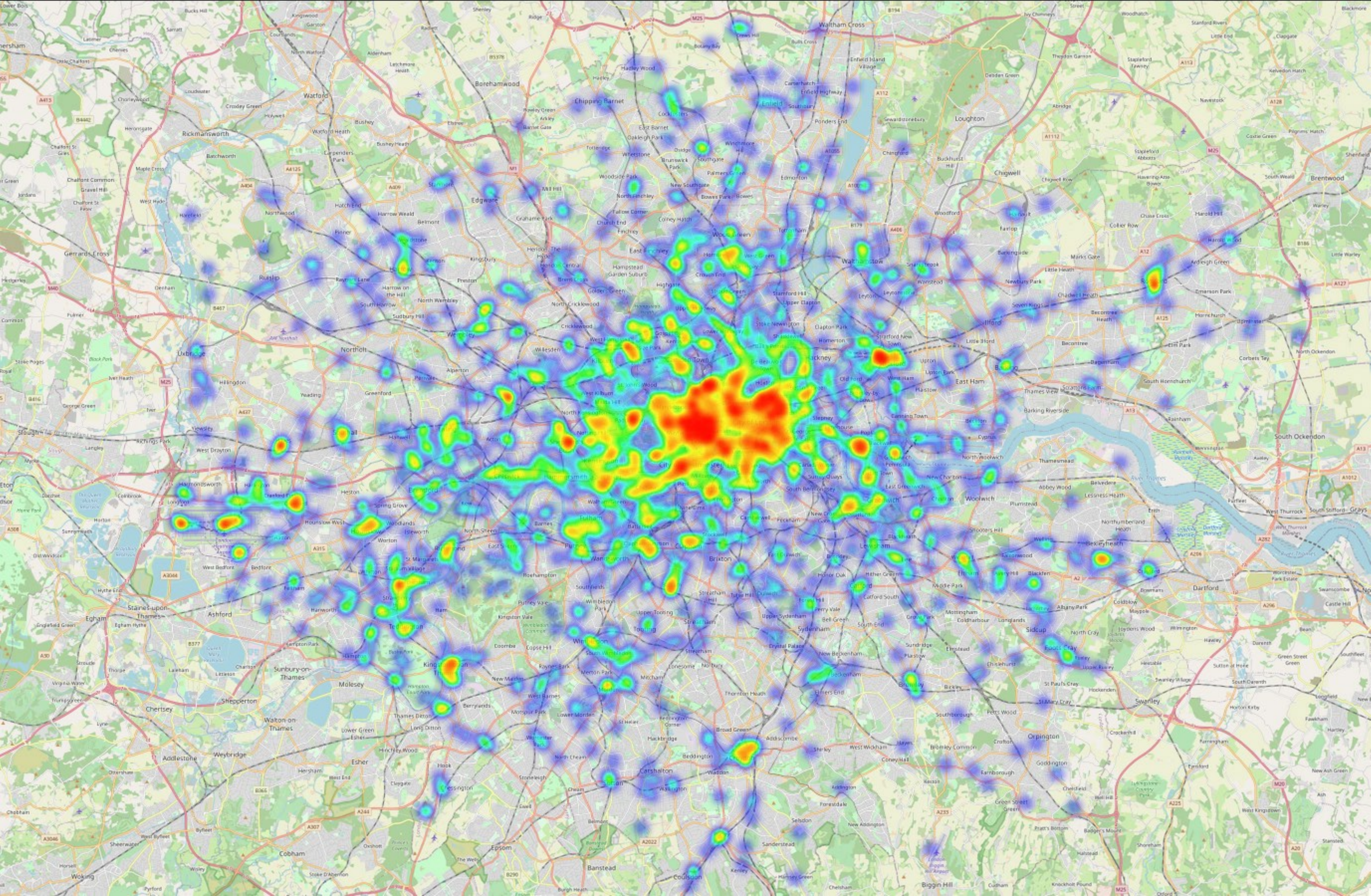}
    \caption[The heatmap relative at the mobility in London]{The heatmap relative at the 14,895 check-ins made by 622 individual during an observation period of three months (April 2012 to July 2012) in London. From the heatmap emerges an high concentration of check-ins in the borough of Westminster and City of London.}
    \label{fig:heatmap_checkins_london}
\end{figure}

For assessing the modeling ability of the proposed model in other scenarios beyond New York City, we simulate the mobility for a set of individuals moving in the area of London for three months. 
Specifically, we instantiate STS-EPR on a weighted spatial tessellation with a granularity of 250 meters.
The dataset $D_{LON}$, relative at the mobility of the individuals in London, is created using the same \textit{modus operandi} as for the New York City dataset $D_{NYC}$; $D_{LON}$ contains 14,895 check-ins (Figure \ref{fig:heatmap_checkins_london}) made by 622 users connected in a social graph $G_{LON}$ which has 1,185 edges.
The weighted spatial tessellation $L_{\mbox{\tiny LON}}(250)$ computed over $D_{LON}$ contains 2,800 relevant locations.

\begin{figure}[!h]
\centering
    \subfigure[]{
\includegraphics[width=0.45\textwidth]{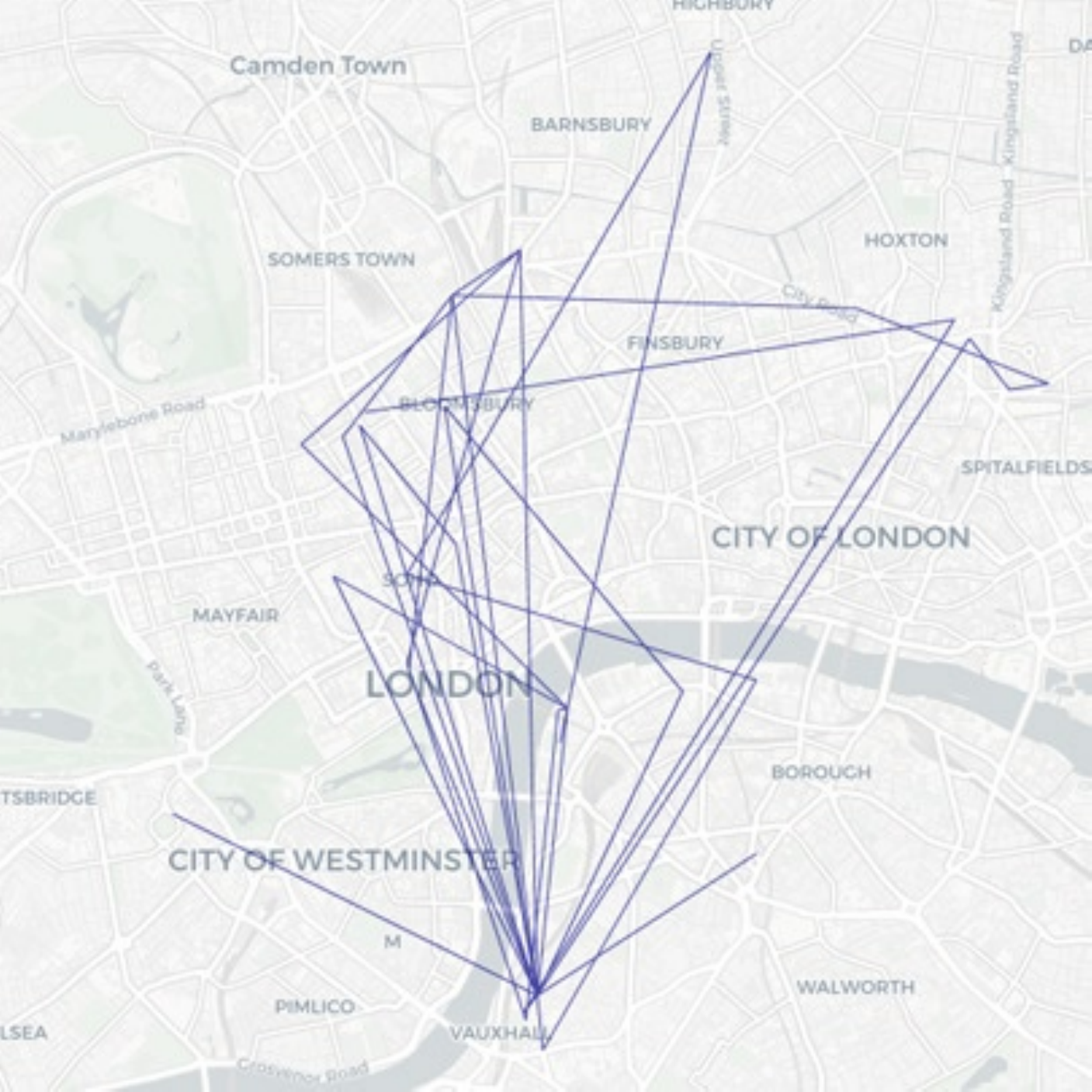}}
\hspace{2mm}
    \subfigure[]{
\includegraphics[width=0.45\textwidth]{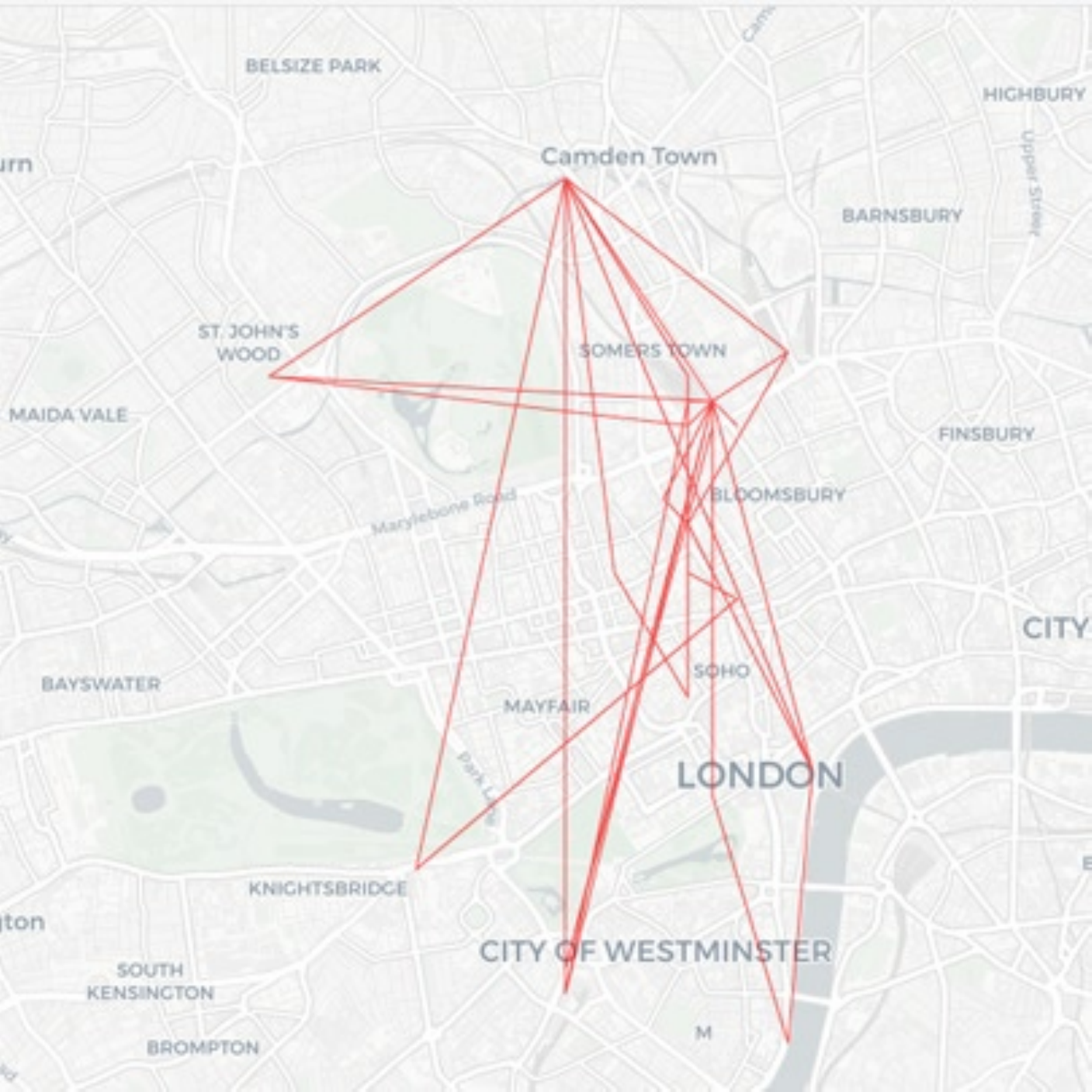}}
\caption [A spatial representation of the trajectory of a real individual and a synthetic individual in London]{A spatial representation of the trajectory of a real individual (a) and a synthetic individual (b) moving in London; the latter is generated using the model STS-EPR with the weighted spatial tessellation $L_{\mbox{\tiny REL}}(250)$ computer over $D_{LON}$.
Figures generated with \textit{scikit-mobility} \cite{scikit_mobility}.
}
\label{fig:real_syn_london}
\end{figure}

First, we verify that the model is not \textit{city-dependent}; we instantiate the model with full knowledge of the London scenario (diary generator $\mbox{MD}_{\mbox{\tiny LON}}$ and weighted spatial tessellation $L_{\mbox{\tiny LON}}(250)$); as shown in Table \ref{tab:results_london}, the model is able to generate trajectories (Figure \ref{fig:real_syn_london}) with realistic mobility patterns.

Next, we simulate three scenarios where we have partial or no information about the displacements and the circadian rhythm of the individuals in London.
In the first scenario (\texttt{None} scenario), we assume to know nothing about the mobility of the individuals in London. 
We use the Mobility Diary Generator $\mbox{MD}_{\mbox{\tiny NYC}}$ computed for New York City, and we generate the weighted spatial tessellation assigning a relevance $w_i$ for a location $r_i$ from a truncated power-law $P(w) \approx (w)^{-\beta} e ^{-w / \lambda}$ where $\beta=1.25$ and $\lambda=104$. 
We fit the parameters over the distribution of the relevance of the locations in New York City.
In the second scenario (\texttt{MD} scenario), we assume to know only the routine of the individuals in London using the Mobility Diary Generator $\mbox{MD}_{LON}$ and the weighted spatial tessellation used by the model is the one fitted on the New York's locations.
In the last scenario ($L_{\mbox{\tiny LON}}$ scenario), we assume to know only the real weighted spatial tessellation $L_{\mbox{\tiny LON}}(250)$ using the Mobility Diary Generator $\mbox{MD}_{NYC}$.

\begin{table}[]
\footnotesize
\renewcommand*{\arraystretch}{1.6}
%\centering
\begin{tabular}{@{}lcccccccc@{}}
\toprule
 & $\Delta r$ & $r_g$ & $L_i$ & $V l$ & $\Delta_t map$ & $t (h)$ & $S^{unc}$ & $mob_{sim}$ \\ \midrule
\multicolumn{1}{l|}{None} &
  {\color[HTML]{000000} \begin{tabular}[c]{@{}c@{}}0.044\\ 0.0033\end{tabular}} &
  {\color[HTML]{000000} \begin{tabular}[c]{@{}c@{}}3.1222\\ 0.726\end{tabular}} &
  \begin{tabular}[c]{@{}c@{}}0.0152\\ 0.0006\end{tabular} &
  \begin{tabular}[c]{@{}c@{}}0.1575\\ 0.0112\end{tabular} &
  \begin{tabular}[c]{@{}c@{}}0.2455\\ 0.0016\end{tabular} &
  \begin{tabular}[c]{@{}c@{}}0.066\\ 0.0011\end{tabular} &
  \begin{tabular}[c]{@{}c@{}}2.2479\\ 0.0779\end{tabular} &
  \begin{tabular}[c]{@{}c@{}}0.2699\\ 0.02\end{tabular} \\
\multicolumn{1}{l|}{{MD}} &
  \begin{tabular}[c]{@{}c@{}}0.0345\\ 0.0021\end{tabular} &
  \begin{tabular}[c]{@{}c@{}}2.3151\\ 0.5801\end{tabular} &
  \textbf{\begin{tabular}[c]{@{}c@{}}0.0102\\ 0.0005\end{tabular}} &
  \begin{tabular}[c]{@{}c@{}}0.1503\\ 0.0123\end{tabular} &
  \begin{tabular}[c]{@{}c@{}}0.1656\\ 0.0042\end{tabular} &
  {\color[HTML]{000000} \begin{tabular}[c]{@{}c@{}}0.0117\\ 0.0006\end{tabular}} &
  {\color[HTML]{000000} \begin{tabular}[c]{@{}c@{}}1.3514\\ 0.2133\end{tabular}} &
  \textbf{\begin{tabular}[c]{@{}c@{}}0.2278\\ 0.0167\end{tabular}} \\
\multicolumn{1}{l|}{$L_{\mbox{\tiny LON}}$} &
  \begin{tabular}[c]{@{}c@{}}0.017\\ 0.0025\end{tabular} &
  \begin{tabular}[c]{@{}c@{}}1.4486\\ 0.3124\end{tabular} &
  \begin{tabular}[c]{@{}c@{}}0.0153\\ 0.0005\end{tabular} &
  \begin{tabular}[c]{@{}c@{}}0.1534\\ 0.0139\end{tabular} &
  {\color[HTML]{000000} \begin{tabular}[c]{@{}c@{}}0.2472\\ 0.0039\end{tabular}} &
  \begin{tabular}[c]{@{}c@{}}0.0655\\ 0.0008\end{tabular} &
  \begin{tabular}[c]{@{}c@{}}2.1396\\ 0.2578\end{tabular} &
  \begin{tabular}[c]{@{}c@{}}0.2847\\ 0.0223\end{tabular} \\
\multicolumn{1}{l|}{\textbf{Full}} &
  \textbf{\begin{tabular}[c]{@{}c@{}}0.0119\\ 0.0008\end{tabular}} &
  \textbf{\begin{tabular}[c]{@{}c@{}}0.9861\\ 0.4517\end{tabular}} &
  \begin{tabular}[c]{@{}c@{}}0.0106\\ 0.0003\end{tabular} &
  \textbf{\begin{tabular}[c]{@{}c@{}}0.1407\\ 0.0078\end{tabular}} &
  \textbf{\begin{tabular}[c]{@{}c@{}}0.1643\\ 0.0045\end{tabular}} &
  \textbf{\begin{tabular}[c]{@{}c@{}}0.0113\\ 0.001\end{tabular}} &
  \textbf{\begin{tabular}[c]{@{}c@{}}1.3445\\ 0.2267\end{tabular}} &
  \begin{tabular}[c]{@{}c@{}}0.2342\\ 0.016\end{tabular} \\ \midrule
\multicolumn{1}{l|}{NYC} & 
  \begin{tabular}[c]{@{}c@{}}0.0366\\ 0.0016\end{tabular} &
  \begin{tabular}[c]{@{}c@{}}0.8855\\ 0.2103\end{tabular} &
  \begin{tabular}[c]{@{}c@{}}0.0089\\ 0.0001\end{tabular} &
  \begin{tabular}[c]{@{}c@{}}0.1947\\ 0.0161\end{tabular} &
  \begin{tabular}[c]{@{}c@{}}0.1665\\ 0.0054\end{tabular} &
  \begin{tabular}[c]{@{}c@{}}0.0072\\ 0.0003\end{tabular} &
  \begin{tabular}[c]{@{}c@{}}2.1176\\ 0.3345\end{tabular} &
  \begin{tabular}[c]{@{}c@{}}0.2568\\ 0.0238\end{tabular} \\ \bottomrule
\end{tabular}
\caption[Table of the results for the experiments of individuals' mobility in London]{Table of the results for the experiments of individuals' mobility in London. Every of the first four rows corresponds to a different scenario described above; the last row reports the results of the experiments in New York City using the tessellation granularity of 250 meters. Every column refers to a standard mobility measure.
Every cell reports the mean Kullback-Leibler score (first row) and the standard deviation (second row). }
\label{tab:results_london}
\end{table}

As shown in Table \ref{tab:results_london}, STS-EPR reproduces the standard mobility measures accurately, with results similar to those obtained in the experiments concerning the area of New York City.
The model can generate realistic trajectories, even with a lack of information. Without including neither the diary generator of the individuals in London nor the weighted spatial tessellation computed over $D_{LON}$ (first row Table \ref{tab:results_london}), the scores are in general worst than the full-knowledge scenario but still good considering that the model use no information about the mobility behavior of the individuals in London.
Including in the model, in a complementary way, the real weighted spatial tessellation and the real Mobility Diary, we obtain more realistic trajectories in terms of spatial and spatio-temporal patterns, respectively.

\begin{figure}[!h]
\centering
    \subfigure[]{\label{fig:jl_london}
\includegraphics[width=0.45\textwidth]{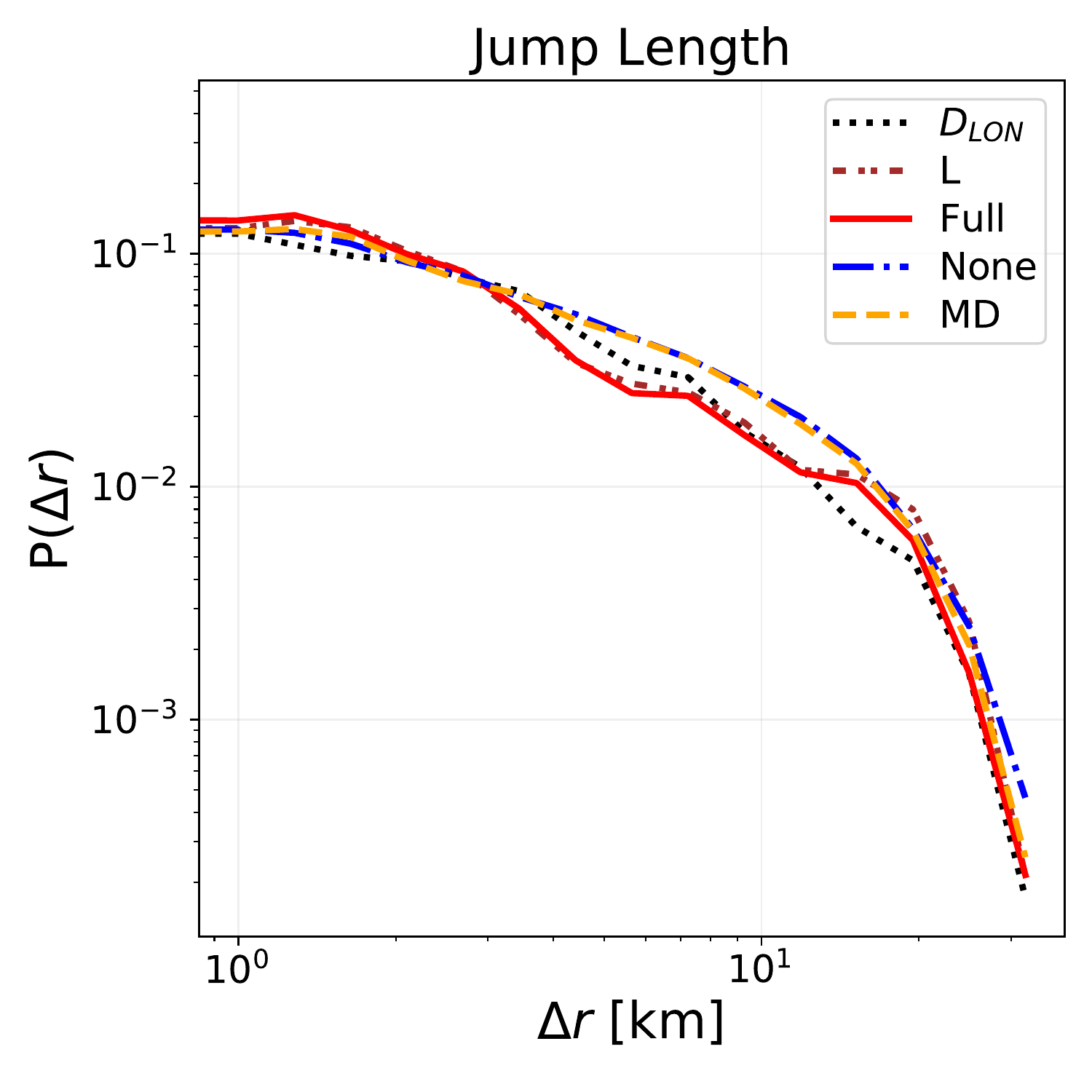}}
\hspace{2mm}
    \subfigure[]{\label{fig:rog_london}
\includegraphics[width=0.45\textwidth]{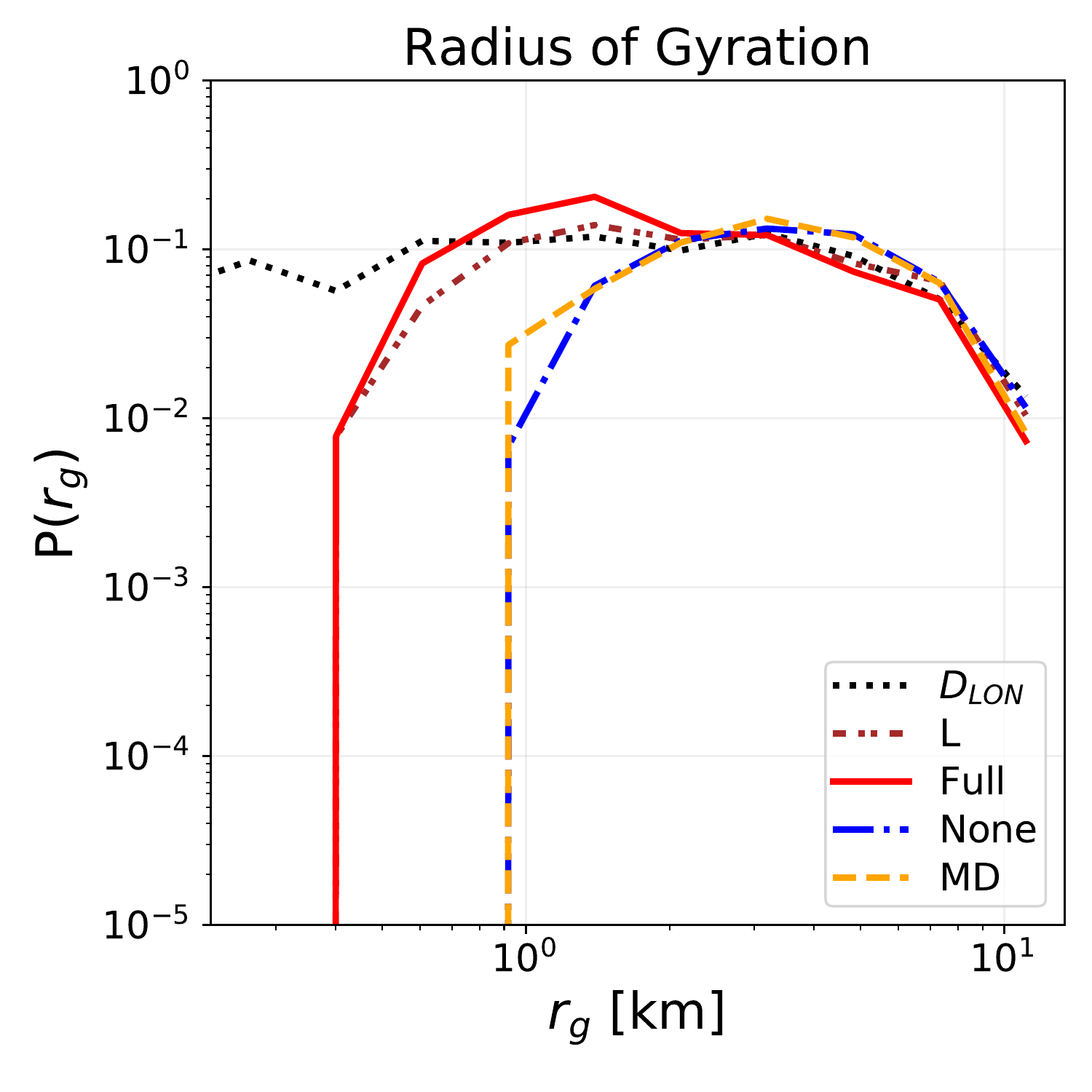}}
\caption[The probability density function for the jump length and radius of gyration computed for the real and synthetic trajectories relative to London]{The probability density function for the jump length (a) and radius of gyration (b) computed for the real and synthetic trajectories.}
\end{figure}

The model replicates the distribution of the jump length associated with the synthetic trajectories (Figure \ref{fig:jl_london}) accurately; instead, the models cannot replicate the shape of the distribution of the radius of gyration (Figure \ref{fig:rog_london}).

\begin{figure}[!h]
\centering
    \subfigure[]{\label{fig:lf_london}
\includegraphics[width=0.45\textwidth]{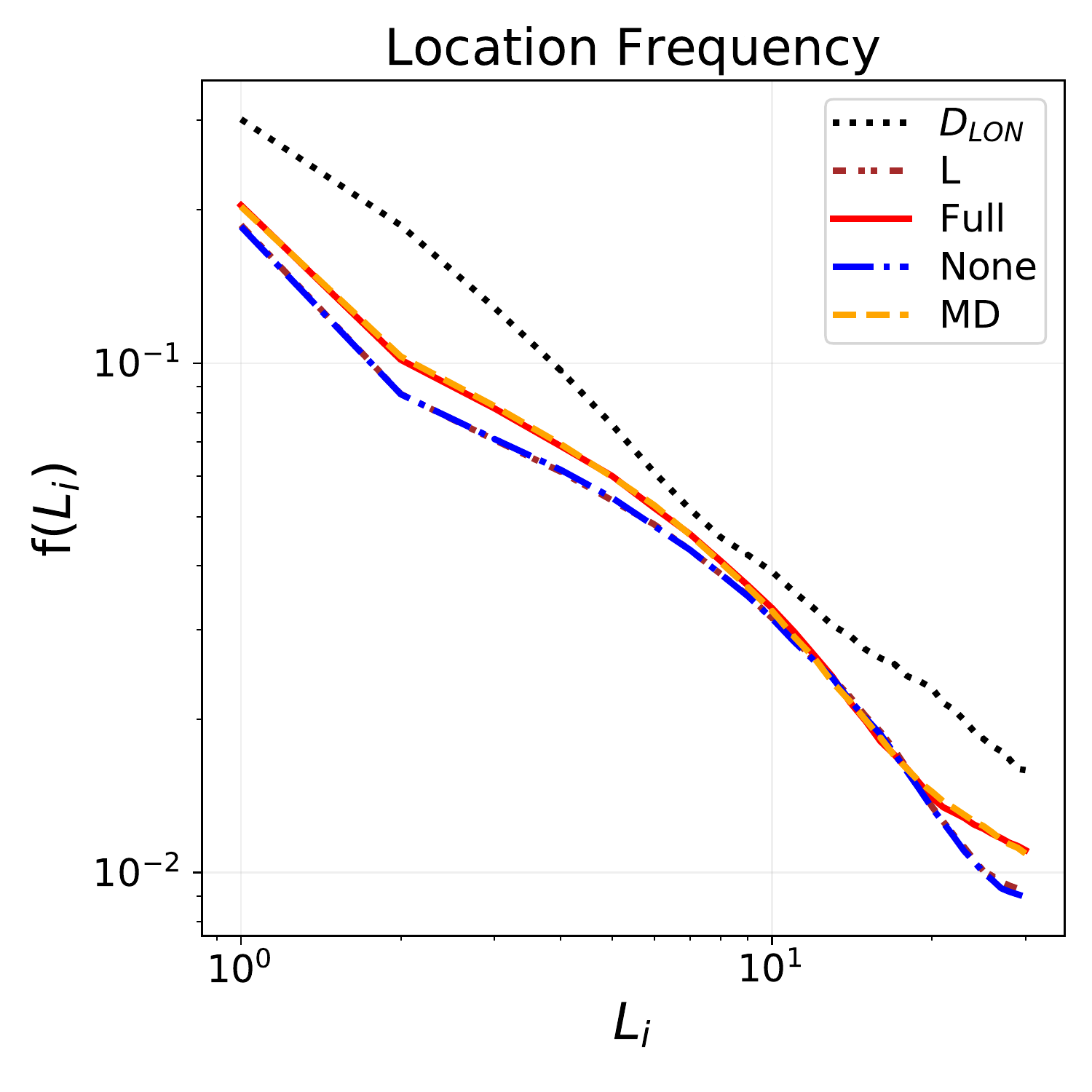}}
\hspace{2mm}
    \subfigure[]{\label{fig:vpl_london}
\includegraphics[width=0.45\textwidth]{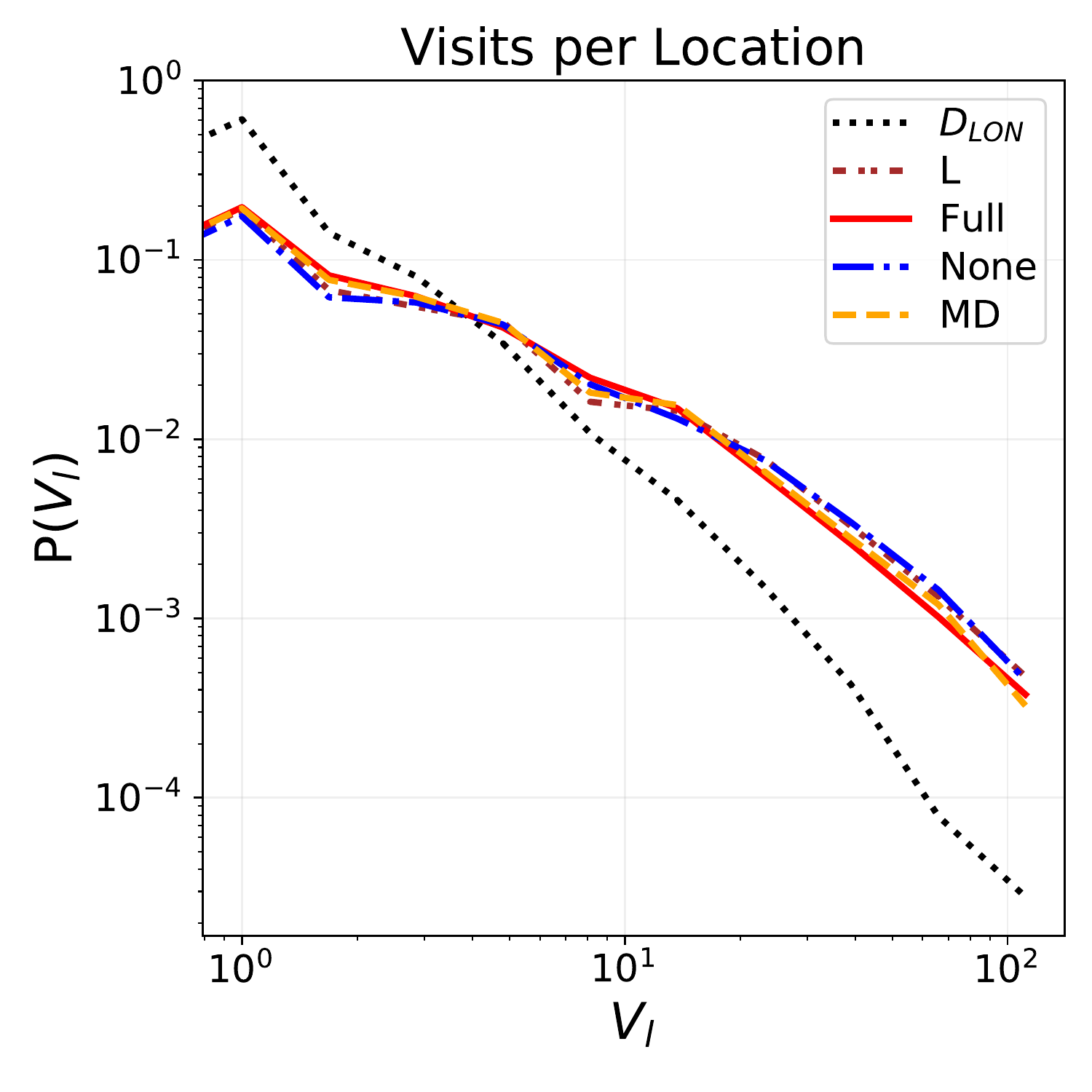}}
\caption[The probability density function for the location frequency and visits per location for the real and synthetic trajectories relative to London]{The probability density function for the location frequency (a) and visits per location (b) computed for the real and synthetic trajectories.}
\end{figure}

Both the measures concerning the frequency and the number of visits in each location are reproduced accurately by the synthetic trajectories; the model in the experiments which uses the Mobility Diary Generator of New York City underestimates the frequency for the first ten locations (Figure \ref{fig:lf_london}); the same holds for the visits per location measure where is present a slightly underestimation of the number of location with a small number of visits. (Figure \ref{fig:vpl_london}).

\begin{figure}[!h]
\centering
    \subfigure[]{
\includegraphics[width=0.47\textwidth]{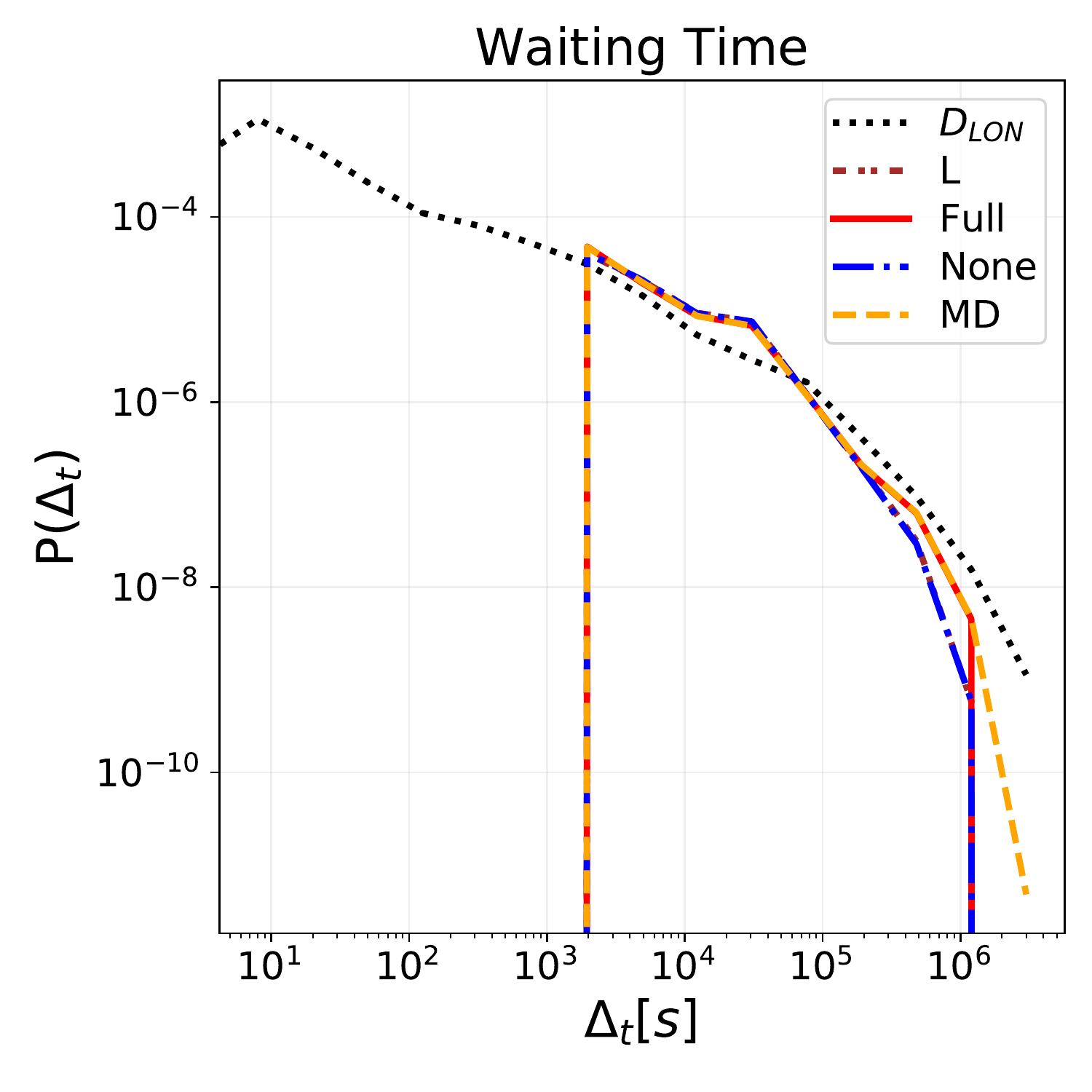}}
\hspace{2mm}
    \subfigure[]{\label{fig:wt_1h_london}
\includegraphics[width=0.45\textwidth]{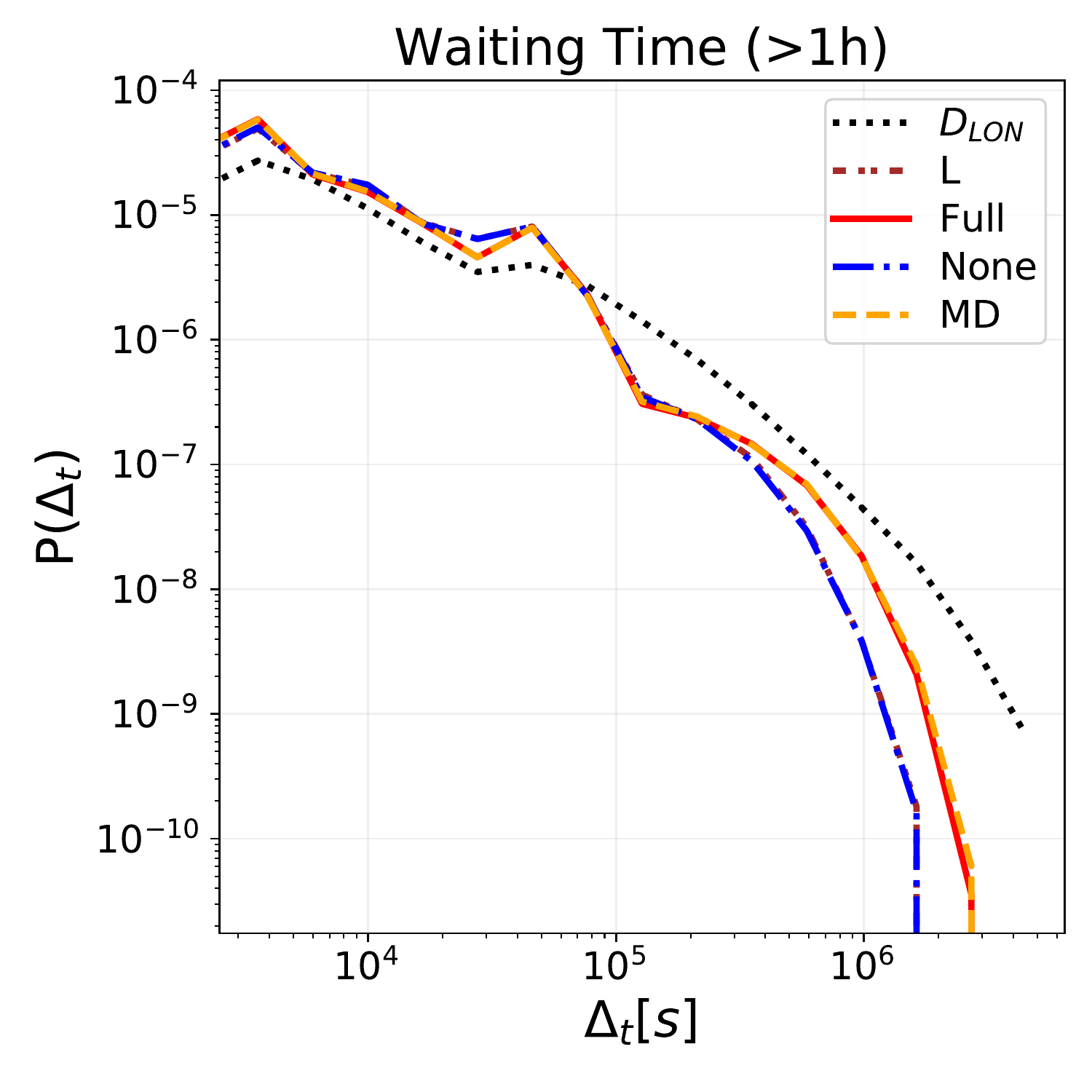}}
    \subfigure[]{\label{fig:wt_remap_london}
\includegraphics[width=0.45\textwidth]{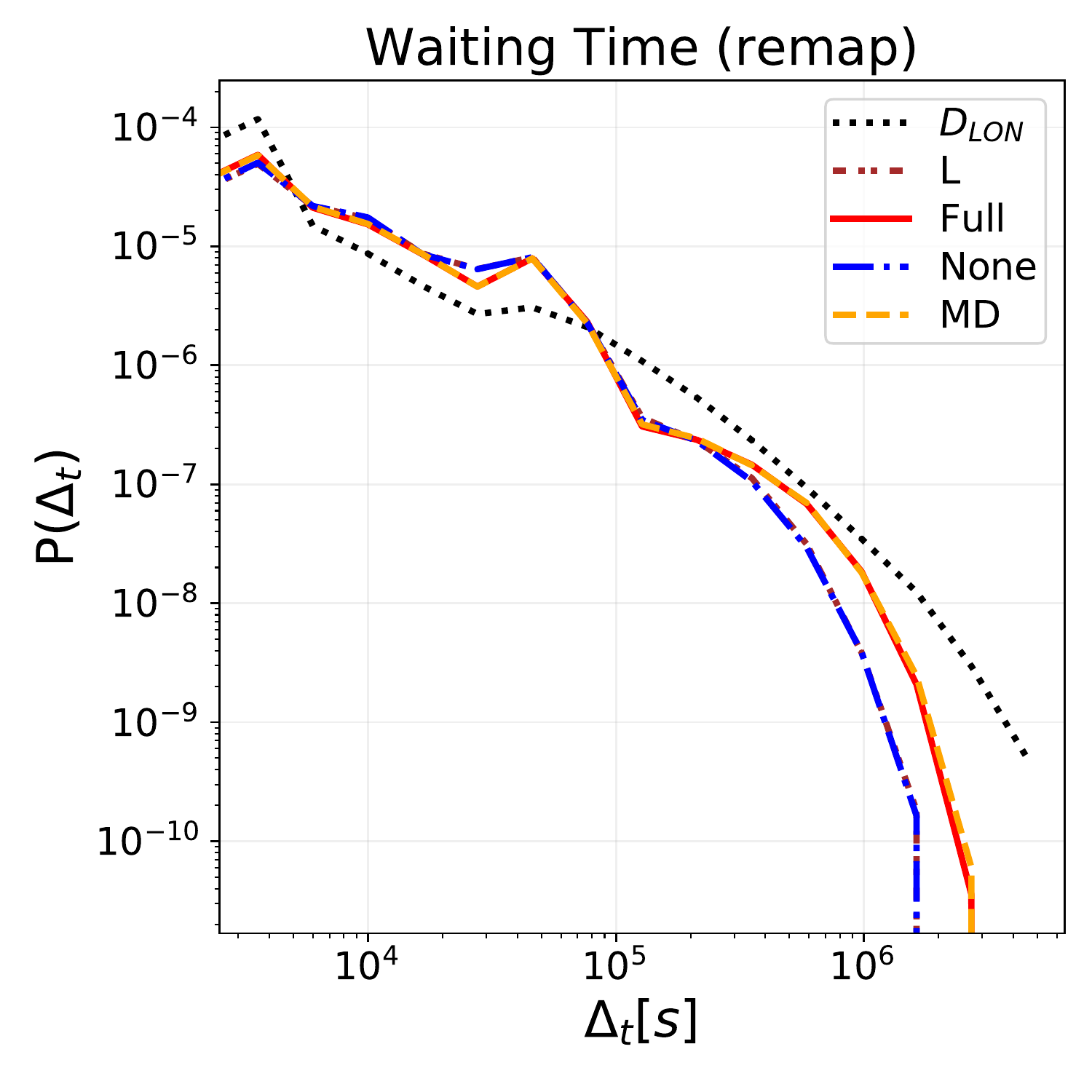}}
\caption[The three options considered for dealing with waiting time values under one hour in the London dataset]{The three options considered for dealing with waiting time values under one hour; in the first case we compare the distributions with all the values (a), in the second options we cut the real distribution (b) and finally we re-map all the values $<1h$ in the value $1h$. The knowledge of the model affects the distribution only for values $>10^5$.}
\label{fig:wt_london}
\end{figure}

\begin{figure}[!h]
\centering
    \subfigure[]{\label{fig:activity_london}
\includegraphics[width=0.45\textwidth]{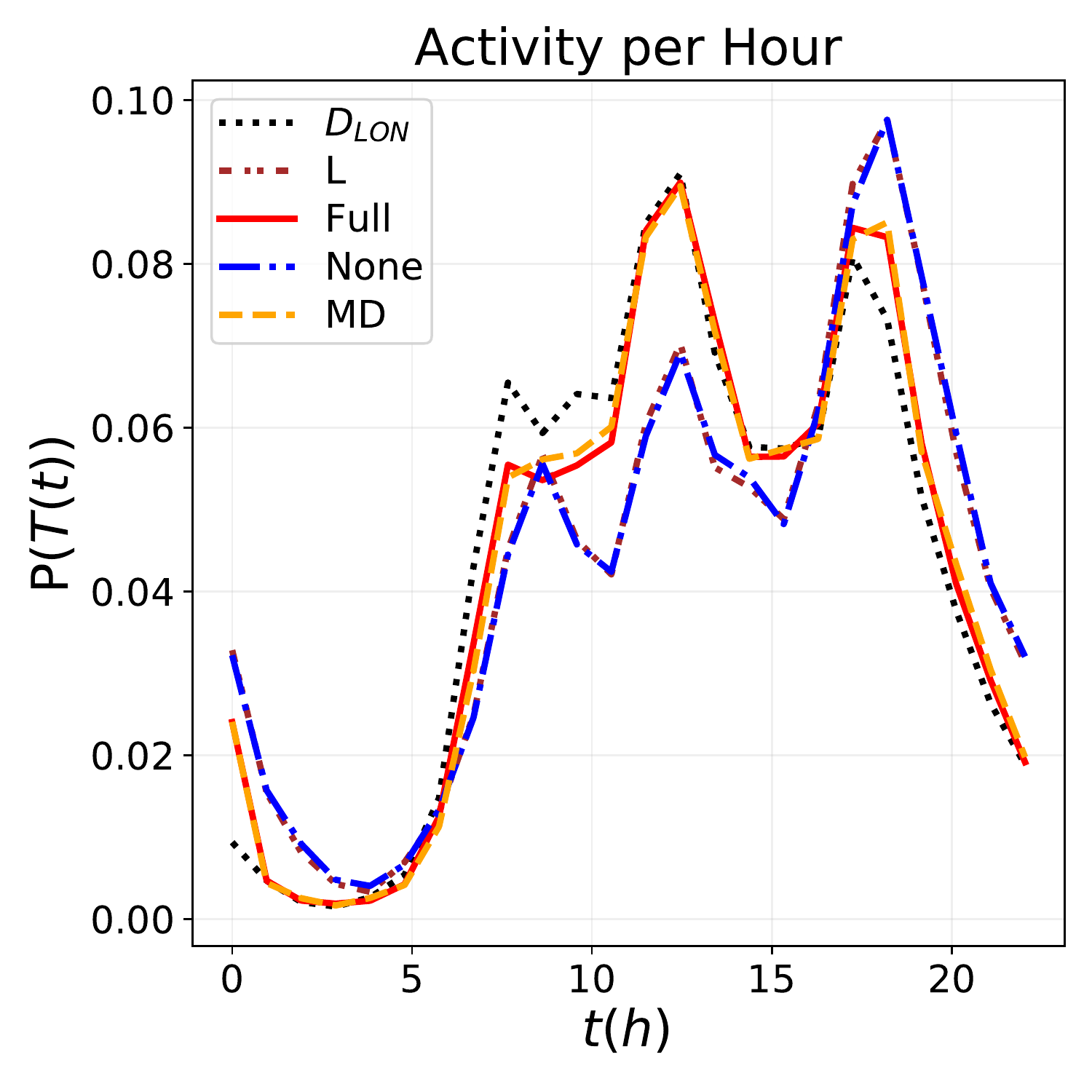}}
\hspace{2mm}
    \subfigure[]{\label{fig:unc_london}
\includegraphics[width=0.45\textwidth]{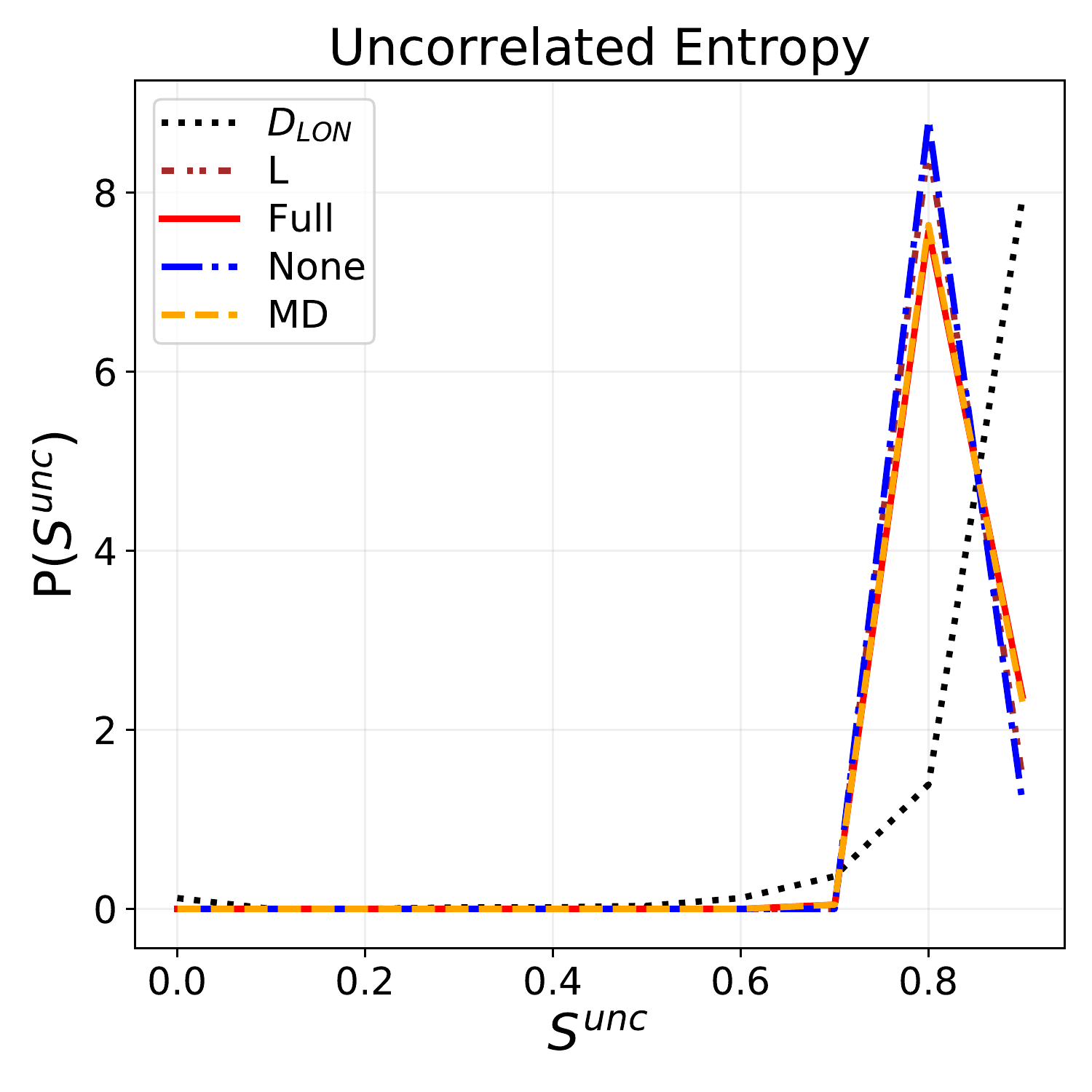}}
\caption[The probability density function for the activity per hour measure and predictability for the real and synthetic trajectories relative to London]{The distribution of the activity per hour measure (a) where is evident the different routine from individuals of New York City and London. Figure (b) shows that the predictability of the agents does not change significantly according to the information known. }
\end{figure}

Figure \ref{fig:wt_london} shows how the waiting time, in the three variants presented before, is affected by the amount of real information known by the model for values $>10^5$.
From Figure \ref{fig:activity_london} is evident the different circadian rhythm between the individuals in New York City and London.
The circadian rhythm of the individuals in New York is characterized by three peaks, while for individuals in London, it is characterized by two peaks.
This can be explained due to the different socio-cultural behaviors of the two studied populations.
The predictability of the synthetic agents is not influenced by the knowledge modeled by STS-EPR (Figure \ref{fig:unc_london}). 
The mobility similarity distribution between the generated trajectories of the agents, changes according to the real information included in the model (Figure \ref{fig:mob_sim_all}); without the use of the real weighted spatial tessellation the model underestimate the mobility similarity between connected users.

\begin{figure}[!h]
\centering
    \subfigure[]{\label{fig:mob_sim_london}
\includegraphics[width=0.45\textwidth]{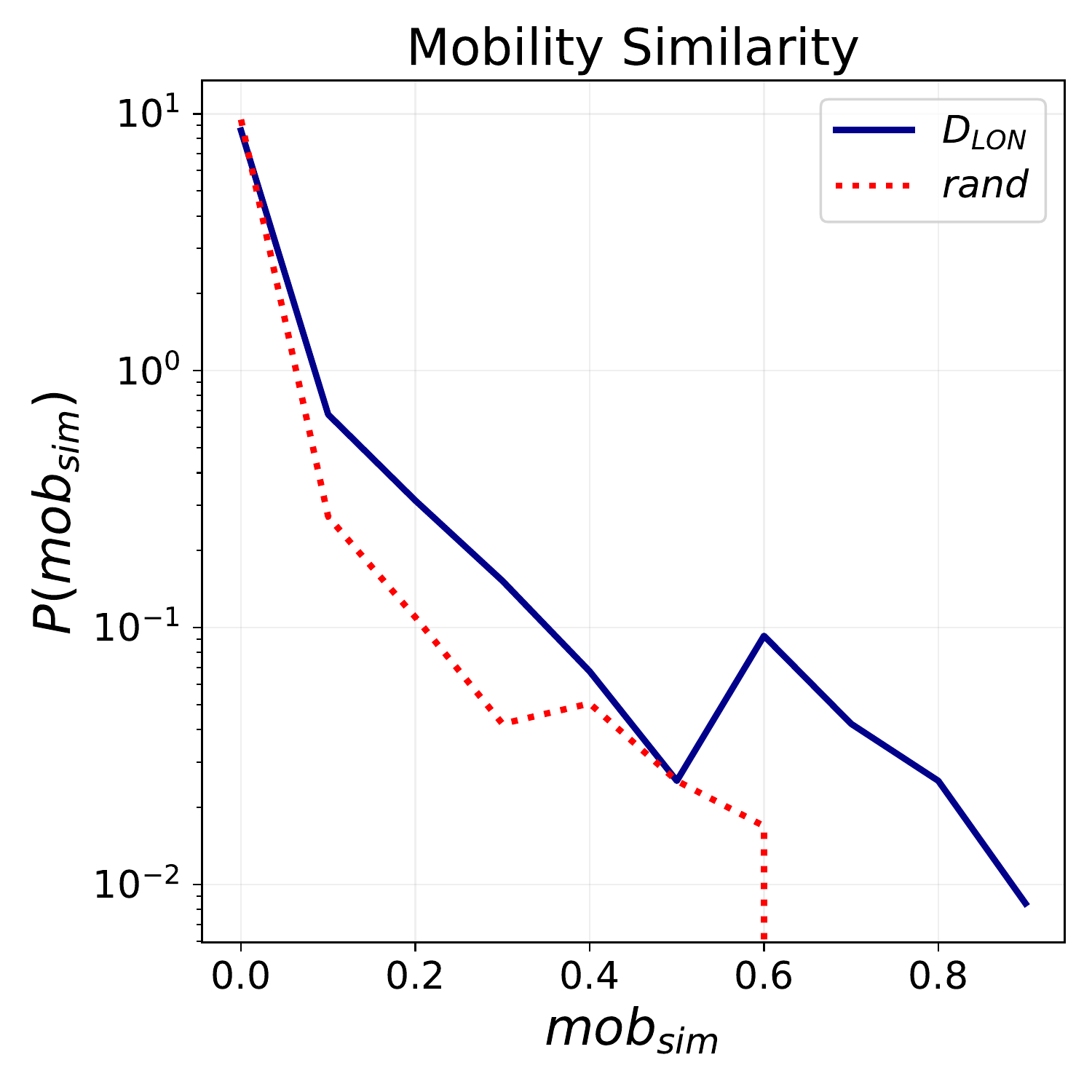}}
\hspace{2mm}
    \subfigure[]{\label{fig:mob_sim_london_all}
\includegraphics[width=0.45\textwidth]{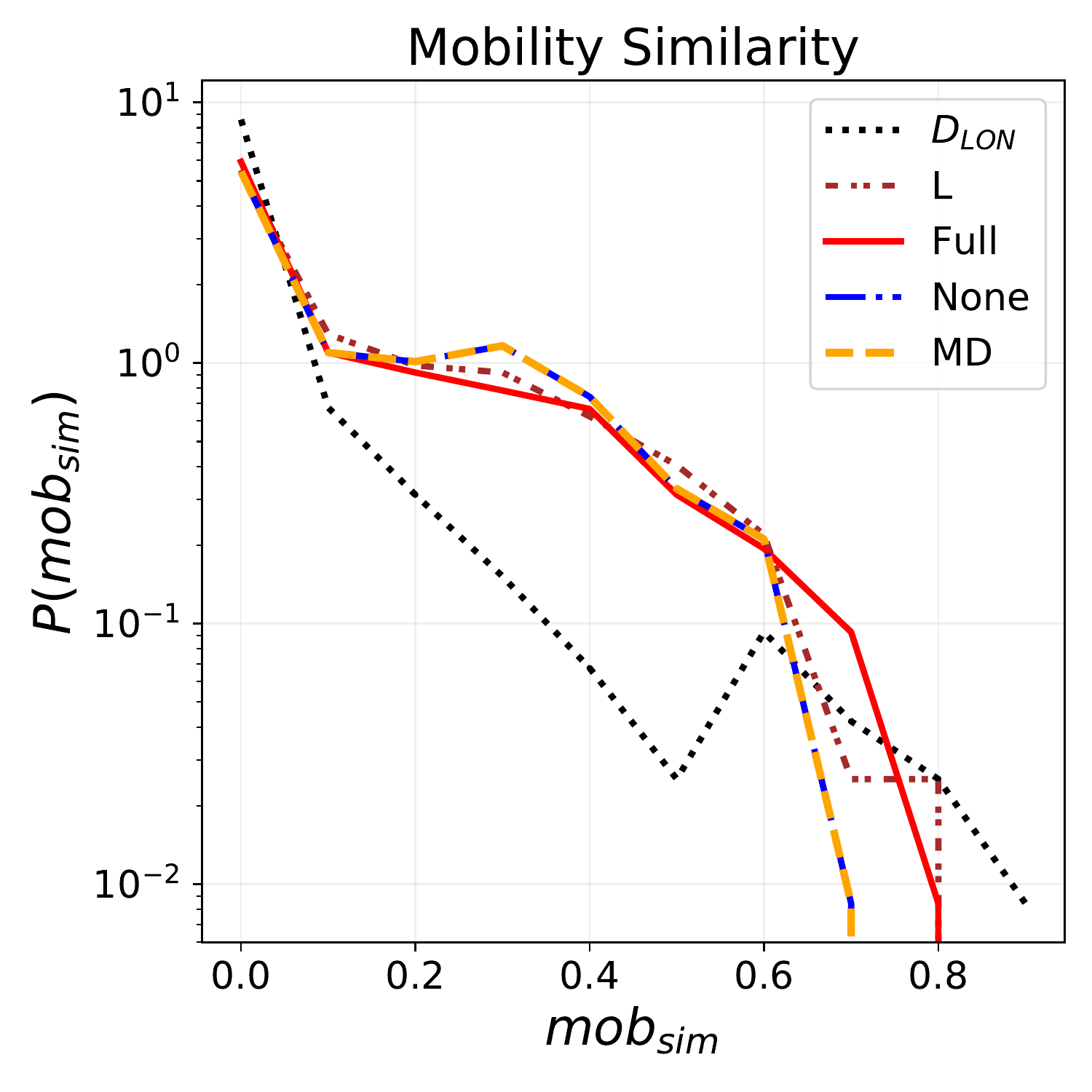}}
\caption[The probability density function of the mobility similarity for the individuals in London]{The distribution of the mobility similarity for the individuals in London considering the actual social graph and a random one (a).
The distribution of the mobility similarity for the generated trajectories; without the use of the real weighted spatial tessellation the model underestimate the mobility similarity between connected users.}
\end{figure}

These experiments demonstrate and validate the applicability of STS-EPR in urban areas beyond New York City.
It can be used in an unseen scenario using a pre-computed Mobility Diary Generator\footnote{A pre-computed Mobility Diary Generator is available at \href{https://github.com/kdd-lab/2019_Cornacchia}{https://github.com/kdd-lab/2019\_Cornacchia}} and a weighted spatial tessellation with relevance picked from the empirical truncated-power law $P(w) \approx (w)^{-\beta} e ^{-w / \lambda}$ where $\beta=1.25$ and $\lambda=104$.

%%%%%%%%%%%%%%%%%%%%%%%%%%%%%%%%%%%%%%%%%%%%%%
%%        Conclusions and Future Works      %%
%%%%%%%%%%%%%%%%%%%%%%%%%%%%%%%%%%%%%%%%%%%%%%
\section{Conclusions and Future Works}
\label{chapter:conclusions}

In this work, we develop a model that considers the social dimension and spatial and temporal dimensions during the generation of the synthetic trajectories.

Starting from GeoSim, we include three mobility mechanisms to improve its modeling ability: a mechanism that takes into account the distance from the current location and the location to explore; the relevance of a location together with the distance from the current location using a gravity-law; an algorithm that captures the tendency of individuals to follow or break their routine.
We also include other novel additional mechanisms: the RSL (Relevance-based Starting Location) principle, which considers the relevance during the assignment of the agents at their starting location; the reachable location concept, where we model the fact that an agent, associated with waiting time, can visit only the locations reachable traveling at a certain speed for the associated amount of time; the concept of the popularity of an agent at a collective level during the contact selection phase, when an agent decides to explore a new location; and we specify how to deal with borderline cases, proposing an action correction phase.

Our experiments on 1,001 individuals connected in a social graph, moving for three months in the urban area of New York City, reveal that STS-EPR can generate realistic trajectories. 
Interestingly, both the mechanisms and the tessellation granularity are crucial to producing realistic trajectories. 
We further validate the modeling ability of STS-EPR simulating the mobility of individuals moving in London, including in the model different levels of knowledge concerning their mobility behaviors. 
From the results obtained, we can conclude that the model can generate realistic mobility trajectories independently from the urban area considered, and even with a lack of crucial information.

The proposed model can be further improved in several directions.
The concept of spatial distance between the current location and the location to visit in the next displacement is considered only in the individual exploration; it can also be considered in the other cases together with the visitation pattern of the individual.
In the proposed models, the social graph is static. An interesting improvement can be to consider a dynamic social graph where the agents can create new links. 
Another consideration is that, for example, during the working hours, an individual tends to interact mainly with its colleagues. In contrast, evening activities will be influenced mostly by its family or friends \cite{opp_net}. The social relationships of an individual change with time and the social graph can be modeled as a time-varying social graph: a graph where the weight of the connections changes according to the time and the social community of the contact.
Currently, in the models, there is no representation of the urban infrastructure like urban roads. An improvement can be to consider the road network of a city and the speed limit associated with each road. In this scenario, the agent reaches its selected location traveling through the road infrastructure, respecting the speed limits associated with the roads on its path.

An interesting improvement can be to use a different weighted spatial tessellation from the squared one used during our experiments. For example, a tessellation where the tiles do not cover the same area, but rather that contains a specific amount of population.
In this way, a high population area will be partitioned in many small tiles; in contrast, a low population area will be represented with a small number of large tiles. Two libraries that allow this partition of the space in hierarchical hexagon tessellation are Uber’s H3\footnote{\href{https://eng.uber.com/h3}{https://eng.uber.com/h3}} and Google S2 Geometry\footnote{\href{https://s2geometry.io}{https://s2geometry.io}}.
From the experiments emerges that the use of a correct Mobility Diary Generator is essential to generate realistic trajectories; as shown in Section \ref{sect:london}, the circadian rhythm varies between populations to their socio-cultural tradition. 
Different factors can shape the circadian rhythm of a population. An interesting improvement is designing a model to generate a plausible Mobility Diary Generator, if it is not available, starting from information about a population, such as the working hours schedule, opening hours of activities (e.g., schools, restaurants, shops).
Also, the relevance of the locations plays a crucial role; when information about the relevance is not available, a solution besides the use of the population density is to assign a relevance according to the empirical power-law distribution presented in Section \ref{sect:london}. 
However, the heatmaps relative to the check-ins in New York City and London show that the relevant locations are clustered in space, and this can not be modeled using only the power-law distribution. A solution can be to create a model in which the relevance of a location is more likely to be high if it surrounded by relevant locations.
Artificial Intelligence techniques, such as Generative Adversarial Networks (GANs), are used to generate synthetic trajectories that follow the distribution of real mobility trajectories used as a train dataset \cite{gan}. 
GANs can be embedded into the mechanistic models to produce more realistic trajectories, capturing the aspects of human movements that can not be modeled from the mechanisms of such generative models \cite{gan}. 
This hybrid model could represent a further step forward in modeling and understanding human mobility.

%%%%%%%%%%%%%%%%%%%%%%%%%%%%%%%%%%%%%%%%%%%%%%
%%                                          %%
%% Backmatter begins here                   %%
%%                                          %%
%%%%%%%%%%%%%%%%%%%%%%%%%%%%%%%%%%%%%%%%%%%%%%

\begin{backmatter}

\section*{Competing interests}
  The authors declare that they have no competing interests.

\section*{Author's contributions}

GC implemented the code, preprocessed the data, made the experiments and the plots, and wrote the article.
LP supervised and directed the work, designed the experiments, selected the baseline models, and wrote the article.
GR suggested experiments about the social aspect of human mobility.

\section*{Acknowledgements}
 This research has been partially supported by EU project H2020-INFRAIA SoBigData++ grant agreement \#871042.
%%  After submission of the .TEX file,                     %%
%%  you will be prompted to submit your .BBL file.         %%

% if your bibliography is in bibtex format, use those commands:
\bibliographystyle{bmc-mathphys} % Style BST file (bmc-mathphys, vancouver, spbasic).
\bibliography{bmc_article}      % Bibliography file (usually '*.bib' )
% for author-year bibliography (bmc-mathphys or spbasic)
% a) write to bib file (bmc-mathphys only)
% @settings{label, options="nameyear"}

% or include bibliography directly:
% \begin{thebibliography}
% \bibitem{b1}
% \end{thebibliography}

%%%%%%%%%%%%%%%%%%%%%%%%%%%%%%%%%%%
%%                               %%
%% Figures                       %%
%%                               %%
%% NB: this is for captions and  %%
%% Titles. All graphics must be  %%
%% submitted separately and NOT  %%
%% included in the Tex document  %%
%%                               %%
%%%%%%%%%%%%%%%%%%%%%%%%%%%%%%%%%%%

%%
%% Do not use \listoffigures as most will included as separate files

%%%%%%%%%%%%%%%%%%%%%%%%%%%%%%%%%%%
%%                               %%
%% Tables                        %%
%%                               %%
%%%%%%%%%%%%%%%%%%%%%%%%%%%%%%%%%%%

%% Use of \listoftables is discouraged.
%%

%%%%%%%%%%%%%%%%%%%%%%%%%%%%%%%%%%%
%%                               %%
%% Additional Files              %%
%%                               %%
%%%%%%%%%%%%%%%%%%%%%%%%%%%%%%%%%%%
\end{backmatter}

\section*{Appendix}

\subsection*{$\mbox{GeoSim}_{d}$}
A modeling limit of GeoSim concerns the spatial patterns of the generated synthetic trajectories.
GeoSim does not take into account the distance from the current location and the location to explore \cite{geosim}, since the destination of the next move is chosen uniformly at random.
Consequently, the probability density functions for both jump size and radius of gyration of the generated trajectories do not follow the proper empirical distribution \cite{gonzalez2008understanding}.

The power-law behavior of the probability density function of the jump length suggests that individuals are more likely to move at small rather than long distances. To take into account this observation in the first extension of GeoSim, namely $\mbox{GeoSim}_{d}$, in the Exploration-Individual action an agent $a$ currently at location $r_j$, selects an unvisited location $r_i$, with $i \in exp_a$, with probability $p(r_i)\propto \frac{1}{d_{ij}}$ where $d_{ij}$ is the geographic distance between location $r_i$ and $r_j$.

\subsection*{$\mbox{GeoSim}_{gravity}$}
Individuals do not consider the distance from a place as the only discriminant factor while selecting the next location to explore.
They are driven by a preferential-exploration force in the selection of the new location to explore \cite{d_epr,ditras}.
The individuals take into account also the relevance of a location at a collective level together with the distance from their current location.
The method used for coupling both the distance and the relevance is the same used in the \textit{d}-EPR model \cite{d_epr}: the use of a gravity law.
The usage of the gravity model is justified by the accuracy of the gravity model to estimate origin-destination matrices even at the country level \cite{d_epr}.

In $\mbox{GeoSim}_{gravity}$, the second proposed extension of GeoSim, an agent $a$ currently at location $r_j$, during the Exploration-Individual action  selects an unvisited location $r_i$, with $i \in exp_a$, with probability $p(r_i)\propto \frac{w_i w_j}{d_{ij}^2}$ where $d_{ij}$ is the geographic distance between location $r_i$ and $r_j$ and $w_i$, $w_j$ represent their relevance.

The relevance of a location can be estimated through the measure visits per location (Sect. \ref{sect:measures}), using a real mobility dataset. In case the real information is not available, the relevance of a location can be estimated using the population density \cite{ditras}.

\end{document}